\documentclass[aps,prd,notitlepage,showpacs,nofootinbib,superscriptaddress]{revtex4-1}
\pdfoutput=1

\usepackage[utf8]{inputenc} 
\usepackage{graphicx}
\usepackage{tabularx}
\usepackage{amsmath}
\usepackage{amssymb}
\usepackage{float}
\usepackage{comment}
\usepackage{slashed}
\usepackage[normalem]{ulem}
\usepackage{caption}
\usepackage{subcaption}
\usepackage[usenames,dvipsnames]{color}
\usepackage[colorinlistoftodos]{todonotes}
\usepackage[colorlinks=true,citecolor=darkred,urlcolor=darkred, pdfborder={0 0 0}]{hyperref}
\usepackage[normalem]{ulem}
\usepackage[colorinlistoftodos]{todonotes}

\newcommand {\ignore}[1]{}

\usepackage{multirow}
\usepackage{subeqnarray}

\usepackage{scalefnt}

\definecolor{mightnightblue}{RGB}{25,25,112}
\definecolor{brown}{rgb}{0.59, 0.29, 0.0}
\definecolor{nicered}{rgb}{0.8,0.1,0.1}
\definecolor{nicegreen}{rgb}{0.1,0.5,0.1}
\definecolor{linkcolor}{rgb}{0,0,0.5}
\definecolor{darkred}{rgb}{0.6,0,0}

\def\gsim{\raise0.3ex\hbox{$\;>$\kern-0.75em\raise-1.1ex\hbox{$\sim\;$}}}
\def\lsim{\raise0.3ex\hbox{$\;<$\kern-0.75em\raise-1.1ex\hbox{$\sim\;$}}}

\usepackage{soul}

\def\21{$\mathrm{SU(2)_L \otimes U(1)_Y}$}

\def\cos{\rm {cos}}
\def\sin{\rm {sin}}
\def\cevns{CE$\nu$NS~}

\definecolor{vdrgreen}{rgb}{0.0, 0.7, 0.0}

\newcommand{\AddrAHEP}{%
  AHEP Group, Institut de F\'{i}sica Corpuscular --
  CSIC/Universitat de Val\`{e}ncia, Parc Cient\'ific de Paterna.\\
 C/ Catedr\'atico Jos\'e Beltr\'an, 2 E-46980 Paterna (Valencia) - Spain}

\begin{document}

\title{\boldmath \color{BrickRed} Recent probes of standard and non-standard neutrino physics with nuclei}

\author{D.K. Papoulias}\email{dipapou@ific.uv.es}\affiliation{\AddrAHEP}
\author{T.S. Kosmas}\email{hkosmas@uoi.gr}\affiliation{Division of Theoretical Physics, University of  Ioannina, GR 45110 Ioannina, Greece}
\author{Y. Kuno}\email{kuno@phys.sci.osaka-u.ac.jp}\affiliation{Department of Physics, Osaka University, 1-1 Machikaneyama, Toyonaka, Osaka 560-0043, Japan}

\begin{abstract}
We review standard and non-standard neutrino physics probes that are based on nuclear measurements. We pay special attention on the discussion of prospects to extract new physics at prominent rare event measurements looking for neutrino-nucleus scattering, such as the coherent elastic neutrino-nucleus scattering (CE$\nu$NS) that may involve lepton flavor violation (LFV) in neutral-currents (NC). For the latter processes several appreciably sensitive experiments are currently pursued or have been planed to operate in the near future, like the COHERENT, CONUS, CONNIE, MINER, TEXONO, RED100, vGEN, Ricochet, NUCLEUS etc. We provide a thorough discussion on phenomenological and theoretical studies, in particular those referring to the nuclear physics aspects in order to provide accurate predictions for the relevant experiments. Motivated by the recent discovery of CE$\nu$NS at the COHERENT experiment and the active experimental efforts for a new measurement at reactor-based experiments, we summarize the current status of the constraints as well as the future sensitivities on nuclear and electroweak physics parameters, non-standard interactions, electromagnetic neutrino properties, sterile neutrinos and simplified scenarios with novel vector $Z^\prime$ or scalar $\phi$ mediators. Indirect and direct connections of \cevns with astrophysics, direct Dark Matter detection and charge lepton flavor violating processes are also discussed.
\end{abstract}

\maketitle
\noindent

\section{Introduction}

During the last few decades, intense research effort has been devoted to multidisciplinary neutrino searches involving physics within and beyond the standard model (SM) in the theory,  phenomenology and experiments that drops in the interplay of particle, nuclear physics, astrophysics and cosmology.

Astrophysical and laboratory searches~\cite{Ejiri:2019ezh} offer unique opportunities to probe great challenges in modern-day physics such as the underlying physics of the fundamental electroweak interactions within and beyond the SM~\cite{Schechter:1980gr,Schechter:1981cv} in the neutral and charged-current sector of semi-leptonic neutrino-nucleus processes~\cite{Kosmas:1996fh,Ejiri:2000ps,Balasi:2015dba}. To meet the sufficient energy and flux requirements, the relevant studies consider different low-energy neutrino sources including (i) Supernova (SN) neutrinos, (ii) accelerator neutrinos (from pion decay at rest, $\pi$-DAR) and (iii) reactor neutrinos, while interesting proposals aiming to use $^{51}$Cr and beta-beam neutrino sources have appeared recently. The detection mechanism of low-energy neutrino interactions with nucleons and nuclei is experimentally hard and limited by the tiny nuclear recoils produced by the scattering process. To this purpose, the nuclear detector materials are carefully selected to fulfill the requirement of achieving a-few-keV or sub-keV threshold capabilities. The detectors developed are based on cutting edge technologies such as scintillating crystals (CsI[Na],  NaI[Tl]), p-type point-contact (PPC) germanium detectors, single-phase or double liquid noble gases (LAr, LXe) charged coupled devices (CCDs), cryogenic bolometers, etc.

The neutral-current coherent elastic neutrino nucleus scattering (CE$\nu$NS) was proposed about four decades ago~\cite{Freedman:1973yd,Tubbs:1975jx,Drukier:1983gj}, while it was experimentally confirmed in 2017 by the COHERENT Collaboration~\cite{Akimov:2017ade} at the Spallation Neutron Source, in good agreement with the SM expectation. The observation of \cevns has opened up a new era, triggering numerous theoretical studies to interpret the available data~\cite{Akimov:2018vzs} in a wide spectrum of new physics opportunities, with phenomenological impact on  astroparticle physics, neutrino oscillations,  dark matter (DM) detection, etc. (see Ref.~\cite{AristizabalSierra:2019hcm} for various applications). In particular, the recent works have concentrated on non-standard interactions (NSI)~\cite{Liao:2017uzy,Dent:2017mpr,AristizabalSierra:2017joc,Denton:2018xmq,Dutta:2019eml,Coloma:2017ncl,Gonzalez-Garcia:2018dep}, electromagnetic properties~\cite{Kosmas:2015sqa,Kosmas:2017tsq,Miranda:2019wdy,Parada:2019gvy}, sterile neutrinos~\cite{Kosmas:2017zbh,Canas:2017umu,Blanco:2019vyp}, CP-violation~\cite{AristizabalSierra:2019ufd} and novel mediators~\cite{Dent:2016wcr,Farzan:2018gtr,Abdullah:2018ykz,Brdar:2018qqj}. Nuclear and atomic effects are explored in Refs.~\cite{Cadeddu:2017etk,Ciuffoli:2018qem,Huang:2019ene,AristizabalSierra:2019zmy,Papoulias:2019lfi,Arcadi:2019uif,Cadeddu:2019qmv}  which may have direct implications to the neutrino-floor~\cite{Papoulias:2018uzy,Boehm:2018sux,Link:2019pbm} and to DM searches~\cite{Ge:2017mcq,Ng:2017aur, Dutta:2019nbn}. Being a rapidly developing field, there are several experimental programs aiming to observe \cevns in the near future, such as the TEXONO~\cite{Wong:2010zzc}, CONNIE~\cite{Aguilar-Arevalo:2016qen}, MINER~\cite{Agnolet:2016zir}, vGEN~\cite{Belov:2015ufh}, CONUS~\cite{conus}, Ricochet~\cite{Billard:2016giu} and NUCLEUS~\cite{Strauss:2017cuu}. 

Future \cevns measurements have good prospects to shed light on the exotic neutrino-nucleus interactions expected in the context of models describing flavor changing neutral-current (FCNC) processes~\cite{Barranco:2005ps} as well as to subleading NSI oscillation effects~\cite{Friedland:2004ah,Friedland:2004pp,Friedland:2005vy,Miranda:2015dra,Farzan:2017xzy} and various open issues in nuclear astrophysics~\cite{Amanik:2004vm,EstebanPretel:2009is}. The main goal of this review article is to provide an up-to-date status of the conventional and exotic neutrino physics probes of \cevns and to summarize the necessary aspects for the interpretation of the experimental data. We focus on the theoretical modelling, calculations and analysis of the data that are relevant at the time of writing and we mainly concentrate on the theoretical and phenomenological physics aspects. For a recent review on the experimental advances of CE$\nu$NS, see Ref.~\cite{Akimov:2019wtg}. 

This review article has been organized as follows: Sect.~\ref{sect:theory} provides the theoretical treatment of low-energy neutrino-nucleus processes for both coherent and incoherent channels and its connection to the more general lepton-nucleus case with a particular emphasis on the nuclear physics aspects. Sect.~\ref{sect:constraints} presents the current status of constraints on SM and exotic physics parameters resulted from the analysis of the COHERENT data and discusses the projected sensitivities from future \cevns measurements at $\pi$-DAR and reactor facilities. In Sect.~\ref{sect:connection} we briefly summarize the most important connections of \cevns with DM searches, charged lepton flavor violation (cLFV) and astrophysics. Finally, the main conclusions are given in Sect.~\ref{sect:conclusions}. 

\section{Theoretical study of neutrino-nucleus interaction}
\label{sect:theory}

At low- and intermediate-energies, the neutrino being a key input to understand open issues in physics within and beyond the SM (see below), necessitated a generation of neutrino experiments for exploring neutrino scattering processes with nucleons and nuclei for both charged-current (inelastic) and neutral-current (coherent elastic and incoherent scattering)  processes. Theoretically, the neutral-current neutrino-nucleus scattering we are interested here, is a well studied process for both coherent~\cite{Papoulias:2015vxa} an incoherent channels~\cite{Bednyakov:2018mjd,Bednyakov:2019dbl}. The accurate evaluation of the required transition matrix elements describing the various interaction channels of the electroweak processes between an initial and a final (many-body) nuclear state, is obtained on the basis of reliable nuclear wavefunctions. From a nuclear theory point of view, such results have been obtained by paying special attention on the accurate contruction of the nuclear ground state in the framework of the quasi-particle random phase approximation (QRPA), using schematic Skyrme \cite{Almosly:2019han} or realistic Bonn C-D pairing interactions~\cite{Chasioti:2009fby}. Focusing on the latter method, the authors of Ref.~\cite{Papoulias:2015vxa} solved iteratively the Bardeen-Cooper-Schrieffer (BCS) equations, achieving a high reproducibility of the available nuclear charge-density-distribution experimental data~\cite{DeJager:1987qc}.

\subsection{Coherent and incoherent neutrino-nucleus cross sections}

In the Donnelly-Walecka theory~\cite{Donnelly:1976fs} all semi-leptonic nuclear processes at low and intermediate energies may be described by an effective interaction Hamiltonian through the leptonic $j_{\mu}^{\text{lept}}$ and hadronic $\mathcal{J}_{\mu}$ current densities,  
\begin{equation}
\hat{H}_{eff}   = \frac{G }{\sqrt{2}} \int
\hat{\ell}_{\mu}^{\mathrm{lept}} ({\bf x}) \hat{\mathcal{J}}^{\mu} ({\bf x}) \, d^3 {\bf x} \, ,
\label{hamiltonian2}
\end{equation}
where $G = G_F$ is the Fermi coupling constant for neutral-current processes and $G = G_F \cos \theta_c$ ($\theta_c$ is the Cabbibo angle) for charged-current  processes. For partial scattering rates, the evaluation of the transition amplitudes $\langle f\vert \hat{H}_{eff} \vert i\rangle$  are treated via a multipole decomposition analysis of the hadronic current (see the Appendix~\ref{app:multiple-operators}). Then, for a given set of an initial $\vert J_{i} \rangle$ and a final $\vert J_{f} \rangle$ nuclear state, the double differential SM cross section becomes~\cite{Donnelly:1978tz}
\begin{equation}
\frac{d^2 \sigma_{i \rightarrow f}}{d \Omega\, d \omega}= \frac{G^{2}}{\pi} F(Z,\varepsilon_f) \frac{\vert \mathbf{k}_f \vert \varepsilon_{f}}{(2 J_{i}+1)}\left(\sum_{J=0}^{\infty} \sigma_{\mathrm{CL}}^{J} + 
\sum_{J=1}^{\infty} \sigma_{\mathrm{T}}^{J} \right),
\label{don_val_crossec}
\end{equation}
with $\varepsilon_{f}$ ($\vert \mathbf{k}_f \vert$) denoting the final energy (momentum) of the outgoing lepton, while $\omega = \varepsilon_i - \varepsilon_f$ stands for the excitation energy of the nucleus where $\varepsilon_i$ is the initial lepton energy. For charged-current processes, the Fermi function $F(Z,\varepsilon_f)$, takes into account the final state interaction of the outgoing charged particle, while for neutral-current processes such as coherent and incoherent neutrino-nucleus scattering it is $F(Z,\varepsilon_f) =1$.

The individual cross sections in Eq.(\ref{don_val_crossec}) receive contributions from the so-called Coulomb $\hat{\mathcal{M}}$, longitudinal $\hat{\mathcal{L}}$, transverse electric $\hat{\mathcal{T}^{el}}$ and transverse magnetic $\hat{\mathcal{T}^{mag}}$ operators for both vector and axial vector components (see the Appendix~\ref{app:multiple-operators}). The cross sections $\sigma_{\mathrm{CL}}^{J}$ and $\sigma_{\mathrm{T}}^{J}$ are expressed in terms of the reduced matrix elements of the eight basic irreducible tensor operators~\cite{Donnelly:1976fs}
\begin{equation}
\begin{aligned}
\sigma_{\mathrm{CL}}^{J} = & (1+a \cos \theta) \vert \langle J_f \vert \vert \hat{\mathcal{M}}_J (\kappa) \vert \vert J_i \rangle \vert^2 
\\
& + (1+a \cos \theta -2 b \sin^2 \theta) \vert \langle J_f \vert \vert \hat{\mathcal{L}}_J (\kappa) \vert \vert J_i \rangle \vert^2 
\\
& + \left[ \frac{\omega}{\kappa} (1+a \cos \theta) + d \right] 2 \Re e \vert \langle J_f \vert \vert \hat{\mathcal{L}}_J (\kappa) \vert \vert J_i \rangle \vert  \vert \langle J_f \vert \vert \hat{\mathcal{M}}_J (\kappa) \vert \vert J_i \rangle \vert^\ast \, , 
\end{aligned}
\end{equation}
\begin{equation}
\begin{aligned}
\sigma_{\mathrm{T}}^{J} = & (1-a \cos \theta + b \sin^2 \theta) \left[ \vert \langle J_f \vert \vert \hat{\mathcal{T}}^{mag}_J (\kappa)  \vert \vert J_i \rangle \vert^2 +  \vert \langle J_f \vert \vert \hat{\mathcal{T}}^{el}_J (\kappa) \vert \vert J_i \rangle \vert^2 \right]
\\
& \mp \left[ \frac{(\varepsilon_i + \varepsilon_f)}{\kappa}  (1-a \cos \theta) - d \right] 2 \Re e \vert \langle J_f \vert \vert \hat{\mathcal{T}}^{mag}_J (\kappa) \vert \vert J_i \rangle \vert  \vert \langle J_f \vert \vert \hat{\mathcal{T}}^{el}_J (\kappa) \vert \vert J_i \rangle \vert^\ast \, .
\end{aligned}
\end{equation}
Here, the $+$ ($-$) sign refers to neutrino (antineutrino) scattering and $\theta$ represents the scattering angle, while the parameters $a$, $b$, $d$ are expressed as
\begin{equation}
a = \frac{\vert \mathbf{k}_f \vert }{\varepsilon_f} = \sqrt{1 - \left( \frac{m_f}{\varepsilon_f} \right)^2}, \quad b = \frac{\varepsilon_i \varepsilon_f a^2 }{\kappa^2}, \quad d = \frac{m_f^2}{\kappa \varepsilon_f} \, .
\end{equation} 
The 4-momentum transfer is trivially obtained from the kinematics of the process and in natural units reads
\begin{equation}
q^2 \equiv q_\mu q^\mu = q_0^2 - \mathbf{q}^2 \,,
\end{equation}
while for later convenience the magnitude of the 3-momentum transfer is defined as
\begin{equation}
\kappa = \vert \mathbf{q} \vert \equiv \vert \vec{q} \vert = \left[ \omega^2 + 2 \varepsilon_i \varepsilon_f (1-a \cos \theta) - m_f^2  \right]^{1/2} \, .
\label{magnitude_3-momentum}
\end{equation}
%

%
\begin{figure}[t]
\centering
\includegraphics[width=0.5 \textwidth]{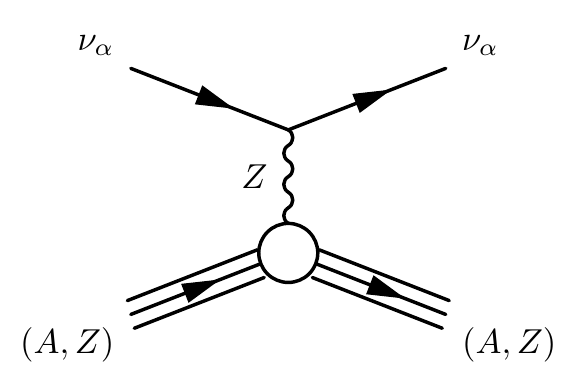}   
\caption{Feynman diagram illustrating the tree-level SM \cevns process.}
\label{fig.Feynman:CENNS}
\end{figure}
%

For sufficiently small momentum transfer, i.e. $q\leq 1/R$ where $R$ is the inverse nuclear radius~\footnote{Typically 25-150 MeV for most nuclei.}, \cevns dominates (see Fig.~\ref{fig.Feynman:CENNS}). In this case, only ground state to ground state ($g.s. \rightarrow g.s.$) transitions occur and lead to the following simplifications: the kinematics of the reaction imply $m_f =0$ and $\vert \mathbf{k}_f \vert = \varepsilon_f$ so that $a=1$ and $d=0$, while the momentum transfer can be cast in terms of the incoming neutrino energy $E_{\nu}$ in the simple form
\begin{equation}
Q^2 = -q^2 = 4 E^{2}_{\nu} \,  \sin^{2} \frac{\theta}{2}\, ,
\label{mom_transf}
\end{equation}
where the usual notation $\varepsilon_i = \varepsilon_f \equiv E_\nu$ has been adopted. Note also that the excitation energy in this case is $\omega = 0$ and  $\kappa = \sqrt{-q^2} = \sqrt{Q^2}$, while angular momentum conservation implies that for \cevns processes the only non-vanishing operator is the Coulomb, $T_{1}^{0}\equiv\hat{\mathcal{M}}_{0}^{0}$ (see the Appendix~\ref{app:multiple-operators} for the definition of the operators $T_i^{J}$). Then the corresponding differential cross section is further simplified and takes the form
\begin{equation}
\left(\frac{d \sigma}{d \cos \theta}\right)_\text{SM}= \frac{G_{F}^{2}}{2 \pi} E_{\nu}^{2} \left(1 + \cos \theta \right) \vert \langle g.s. \vert \vert \hat{\mathcal{M}}_0^0(Q) \vert \vert g.s. \rangle \vert^{2},
\label{dsdcos:BCS}
\end{equation}
where the matrix element for $g.s. \rightarrow g.s.$ transitions is explicitly written in terms of the nuclear form factors for protons $F_p(Q^2)$ and neutrons $F_n(Q^2)$, as
\begin{equation}
\langle g.s. \vert \vert \hat{\mathcal{M}}_0^0(Q) \vert \vert g.s. \rangle=\frac{1}{2} \left[\left(1-4 \sin^{2}\theta_{W}\right)Z\,F_{p}(Q^2)-N\,F_{n}(Q^2)\right]\, .
\label{Coul-ME}
\end{equation}
At \cevns experiments the detection mechanism is sensitive to the tiny nuclear recoils generated in the aftermath of the scattering process. It is therefore reasonable to express the differential cross section with respect to the nuclear recoil energy $T_N$, which in the low energy approximation $T_N \ll E_{\nu}$, reads
\begin{equation}
\left(\frac{d \sigma}{dT_N}\right)_\text{SM}=\frac{G_{F}^{2}\,M}{4 \pi} \left(1- \frac{M\,T_N}{2 E_{\nu}^2} \right) \vert \langle g.s. \vert \vert \hat{\mathcal{M}}_0^0(Q) \vert \vert g.s. \rangle \vert^{2}\, ,
\label{dsdT:BCS}
\end{equation}
where $T_N=Q^{2}/2M $ and $M$ is the mass of the nuclear isotope. The calculations of Ref.~\cite{Papoulias:2015vxa} involved the BCS form factors for protons (neutrons)
\begin{equation}
F_{N_{n}}=\frac{1}{N_{n}}\sum_{j} \sqrt{2j+1} \langle g.s. \vert \vert j_{0}(\kappa r) \vert \vert g.s. \rangle \left(v^{j}_{p(n)}\right)^{2} \, ,
\label{eq:BCS}
\end{equation}
with $N_n=Z$ or $N$ and $v^{j}_{p(n)}$ represents the  occupation probability amplitude of the $j$-th single nucleon level.

The method described above, involves realistic nuclear structure calculations making it more reliable compared to the use of phenomenological form factors, especially for accelerator neutrino sources (see the discussion in Subsect.~\ref{other-methods}). For the reader's convenience, Eq.(\ref{dsdT:BCS}) is also expressed through the vector weak nuclear charge $\mathcal{Q}_W^V$ in the approximation of equal proton and neutron form factors, as~\cite{Lindner:2016wff}
\begin{equation}
\left(\frac{d \sigma}{dT_N}\right)_\text{SM}=\frac{G_{F}^{2}\,M}{\pi} (\mathcal{Q}_W^V)^2 \left(1- \frac{M\,T_N}{2 E_{\nu}^2} \right) F(Q)^{2}\, ,
\label{dsdT:druckier}
\end{equation}
where the vector $\mathcal{Q}_W^V$ weak charge is given by~\cite{Barranco:2005yy} 
\begin{equation}
\begin{aligned}
\mathcal{Q}_W^V =  \left[  2(g_{u}^{L} + g_{u}^{R}) + (g_{d}^{L} + g_{d}^{R}) \right] Z  + \left[ (g_{u}^{L} + g_{u}^{R}) +2(g_{d}^{L} + g_{d}^{R}) \right] N   \, , 
\end{aligned}
\label{eq:weak-charge}
\end{equation}
with the left- and right-handed couplings of $u$ and $d$ quarks to the
$Z$-boson being
\begin{equation}
\begin{aligned}
g_{u}^{L} =& \rho_{\nu N}^{NC} \left( \frac{1}{2}-\frac{2}{3} \hat{\kappa}_{\nu N} \hat{s}^2_Z \right) + \lambda^{u,L} \, ,\\
g_{d}^{L} =& \rho_{\nu N}^{NC} \left( -\frac{1}{2}+\frac{1}{3} \hat{\kappa}_{\nu N} \hat{s}^2_Z \right) + \lambda^{d,L} \, ,\\
g_{u}^{R} =& \rho_{\nu N}^{NC} \left(-\frac{2}{3} \hat{\kappa}_{\nu N} \hat{s}^2_Z \right) + \lambda^{u,R} \, ,\\
g_{d}^{R} =& \rho_{\nu N}^{NC} \left(\frac{1}{3} \hat{\kappa}_{\nu N} \hat{s}^2_Z \right) + \lambda^{d,R} \, .
\end{aligned}
\end{equation}
The latter expressions include the radiative corrections from the PDG~\cite{Tanabashi:2018oca}: $\rho_{\nu N}^{NC} = 1.0082$, $\hat{\kappa}_{\nu N} = 0.9972$, $\lambda^{u,L} = -0.0031$, $\lambda^{d,L} = -0.0025$ and $\lambda^{d,R} =2\lambda^{u,R} = 3.7 \times 10^{-5}$ while concerning the weak mixing-angle the adopted value is $\hat{s}_Z^2 \equiv \sin^2 \theta_W= 0.2382$. Regarding the incoherent neutrino-nucleus cross section and for the 
sake of completeness we note that apart from the Donnelly-Walecka method given in Eq.(\ref{don_val_crossec}) a usefull formalism has been recently given in Ref.~\cite{Bednyakov:2018mjd}. 

\begin{figure}[h]

\begin{minipage}{\textwidth}
\includegraphics[width=0.49\textwidth]{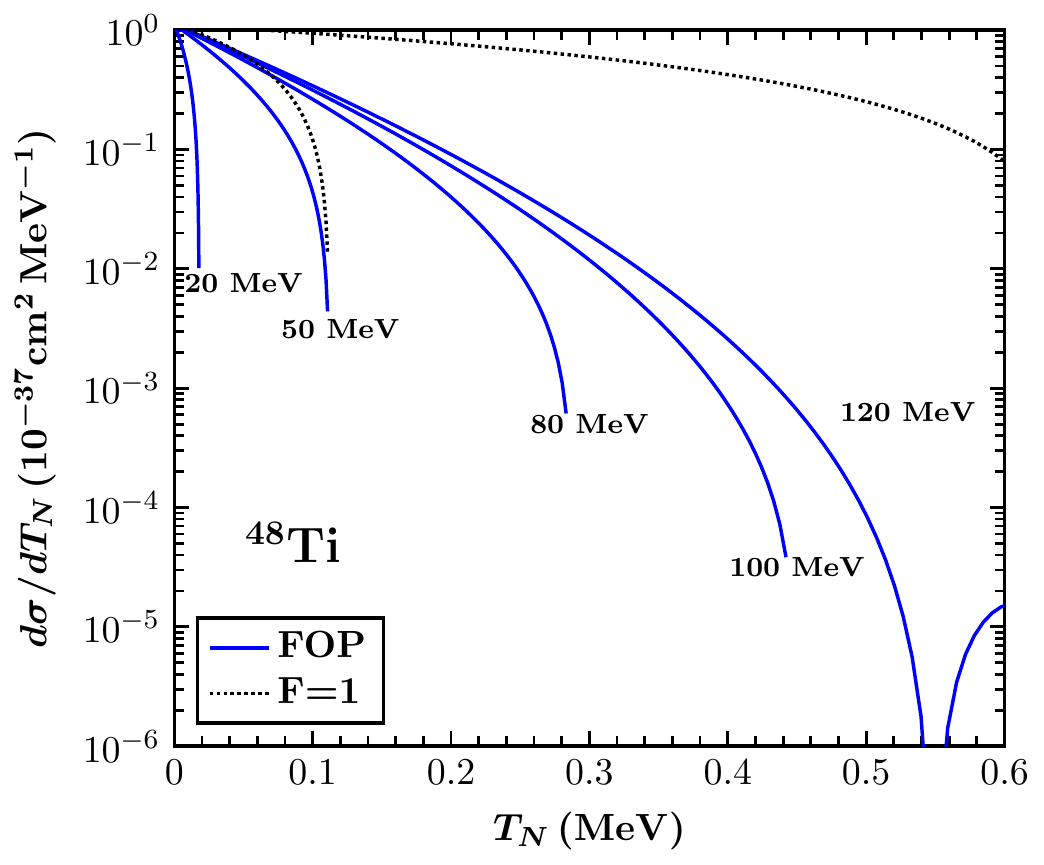} 
\includegraphics[width=0.49\textwidth]{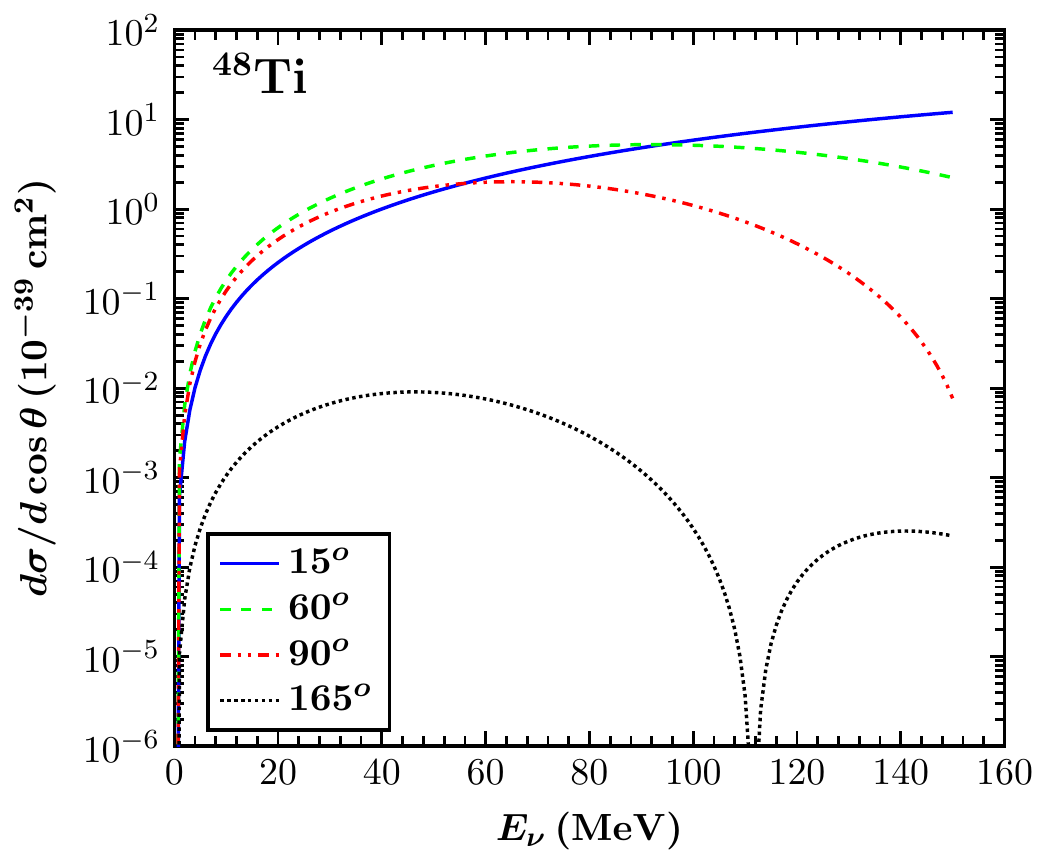}
\end{minipage}
\begin{minipage}{\textwidth}
\includegraphics[width=0.49\textwidth]{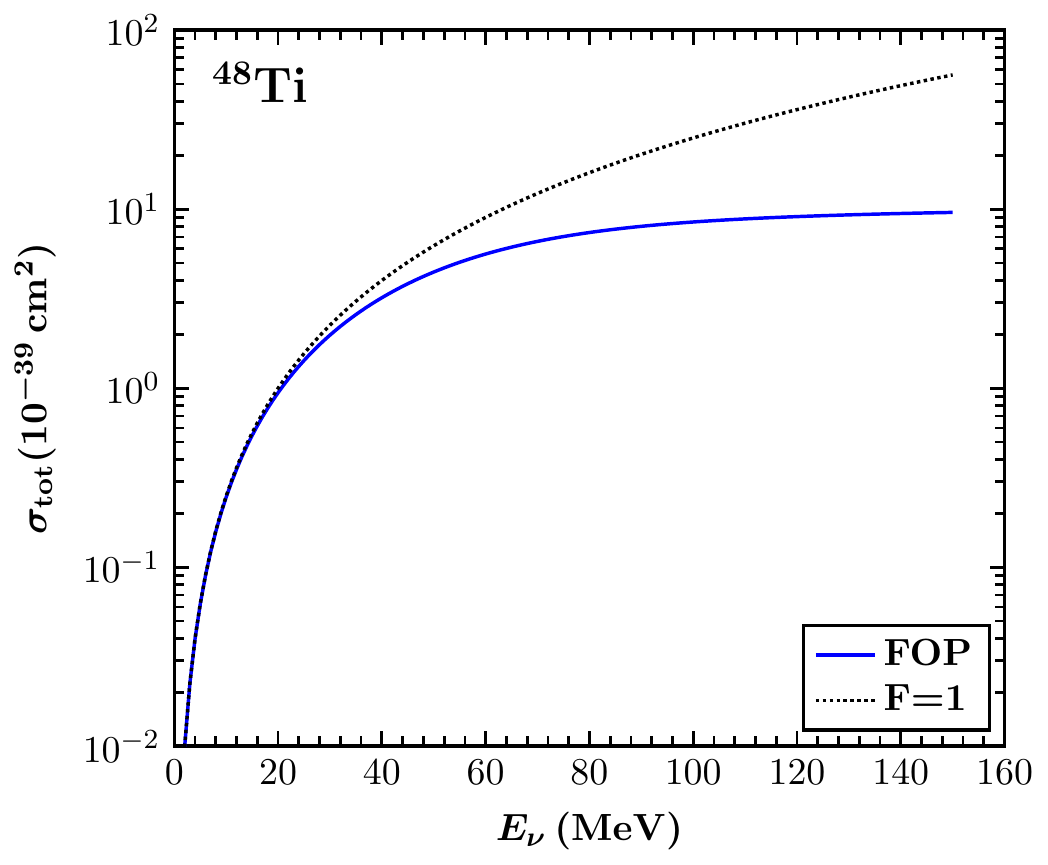}
\caption{Upper Left: The differential cross section $d\sigma /dT_N$ in terms of the nuclear recoil energy $T_N$ for differenct incident neutrino energies. The results are compared to the case of point-like nucleus ($F=1$) for $E_{\nu}=50$ and $E_{\nu}=120$ MeV.  Upper Right: The differential cross section $d\sigma /d \cos \theta$ as a function of the incoming neutrino energy $E_{\nu}$, for different scattering angles (for backward scattering the cross section vanishes). Bottom: The \cevns cross section $\sigma_{\mathrm{tot}}$  as a function of the neutrino energy. An asymptotic behavior is found at neutrino energies $E_{\nu} \geq 80$ MeV or higher. Taken from Ref.~\cite{Papoulias:phd}.}
\label{HNPS:fig.1}
\end{minipage}
\end{figure}

The differential cross sections $d \sigma/dT_N$ and $d \sigma /d \cos \theta$ are shown in the upper left and upper right panel of Fig.~\ref{HNPS:fig.1}, from where it can be seen that large differences appear if the form factor dependence is neglected. On the other hand at low neutrino energies, i.e. $E_{\nu}\leq 20 \,\mathrm{MeV}$ (relevant for reactor and solar neutrinos), the agreement of these two approximations is rather good. It is worth mentioning that forward scattering ($\theta=0$) leads to maximum $d\sigma /d \cos \theta$, as well as that for this particular case the form factor is by definition equal to unity due to the zero momentum transfer, see Eq.(\ref{mom_transf}). Finally  the bottom panel illustrates a comparison of the \cevns cross section by incorporating the nuclear form factors and by assuming $F=1$.

%

\subsection{Theoretical methods for obtaining the nuclear form factors}
\label{other-methods}

Electron scattering data provide high precision measurements of the proton charge density distribution~\cite{Angeli:2013epw}. The absence of similar data for neutron densities, restricts us to rely on the approximation of $\rho_p(\mathbf{r}) = \rho_n(\mathbf{r})$ and thus assume $F_p(Q)= F_n(Q) \equiv F(Q)$ [see Eq.(\ref{dsdT:druckier})]. In the context of nuclear theory, it is possible to treat separately the proton and neutron nuclear form factors by employing non-trivial techniques. The most reliable methods for this purpose include the large-scale Shell-Model~\cite{Kortelainen:2006rd,Toivanen:2009zza}, the QRPA~\cite{Papoulias:2013gha}, Microscopic Quasiparticle Phonon Model (MQPM)~\cite{Pirinen:2018gsd}, the deformed Shell-Model (DSM) method~\cite{Papoulias:2018uzy} and others. Recently, crucial information on important nuclear parameters has been extracted from the analysis of the recent COHERENT data in Refs.~\cite{Cadeddu:2017etk,Ciuffoli:2018qem,Huang:2019ene,Papoulias:2019txv}.

The point-nucleon charge density distribution $\rho(\mathbf{r})$, is defined as the expectation value of the density operator~\cite{Kosmas:1992yxv}
\begin{equation}
\hat{\rho}(\mathbf{r}) = \sum_{j=1}^{A} \frac{1}{2} (1 \pm \tau_{3j}) \delta(\mathbf{r} - \mathbf{r}_j)\, ,
\label{point-nucleon_charge_density}
\end{equation}
where the $+$ ($-$) sign refers to the point-proton (neutron) charge density distribution. Assuming the nuclear ground state to be approximately described by a Slater determinant constructed from single-particle wavefunctions, the distributions of Eq.(\ref{point-nucleon_charge_density}) are given by summing in quadrature the point-nucleon wavefunctions. According to Ref.~\cite{Kosmas:1992yxv}, for closed (sub)shell nuclei the charge density distribution is assumed to be spherically symmetric while the interesting radial component ($r = \vert \mathbf{r} \vert$)  of the proton charge density distribution, $\rho_p(r)$, can be cast in the form
\begin{equation}
\rho_p(r) = \frac{1}{4 \pi} \sum_{\begin{subarray}{c}  (n,l)j \\ \text{occupied} \end{subarray}} (2j+1) \vert R_{nlj}(r) \vert^2 \, ,
\end{equation}
where $R_{nlj}(r)$ denotes the radial component of the single-particle wavefunction with quantum numbers $n$, $l$ and $j$. The nuclear form factor depends on the three momentum transfer squared $\mathbf{q}^2 \equiv \vert \mathbf{q} \vert^2$ and can be obtained via a Fourier transformation
\begin{equation}
F_{p\, (n)}(\mathbf{q}^2)=\frac{4 \pi}{N_n}\int \rho_{p \, (n)}(r) j_{0}(\vert \mathbf{q} \vert r)\, r^2 \, dr \, , \quad N_n = Z \, \, \text{or} \, \,  N
\label{definition-ff}
\end{equation}
where $j_{0}(x) = \sin x /x$ denotes the zero-order Spherical Bessel function of first kind. The nuclear form factors lead to a suppression of the \cevns cross section and subsequently to a suppression of the expected event rates (see Ref.~\cite{Cadeddu:2017etk} for a comparison with the COHERENT data). The uncertainties of the nuclear form factors are explored in Ref.~\cite{AristizabalSierra:2019zmy} where it is pointed out that studies looking for physics beyond the SM can be seriously affected by the uncertainty of the neutron form factor~\cite{Amanik:2009zz}. It is therefore important to treat with special care the nuclear form factors since new physics could be claimed or missed, if their uncertainties are not properly taken into account. In addition to the form factors obtained in the  framework of the nuclear BCS method of Eq.(\ref{eq:BCS}), below we present a summary of various form factor approximations widely considered in the literature\\

i) \emph{Form factors from available electron-scattering experimental data}\\
The proton nuclear form factors $F_{p}(\mathbf{q}^2)$, may  be evaluated through a model independent analysis (e.g. by employing a Fourier-Bessel expansion) of the electron scattering data~\cite{DeJager:1987qc}, having however the disadvantage of assuming  $F_{p}(\mathbf{q}^2) = F_{n}(\mathbf{q}^2)$.   \\ 
%

ii) \emph{Fractional occupation probabilities (FOP) in a simple Shell-Model}\\
For Harmonic Oscillator (h.o.) wavefunctions the nuclear form factor $F_p(\mathbf{q}^2)$ for protons can be expressed in polynomial form~\cite{Kosmas:1988tz,Kosmas:1990tc}
\begin{equation}
F_{p}(\mathbf{q}^2)=\frac{1}{Z} e^{-(\vert \mathbf{q} \vert\,b)^{2}/4} \Phi \left(\vert \mathbf{q} \vert \,b,Z \right), \qquad \Phi\left(\vert \mathbf{q} \vert \,b,Z \right)= \sum_{\lambda=0}^{N_{\mathrm{max}}} \theta_{\lambda} (\vert \mathbf{q} \vert \,b)^{2 \lambda} \, ,
\label{forfac}
\end{equation}
with $N_{\mathrm{max}}=(2n+ l)_{\mathrm{max}}$ denoting for the number of quanta of the highest occupied proton (neutron) level. In a similar manner, the radial nuclear charge density distribution $\rho_p(r)$  is written in terms of the polynomials $\Pi\left(r/b,Z\right)$ in the following compact form~\cite{Kosmas:1988tz,Kosmas:1990tc}
\begin{equation}
\rho_{p}(r)=\frac{1}{\pi^{3/2} b^{3}}e^{-(r/b)^{2}} \, \Pi\left(\frac{r}{b},Z\right), \qquad \Pi\left(\chi,Z\right)=\sum_{\lambda=0}^{N_{\mathrm{max}}} f_{\lambda} \chi^{2 \lambda},
\label{rho(r)}
\end{equation}
with the definition $\chi=r/b$ ($b$ stands for the h.o. size parameter). The explicit expressions for calculating the coefficients $\theta_\lambda$ and  $f_\lambda$ are given in the Appendix~\ref{app:coeff-FOP}.
\begin{figure}[t]
\centering
\includegraphics[width=0.8\textwidth]{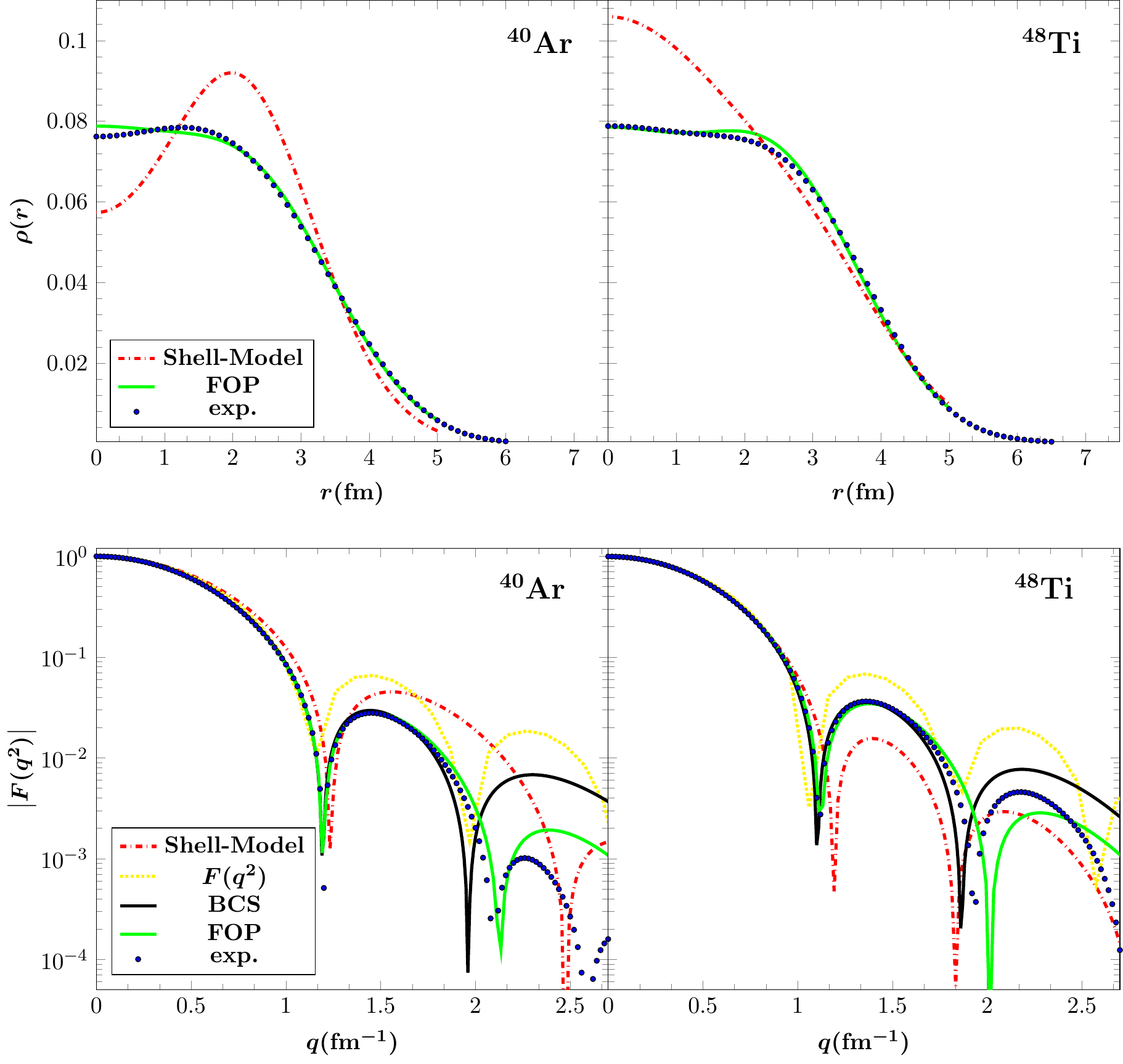}
\caption{Charge density distribution (top) and  nuclear form factor (bottom). The results refer to  the $^{40}$Ar and $^{48}$Ti isotopes and are compared for different nuclear methods. Figure adapted from Ref.~\cite{Papoulias:2015vxa}  under the terms of the Creative Commons Attribution 4.0 International license.}
\label{fig.2}
\end{figure}

\begin{figure}[t]
\begin{center}
\includegraphics[width=0.5\textwidth]{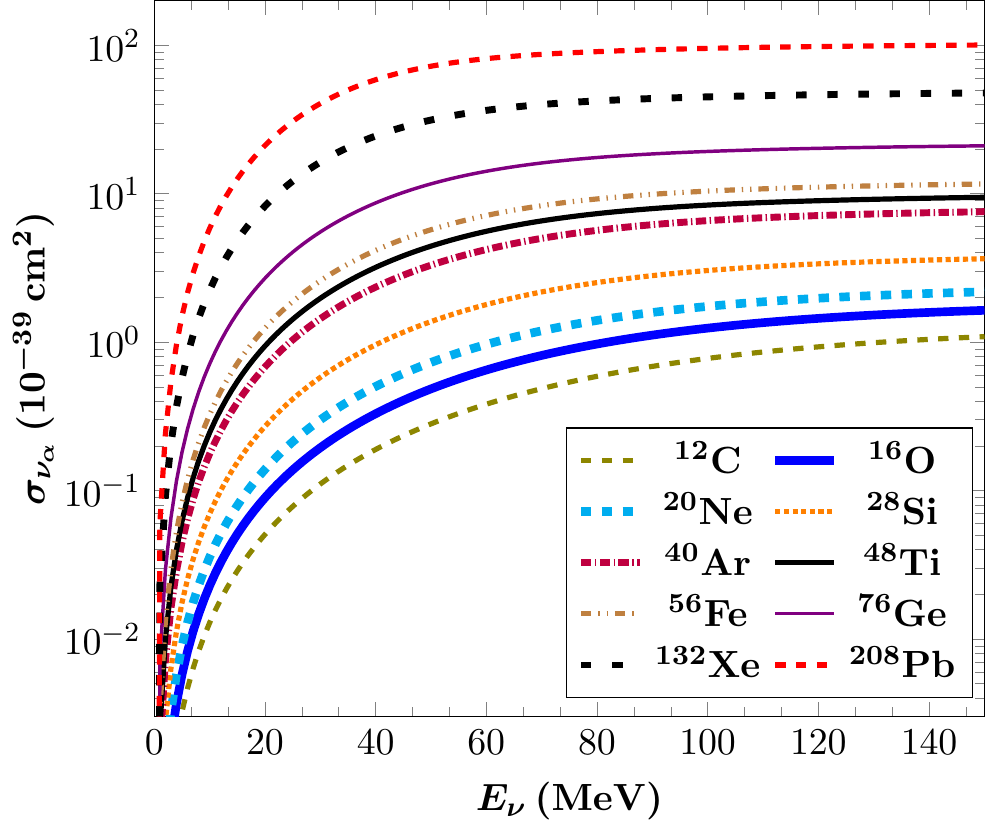}
\end{center}
\caption{Integrated \cevns cross sections  $\sigma_{\nu_{\alpha}(\bar{\nu}_{\alpha})}(E_\nu)$   for a set of nuclear targets ranging from light to heavy isotopes. Figure adapted from Ref.~\cite{Papoulias:2015vxa}  under the terms of the Creative Commons Attribution 4.0 International license.}
\label{AHEP:Figure-3} 
\end{figure}

The occupation probabilities entering Eqs.(\ref{definition-ff}) and (\ref{forfac}) are assumed equal to unity (zero) for the states below (above) the Fermi surface, e.g. the filling numbers of the states for closed (sub)shell nuclei are those predicted by the simple Shell-Model. Going one step further, Ref.~\cite{Kosmas:1992yxv} introduced depletion/occupation numbers to describe the occupation probabilities of the surface levels, which satisfy the relation
\begin{equation}
\sum_{\begin{subarray}{c} (n, l)j\\ \mathrm{all} \end{subarray} } \alpha_{n l j}(2j+1) =N_{n}\, .
\label{depletion}
\end{equation} 
In this framework, there is a number of \emph{active} surface nucleons (above or below the Fermi level) with non-vanishing occupation probability $0 \leq \alpha_{n l j}\leq 1$  and a number of \emph{core} levels with $\alpha_{n l j}=1$. The parameters are properly adjusted so that a high reproducibility of the experimental data is achieved~\cite{DeJager:1987qc}. By introducing four parameters $\alpha_{i}, \, \, i=1,2,3,4$ in Eq.(\ref{depletion}) the polynomial $\Pi(\chi,Z)$ of Eq.(\ref{rho(r)}) reads
\begin{equation}
\begin{aligned}
\Pi(\chi,Z,\alpha_{i})= & \Pi(\chi,Z_{2}) \frac{\alpha_{1}}{Z_{1}-Z_{2}} + \Pi(\chi,Z_{1}) \left[\frac{\alpha_{2}}{Z_{c}-Z_{1}}-\frac{\alpha_{1}}{Z_{1}-Z_{2}} \right] \\
& + \Pi(\chi,Z_{c}) \left[\frac{Z^{\prime}-Z}{Z^{\prime}-Z_{c}}-\frac{\alpha_{2}}{Z_{c}-Z_{1}}-\frac{\alpha_{3}}{Z^{\prime}-Z_{c}} \right] \\
& + \Pi(\chi,Z^{\prime}) \left[\frac{Z-Z_{c}}{Z^{\prime}-Z_{c}} + \frac{\alpha_{3}}{Z^{\prime}-Z_{c}}- \frac{\alpha_{4}}{Z^{\prime \prime}-Z^{\prime}} \right] \\
& + \Pi(\chi,Z^{\prime \prime}) \left[\frac{\alpha_{4}}{Z^{\prime \prime}-Z^{\prime}} - \frac{\lambda}{Z^{\prime \prime \prime}-Z^{\prime \prime} }\right] + \Pi(\chi,Z^{\prime \prime \prime}) \frac{\lambda}{Z^{\prime \prime \prime}-Z^{\prime \prime}}\, ,
\end{aligned}
\label{fractional}
\end{equation}
with $\lambda=\alpha_{1}+\alpha_{2}-\alpha_{3}-\alpha_{4}$ (see Ref.~\cite{Papoulias:2015vxa} for the fitted values).\\

iii) \emph{Use of effective expressions for the nuclear form factors}\\
Besides calculations in the spirit of a nuclear structure model, a reliable description of the nuclear form factors (at least for low-energy reactor and solar neutrinos) may be obtained through the use of phenomenological approximations of the charge density distribution. The Helm-type density distribution is a convolution of a uniform nucleonic density with cut-off radius $R_0$ (accounting for the interior density)  with a Gaussian falloff with folding width $s$ (surface thickness). The corresponding Helm form factor takes the analytical form as~\cite{Helm:1956zz}
\begin{equation}
F_{\text{Helm}}(Q^2) = 3 \frac{j_1(Q R_0)}{q R_0}\, e^{-(Q s)^2/2} \, ,
\label{eq:helm}
\end{equation} 
where $j_1(x) = \frac{\sin x}{x^2} - \frac{\cos x}{x}$ is the 1st-order Spherical Bessel function.
The first three moments can be analytically expressed as~\cite{Piekarewicz:2016vbn}
\begin{equation}
\begin{aligned}  
\left\langle R^2_n \right\rangle & = \frac { 3 } { 5 } R _ { 0 } ^ { 2 } + 3 s ^ { 2 }  \\  
\left\langle R^4_n \right\rangle & = \frac { 3 } { 7 } R _ { 0 } ^ { 4 } + 6 R _ { 0 } ^ { 2 } s ^ { 2 } + 15 s ^ { 4 }  \\  
\left\langle R^6_n \right\rangle & = \frac { 1 } { 3 } R _ { 0 } ^ { 6 } + 9 R _ { 0 } ^ { 4 } s ^ { 2 } + 63 R _ { 0 } ^ { 2 } s ^ { 4 } + 105 s ^ { 6 } \, .
\end{aligned}
\end{equation}
The surface thickness parameter is fixed to $s=0.9$ by fitting muon spectroscopy data~\cite{Fricke:1995zz}, having also the advantage of improving the matching between the Helm and the symmetrized Fermi (SF) distributions~\cite{Lewin:1995rx}.
The SF approximation follows from a Woods-Saxon charge density distribution and is expressed through the half density radius $c$ and the diffuseness parameter $a$, as~\cite{Sprung:1997}
\begin{equation}
\begin{aligned} 
F_{\text{SF}} \left( Q ^ { 2 } \right) =  \frac { 3 } { Q c \left[ ( Q c ) ^ { 2 } + ( \pi Q a ) ^ { 2 } \right] }  \left[ \frac { \pi Q a } { \sinh ( \pi Q a ) } \right]   \left[ \frac { \pi Q a \sin ( Q c ) } { \tanh ( \pi Q a ) } - Q c \cos ( Q c ) \right] \, ,
\end{aligned} 
\end{equation}
with 
\begin{equation}
c = 1.23 A^{1/3} - 0.60 \, \text{(fm)}, \quad a=0.52 \, \text{(fm)} \, ,
\label{eq:SF-vals}
\end{equation}
while the surface thickness is written as $t = 4a \ln 3 $~\cite{Cadeddu:2017etk}. The corresponding first three moments of the SF form factor read~\cite{Piekarewicz:2016vbn}
\begin{equation}
\begin{aligned} \left\langle R^2_n \right\rangle & = \frac { 3 } { 5 } c ^ { 2 } + \frac { 7 } { 5 } ( \pi a ) ^ { 2 } \\ \left\langle R^4_n \right\rangle & = \frac { 3 } { 7 } c ^ { 4 } + \frac { 18 } { 7 } ( \pi a ) ^ { 2 } c ^ { 2 } + \frac { 31 } { 7 } ( \pi a ) ^ { 4 } \\  \left\langle R^6_n \right\rangle & = \frac { 1 } { 3 } c ^ { 6 } + \frac { 11 } { 3 } ( \pi a ) ^ { 2 } c ^ { 4 } + \frac { 239 } { 15 } ( \pi a ) ^ { 4 } c ^ { 2 } + \frac { 127 } { 5 } ( \pi a ) ^ { 6 } \, .
\end{aligned}
\end{equation}
The Klein-Nystrand (KN) distribution is obtained from the convolution of a Yukawa potential with range $a_k = 0.7$ fm over a Woods-Saxon distribution  (hard sphere with radius $R_A$). The resulting KN form factor reads~\cite{Klein:1999qj}
\begin{equation}
F_{\text{KN}} = 3 \frac{j_1(Q R_A)}{Q R_A} \left[ 1 + (Q a_k )^2 \right]^{-1} \, ,
\end{equation}
and is adopted by the COHERENT Collaboration, while in this case root mean square (rms) radius reads
\begin{equation}
\langle R^2_n \rangle_\text{KN} = 3/5 R_A^2 + 6 a_k^2 \, .
\end{equation}

Fig.~\ref{fig.2} presents the charge density distribution in the top panel and the corresponding nuclear form factors for $^{40}$Ar (interesting for LAr \cevns detectors) and $^{48}$Ti (interesting for $\mu^- \to e^-$ conversion in nuclei) in the lower panel, while the results are compared for the various methods used. A comparison of the form factors for $^{127}$I and $^{133}$Cs that are of interest for COHERENT, evaluated with the DSM method (not covered here), with those of the Helm, SF and KN parametrizations, is given in Ref.~\cite{Papoulias:2019lfi}. By incorporating realistic nuclear structure calculations on the basis of the BCS method, the SM \cevns cross section is given in Fig.~\ref{AHEP:Figure-3}  for a set of different isotopes throughout the periodic table.  For heavier isotopes the form factor suppression is more pronounced and therefore the cross sections flatten more quickly, since the nuclear effects become significant even at low-energies.


\section{Constraints within and beyond the SM from CE$\nu$NS}
\label{sect:constraints}

\begin{figure}[h]
\includegraphics[width= 0.51 \textwidth]{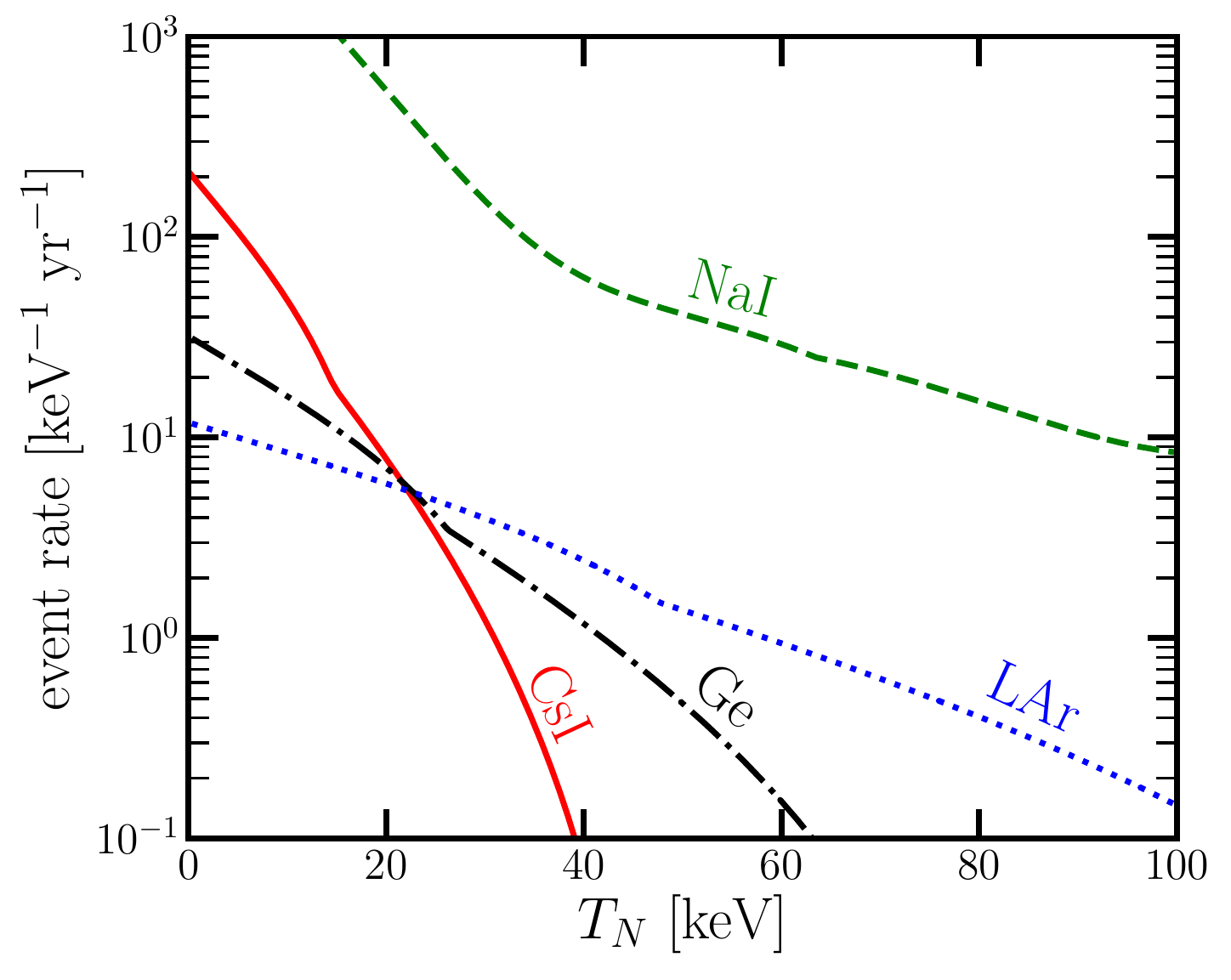}
\includegraphics[width= 0.45 \textwidth]{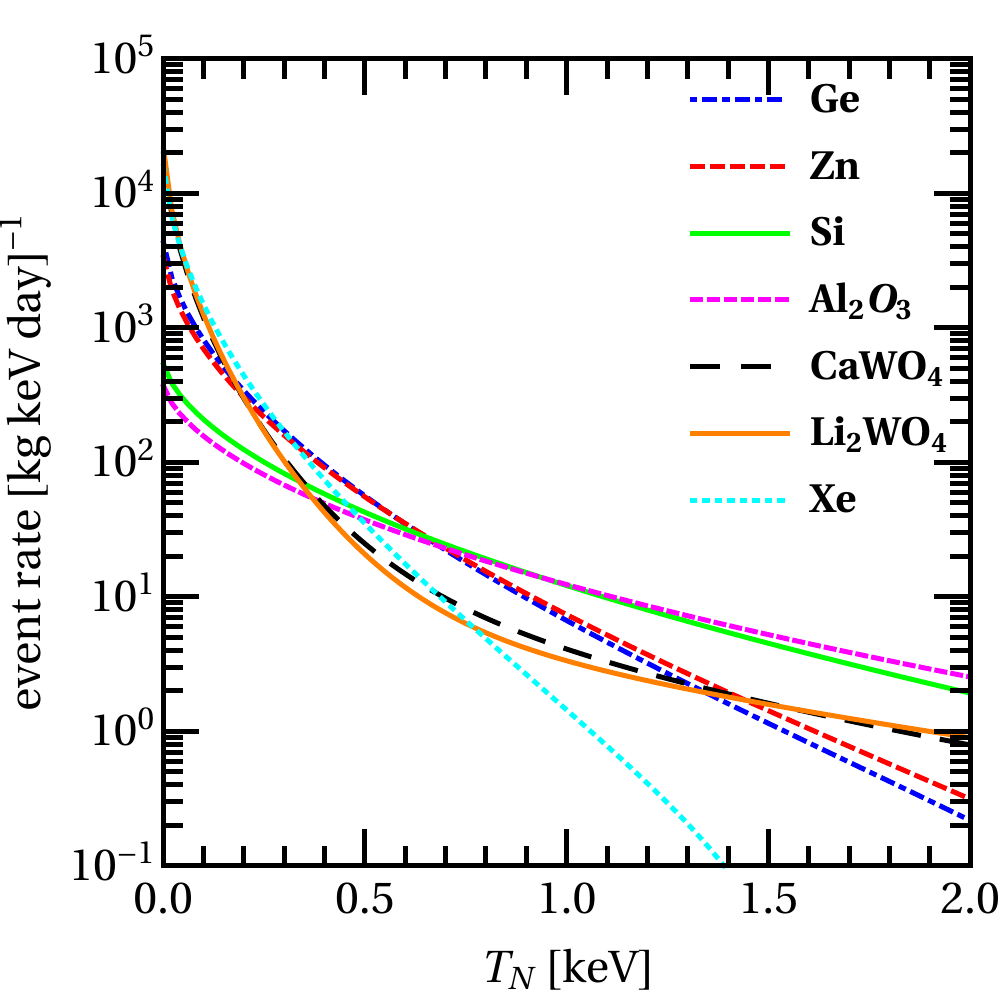}

\caption{Expected \cevns event rate for the different detector subsystems of the COHERENT experiment (left) and for the different target nuclei relevant to reactor based experiments (right). For the case of COHERENT the results are shown according to the setups of Table~\ref{tab:experiments}, while for reactor based experiments the calculation assumes 1~kg of each target located at 20~m from a 4 GW reactor NPP. The impact of QF and efficiency is ignored.}
\label{fig:COHERENT_detectors}
\end{figure}

The observation of \cevns by the COHERENT experiment with a $\pi$-DAR neutrino source is a portal to new physics triggering a considerable number of phenomenological studies at low-energies. New constraints have been put on neutrino,  electroweak and nuclear physics parameters,   that we devote an effort to summarize below. The experimental confirmation of \cevns has also prompted a great rush in the experimental physics community and several projects are aiming to measure \cevns using reactor neutrinos from nuclear power plants (NPP). It should be stressed that given the large potential of improvement in  detector  technology  and  control  of  systematics,  it is feasible to further explore the low-energy and precision neutrino frontier. The CONUS experiment is currently running at the Brokdorf NPP (Germany) and has already released preliminary results while the COHERENT experiment has released new results from the engineering run with a LAr detector~\cite{Akimov:2019rhz}.  Moreover, a number of prominent experiments are in preparation such as: the MINER experiment at the TRIGA Nuclear Science Center at Texas A\&M University (USA), the CONNIE project at the Angra NPP (Brazil), the NUCLEUS and Ricochet experiments at the Chooz NPP (France)~\footnote{BASKET~\cite{ALIANE2019162784} is a synergy of Ricochet and NUCLEUS that is developing a $\mathrm{Li_2 WO_4[Mo]}$ Scintillating bolometer.}, the TEXONO program at the Kuo-Sheng NPP (Taiwan), the vGEN and RED100 experiments at the Kalinin NPP (Russia),   the Coherent Captain-Mills (CCM) project at Los Alamos Neutron Science Center (LANSCE) as well as new proposals for a \cevns measurement by employing a $^{51}$Cr source~\cite{Bellenghi:2019vtc} and new posibilities in China~\footnote{See e.g. talk by Ran Han: \href{https://zenodo.org/record/3464506}{10.5281/zenodo.3464505}} (an exhaustive review of the \cevns experimental developments is given in Ref.~\cite{Akimov:2019wtg}). Finally, it has been recently discussed the possibility of measuring \cevns at the European Spallation Source (ESS)~\cite{Baxter:2019mcx}. Table~\ref{tab:experiments} lists a summary of the current and future experimental projects, while Fig.~\ref{fig:COHERENT_detectors} demonstrates the differential event rate for the various target nuclei at  $\pi$-DAR (see Ref.~\cite{Rich:2017lzd}) and at the various reactor \cevns experiments neglecting detector efficiencies and quenching factors (QF). 

\begin{figure}[h]
\includegraphics[width= 0.49 \textwidth]{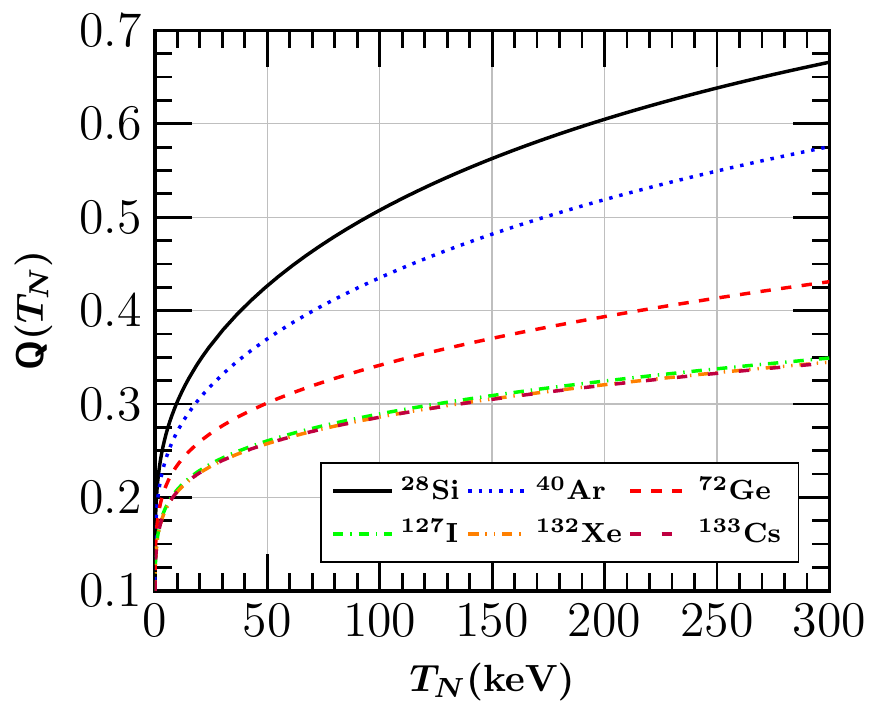}
\includegraphics[width= 0.49 \textwidth]{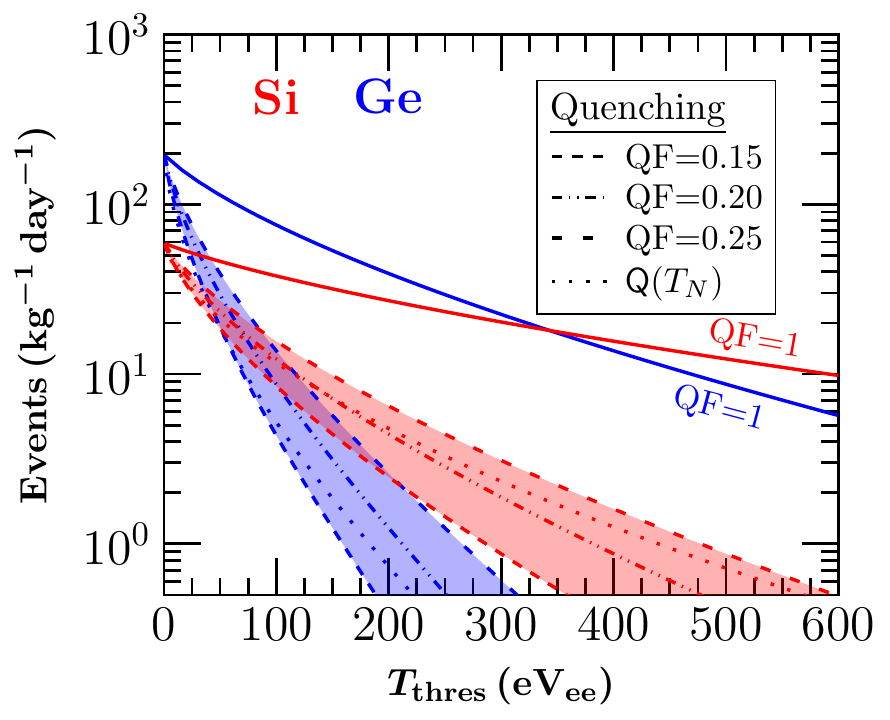}

\caption{Left: The QF as a function of the nuclear recoil energy. Right: the impact of the QF on the expected number of events at a reactor based experiment assuming a typical neutrino flux of $10^{13} \, \mathrm{cm^{-2} s^{-1}}$ for \cevns off Silicon and Germanium detectors. The figure in the left panel has been adapted from Ref.~\cite{Miranda:2019wdy}  under the terms of the Creative Commons Attribution 4.0 International license.}
\label{fig:quenching}
\end{figure}

In reality however, these experiments are sensitive to an ionization energy (e.g. electron equivalent energy $\mathrm{eV_{ee}}$) since a large portion of the nuclear recoil energy $\mathrm{eV_{nr}}$ is lost to heat (conversion to phonons). The energy discrepancy has to be determined experimentally and is taken into account in terms of the QF. The latter quantity is crucial for such processes and depends on the nuclear recoil energy as well as on the target nucleus in question. Theoretically it follows the empirical form arising from the Lindhard theory~\cite{Lewin:1995rx} 
\begin{equation}
\mathsf{Q}(T_N) = \frac{\kappa g(\gamma)}{1+\kappa g(\gamma)}\, ,
\end{equation}
with $g(\gamma)= 3 \gamma^{0.15} + 0.7 \gamma^{0.6}+\gamma$ and $\gamma  = 11.5 \, T_N(\text{keV}) Z^{-7/3}$, $\kappa = 0.133 \, Z^{2/3} A^{-1/2}$. 
The left and right panels of Fig.~\ref{fig:quenching} quantify the effect of the QF in the case of \cevns.

\begin{table}[t!]
\begin{tabular}{lccccll}
\toprule
Experiment                       & $T_\text{th}$ & Baseline (m) & Target                                                  & Mass (kg)       & Technology                                                                            & Source                                                                   \\\\
\hline
\multirow{4}{*}{COHERENT~\cite{Akimov:2018ghi}} & 6.5 keV       & 19.3           & CsI{[}Na{]}                                             & 14.57           & \begin{tabular}[c]{@{}l@{}}Scintillating\\ crystal\end{tabular}                       & \multirow{4}{*}{\begin{tabular}[c]{@{}l@{}}$\pi$-DAR\\ SNS\end{tabular}} \\
                          & 5 keV         & 22             & Ge                                                      & 10              & HPGe PPC                                                                              &                                                                          \\
                          & 20 keV        & 29             & LAr                                                     & $2 \times 10^3$ & Single phase                                                                          &                                                                          \\
                          & 13 keV        & 28             & NaI{[}Tl{]}                                             & 185*/3388       & \begin{tabular}[c]{@{}l@{}}Scintillating\\ crystal\end{tabular}                       &                                                                          \\
                          \hline
CCM~\cite{CCM}                       & 10-20 keV     & 20-40          & LAr                                                     & $10^4$          & Scintillation                                                                         & \begin{tabular}[c]{@{}l@{}}$\pi$-DAR\\ Lujan\end{tabular}                \\
\hline
CONUS~\cite{Hakenmuller:2019ecb}                     & 300 eV        & 17             & Ge                                                      & 4               & HPGe                                                                                  & NPP 3.9 GW                                                               \\
MINER~\cite{Agnolet:2016zir}                     & 10 eV         & 1              & Ge/Si                                                   & 30              & cryogenic                                                                             & NPP 1 MW                                                                 \\
CONNIE~\cite{Aguilar-Arevalo:2019jlr}                    & 28 eV         & 30             & Si                                                      & 1               & Si CCDs                                                                               & NPP 3.8 GW                                                               \\
Ricochet~\cite{Billard:2016giu}                  & 50-100 eV     & \textless{}10  & Ge/Zn                                                   & 10              & Ge, Zn bolometers                                                                     & NPP 8.54 GW                                                              \\
NUCLEUS~\cite{Angloher:2019flc}                   & 20 eV         & \textless{}10  & \begin{tabular}[c]{@{}l@{}}$\mathrm{CaWO_4}$ \\ $\mathrm{Al_2 O_3}$\end{tabular} & $10^{-3}$       & \begin{tabular}[c]{@{}l@{}}Cryogenic $\mathrm{CaWO_4}$ \\ $\mathrm{Al_2 O_3}$ calorimeter\\ array\end{tabular} & NPP 8.54 GW                                                              \\
RED100~\cite{Akimov:2017hee}                    & 500 eV        & 19             & Xe                                                      & 100             & LXe dual phase                                                                        & NPP 3 GW                                                                 \\
vGEN                  & 350 eV        & 10             & Ge                                                      & 4 $\times$ 0.4  & Ge PPC                                                                                & NPP 3 GW                                                                 \\
TEXONO~\cite{Wong:2005vg}                    & 150-200 eV    & 28             & Ge                                                      & 1               & p-PCGe                                                                                & NPP 2 $\times$ 2.9 GW       \\
\botrule                                         
\end{tabular}
\caption{Current and future experimental proposals for \cevns searches.}
\label{tab:experiments}
\end{table}

\subsection{Electroweak and nuclear physics}

\begin{figure}[t]
\centering
\includegraphics[width= 0.505\linewidth]{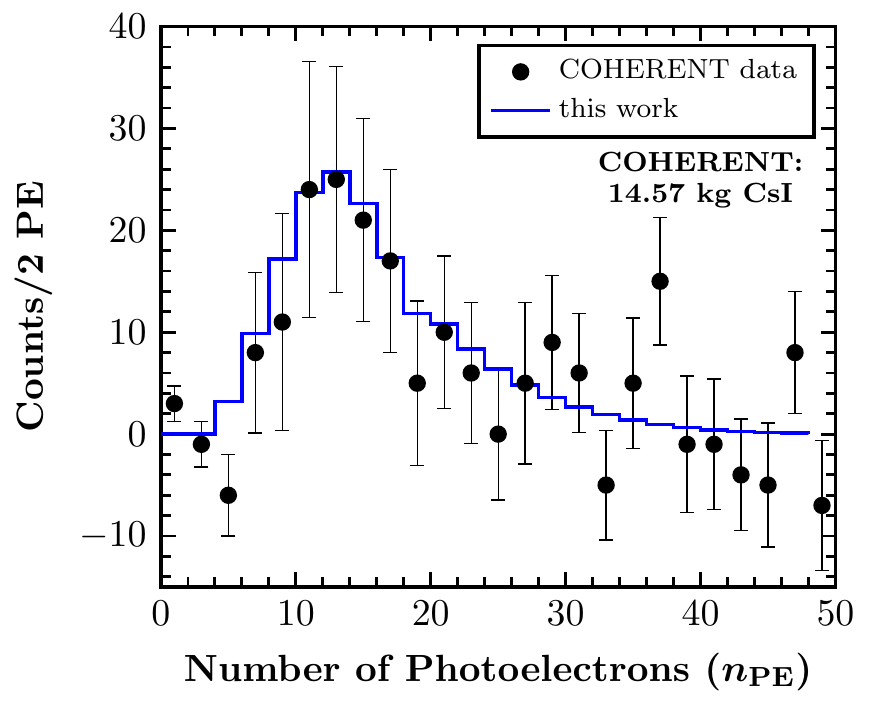}
\includegraphics[width= 0.485\linewidth]{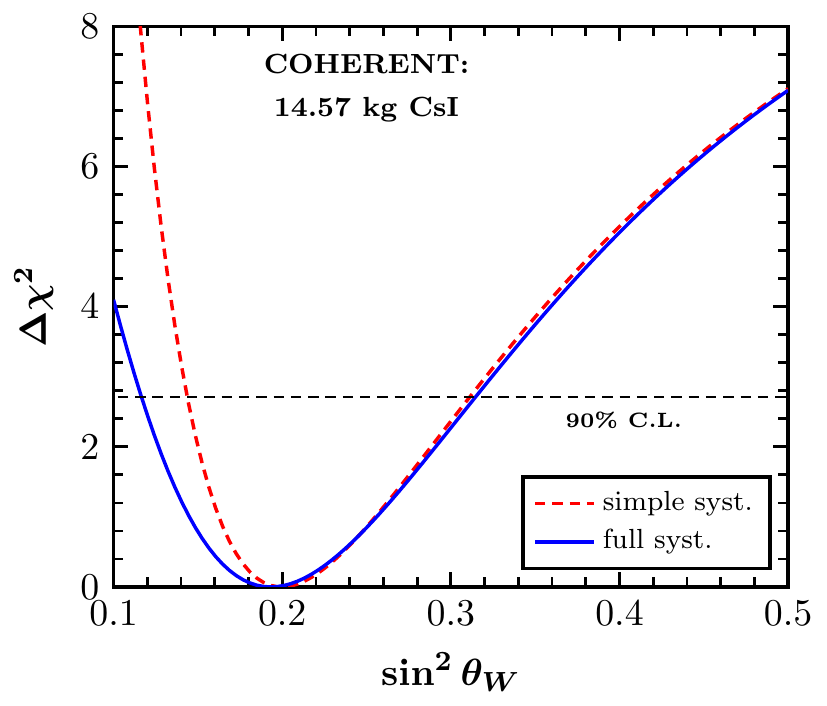}
\caption{Comparison between the simulated number of \cevns events and the experimental data by COHERENT (left) and sensitivity of COHERENT on the weak mixing angle (right). Figure adapted from Ref.~\cite{Kosmas:2017tsq}  under the terms of the Creative Commons Attribution 4.0 International license. }
\label{fig:events-CsI}
\end{figure}

The left panel of Fig.~\ref{fig:events-CsI} shows the expected number of events at the CsI[Na] detector of COHERENT and gives a comparison with the experimental data, from where it can be seen that a good agreement is reached.  In Ref.~\cite{Kosmas:2017tsq} the authors analyzed the \cevns data and obtained a low-energy determination of the weak mixing angle, as illustrated in the right panel of  Fig.~\ref{fig:events-CsI}. The obtained constraint at 90\% C.L. reads~\cite{Kosmas:2017tsq}
\begin{equation}
\sin^2 \theta_W =  0.197^{+0.128}_{-0.080} \, .
\end{equation}
An interesting analysis combining atomic parity violating (APV) and \cevns data was performed in Ref.~\cite{Cadeddu:2018izq}, while the prospects regarding the future reactor-based \cevns experiments such as those presented in Table~\ref{tab:experiments}, have been extracted in Ref.~\cite{Canas:2018rng}. On the other hand, an improved determination of the CsI[Na] quenching factor  can in principle lead to a significantly better agreement between the experimental results and the theoretical simulations~\cite{Collar:2019ihs}, as well as to an improved sensitivity on the weak mixing angle~\cite{Papoulias:2019txv}.

\begin{figure*}[t]
\centering
\includegraphics[width=0.95\linewidth]{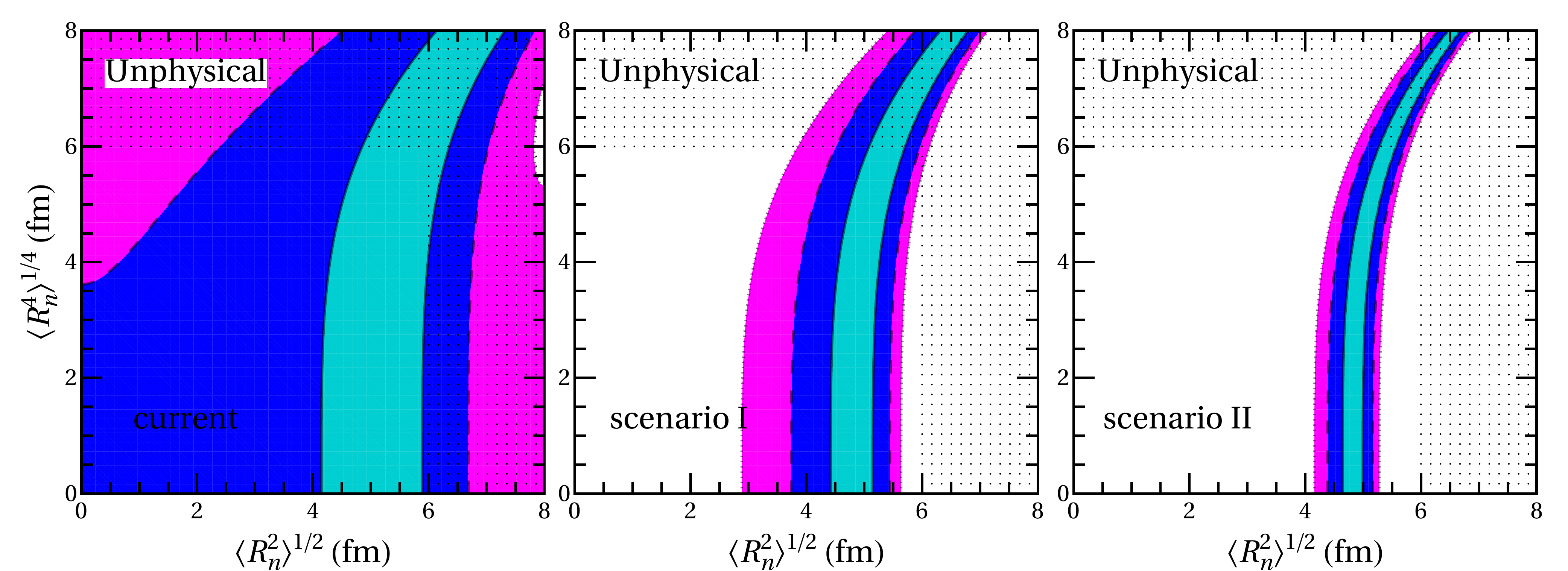}
\caption{Sensitivity contours in the $\langle R_n^2 \rangle^{1/2} $--$\langle R_n^4 \rangle^{1/4}$ plane from the COHERENT data assuming the current and possible future detector specifications (see the text). The allowed regions are shown at $1\sigma$ (turquoise), 90\% C.L. (blue) and 99\% C.L. (magenta). Taken from Ref.~\cite{Papoulias:2019lfi}.}
\label{fig:RS}
\end{figure*}

%
The discussion made in the previous Section emphasized how the \cevns cross section depends on the nuclear physics effects which are incorporated through the momentum variation of the relevant nuclear form factor. The authors of Ref.~\cite{AristizabalSierra:2019ufd}  demonstrated how the intrinsic nuclear structure uncertainties may have a significant impact to searches beyond the SM such those regarding NSIs, sterile neutrinos and neutrino generalized interactions (GNIs). Starting from the form factor of  Eq.(\ref{definition-ff}) and expanding in terms of even moments of the charge density distribution one arrives to a model independent expression~\cite{Patton:2012jr}
\begin{equation}
F_{p,n}(Q^2) \approx 1 - \frac{Q^2}{3!} \langle R_{p,n}^2 \rangle +  \frac{Q^4}{5!} \langle R_{p,n}^4 \rangle -\frac{Q^6}{7!} \langle R_{p,n}^6 \rangle + \cdots  \, ,
\label{eq:form-factor-expansion}
\end{equation}
with the $k$-th radial moment defined as
\begin{equation}
\langle R_{p,n}^k \rangle =\frac{\int \rho_{p,n}(\vec{r}) \, r^k \, d^3 \vec{r}}{\int \rho_{p,n}(\vec{r}) \, d^3 \vec{r}} \, ,
\end{equation}
allowing the study of contributions of higher-order moments to nuclear form factors~\cite{Ciuffoli:2018qem}. A sensitivity analysis of the two first moments with current and future COHERENT data is depicted in Fig.~\ref{fig:RS} where the allowed regions are presented at 1$\sigma$, 90\% and 99\% C.L. The calculation in this case was restricted in the physical region [0,6] fm in order to obey the 
upper limit on $ R_n(^{208} \text{Pb})=5.75 \pm 0.18$ fm from the PREM experiment~\cite{Horowitz:2012tj}. The future scenarios considered assume improved statistical uncertainties and more massive detectors in accord with the next generation COHERENT experiments~\cite{Akimov:2018ghi} (see Ref.~\cite{Papoulias:2019lfi} for details), while as demonstrated in Ref.~\cite{Patton:2012jr} multi-ton scale detectors will provide significant improvements.

%
%
%

\begin{figure}[t]
\includegraphics[width= 0.4 \textwidth]{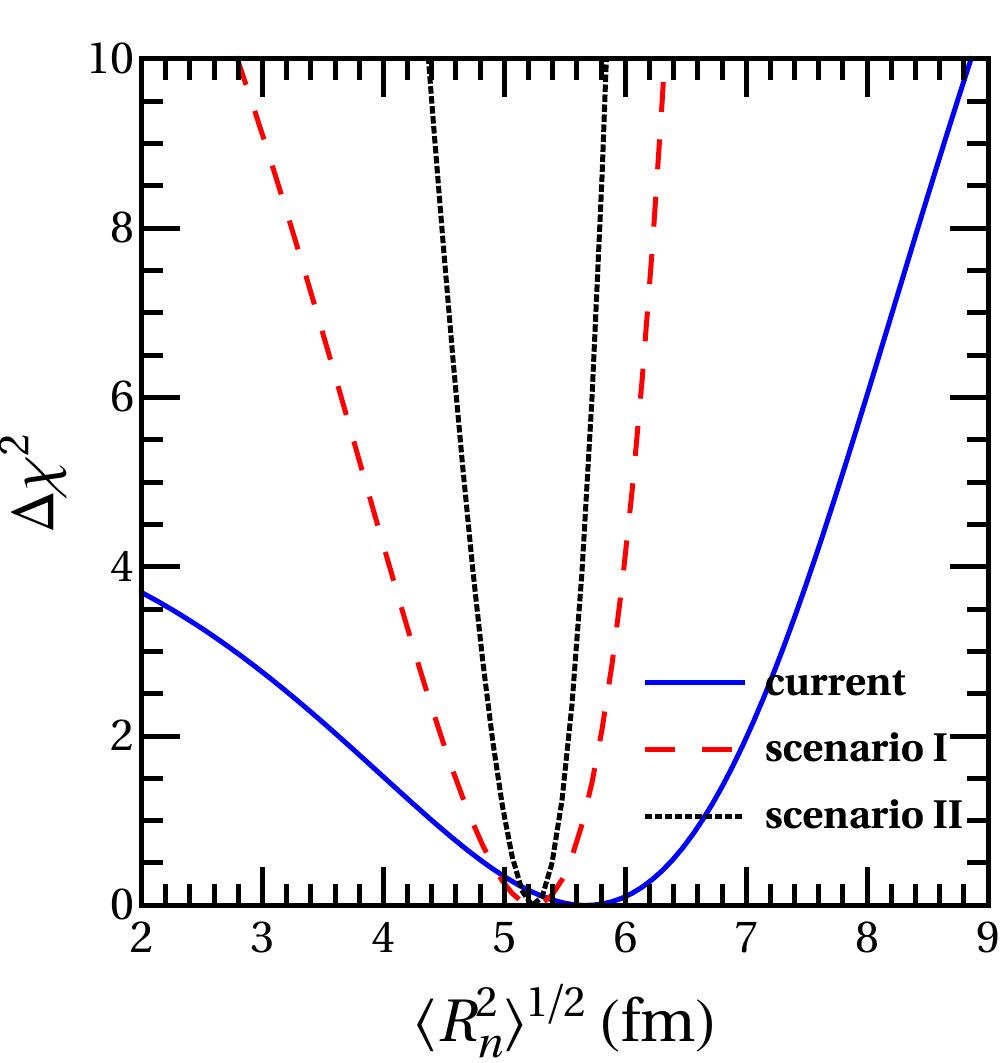}
\caption{COHERENT sensitivity on the average  nuclear rms radius of CsI  assuming the current and possible future detector spesifications. Taken from Ref.~\cite{Papoulias:2019lfi}.}
\label{fig:rms}
\end{figure}

The average CsI neutron rms radius has been explored in Refs.~\cite{Cadeddu:2017etk, Papoulias:2019lfi, Huang:2019ene} using the energy spectrum of the available \cevns data. The corresponding sensitivity profiles are presented in Fig.~\ref{fig:rms}, leading to the best fits  at 90\% C.L.~\cite{Papoulias:2019lfi}
\begin{equation}
\begin{aligned}
\langle R_n^2 \rangle^{1/2} =& 5.64^{+ 0.99}_{-1.23} \, \text{fm} \quad  \text{(current)} \, ,\\
\langle R_n^2 \rangle^{1/2} =& 5.23^{+ 0.42}_{-0.50} \, \text{fm} \quad \text{(scenario I)} \, , \\
\langle R_n^2 \rangle^{1/2} =& 5.23^{+ 0.22}_{-0.22} \, \text{fm} \quad \text{(scenario II)} \, ,
\end{aligned}
\end{equation}
while the potential of improvement through a more accurate determination of the QF is promising (see e.g. Ref.~\cite{Papoulias:2019txv}).
An independent analysis combining APV and \cevns data was perfomed in Ref.~\cite{Cadeddu:2019eta} leading to essentially similar results.
Finally, it is worthwhile to mention the reported upper bound on the neutron skin $\Delta R_{np} = \Delta R_n - \Delta R_p = 0.7^{+0.9}_{-1.1}$~fm~\cite{Cadeddu:2017etk}.

\subsection{Nonstandard and generalized neutrino interactions}\label{sec:NSI1}

Non-standard interactions (NSI)~\cite{Dev:2019anc} apperar in several appealing SM extensions~\cite{Babu:2019mfe} involving four-fermion contact interaction, various seesaw realizations~\cite{Deppisch:2005zm,Malinsky:2008qn,Forero:2011pc}, left-right symmetry~\cite{Das:2012ii}, gluonic operators~\cite{Petrov:2013vka}, etc., constituting an interesting model independent probe of new physics. NSIs may have implications to SN~\cite{Amanik:2006ad}, neutrino oscillations~\cite{Farzan:2017xzy} and \cevns~\cite{Barranco:2005yy, Scholberg:2005qs}, while recently NSI terms were explored in the context of GNI~\cite{AristizabalSierra:2018eqm} and effective field theory (EFT) operators~\cite{Altmannshofer:2018xyo, Bischer:2019ttk}. Finally the RG issue has been partly addressed in the context of NSI in Ref.~\cite{Davidson:2019iqh}.

For sufficiently low energies vector-type NSIs  arise from the effective four-fermion operators~\cite{Miranda:2015dra}
\begin{equation}
\mathcal{O}_{\alpha \beta}^{q V} = \left(\bar{\nu}_\alpha \gamma^\mu L \nu_\beta \right) \left(\bar{q} \gamma_\mu P q\right) + \mathrm{H.c.} \, ,
\end{equation}
leading to new contributions to the \cevns rate from exotic processes of the form
\begin{equation}
 \nu_{\alpha}(\bar{\nu}_{\alpha}) + (A,Z) \rightarrow \nu_{\beta}(\bar{\nu}_{\beta})  + (A,Z) \, ,
\label{AHEP:neutrin-NSI}
\end{equation} 
where $\alpha, \beta = \{e,\mu,\tau$\}  ($\alpha \neq \beta$), $q$ denotes a first-generation quark $q=\{u,d\}$ and $P = \{L,R\}$ is the chiral projection operator. For the case of \cevns the new interactions are taken into account through the NSI charge with the substitution $\mathcal{Q}_W^V \rightarrow \mathcal{Q}_{\mathrm{NSI}}^V$ in Eq.(\ref{eq:weak-charge}). The latter contains flavor-preserving non-universal ($\epsilon_{\alpha \alpha}^{qV}$) and flavor changing ($\epsilon_{\alpha \beta}^{qV}$) terms and is expressed as
\begin{equation}
\begin{aligned}
\mathcal{Q}_{\mathrm{NSI}}^V = & (2 \epsilon_{\alpha \alpha}^{uV} + \epsilon_{\alpha \alpha}^{dV} + g^V_p) Z + (\epsilon_{\alpha \alpha}^{uV} + 2 \epsilon_{\alpha \alpha}^{dV} + g^V_n) N \\
& + \sum_{\alpha \neq \beta} \left[ (2 \epsilon_{\alpha \beta}^{uV} + \epsilon_{\alpha \beta}^{dV}) Z + (\epsilon_{\alpha \beta}^{uV} + 2 \epsilon_{\alpha \beta}^{dV} ) N \right] \, ,
\end{aligned}
\end{equation}
implying that the NSI \cevns cross section becomes flavor dependent.

\begin{figure*}[t]
\centering
\includegraphics[width= 0.9\linewidth]{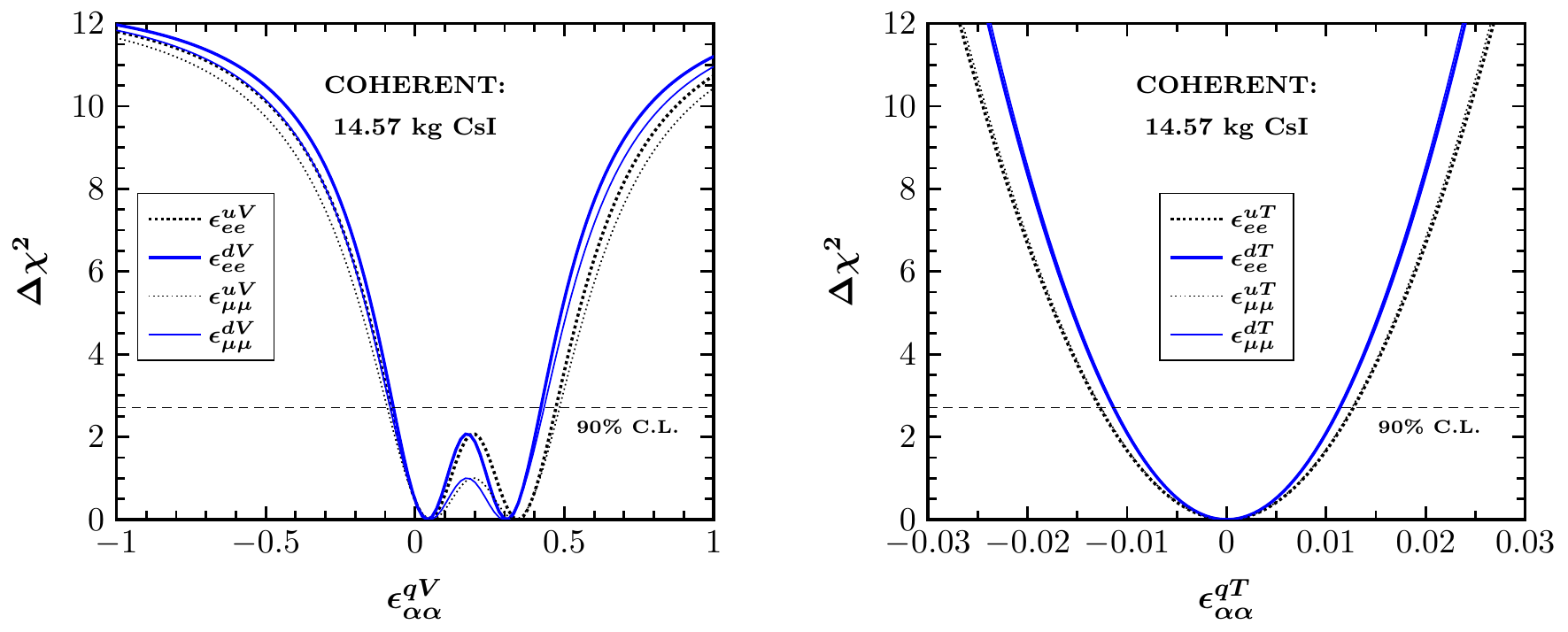}
\caption{Sensitivity of COHERENT to vector (left) and tensor (right) NSI parameters.
Thick (thin) curves correspond to the $\nu_e$ ($\nu_\mu + \bar{\nu}_\mu$) beam. Figure adapted from Ref.~\cite{Kosmas:2017tsq}  under the terms of the Creative Commons Attribution 4.0 International license.}
\label{fig:deltachi-NSI}
\end{figure*}
%
%
\begin{figure*}[t]
\centering
\includegraphics[width= 0.9\linewidth]{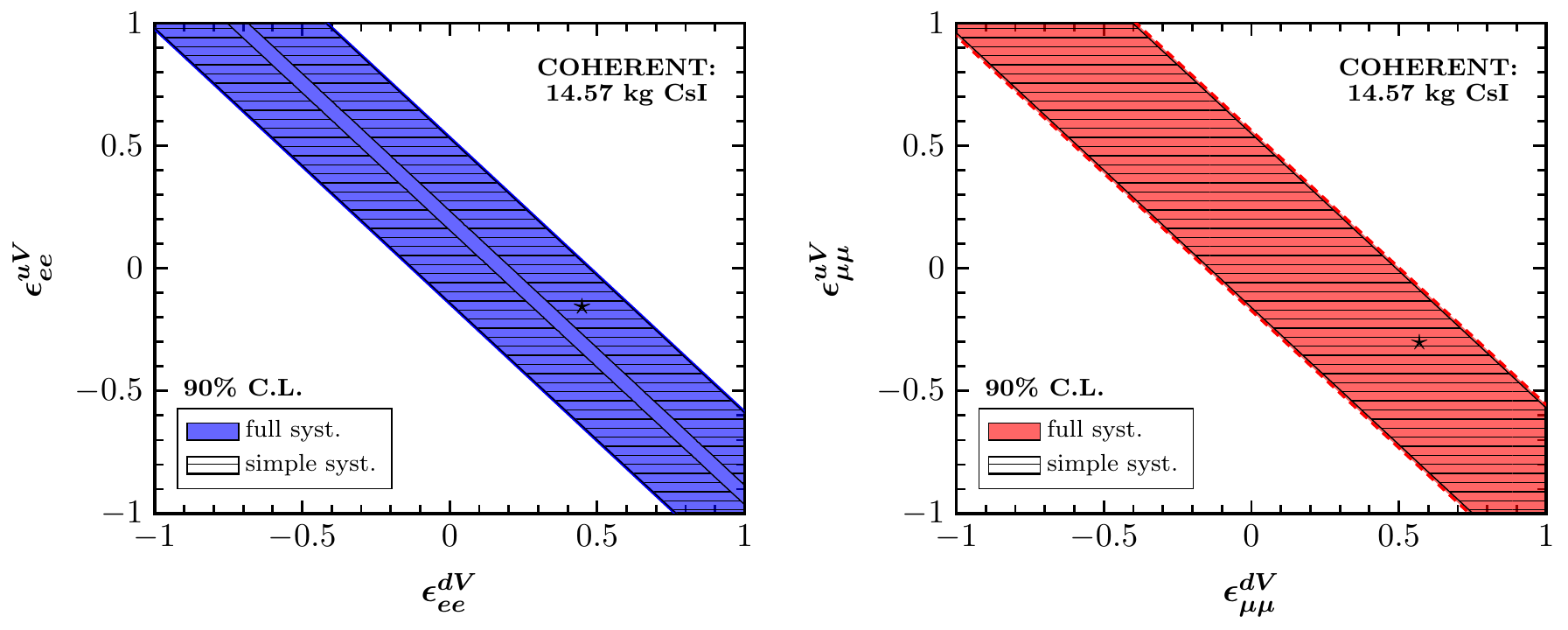}
\includegraphics[width= 0.9\linewidth]{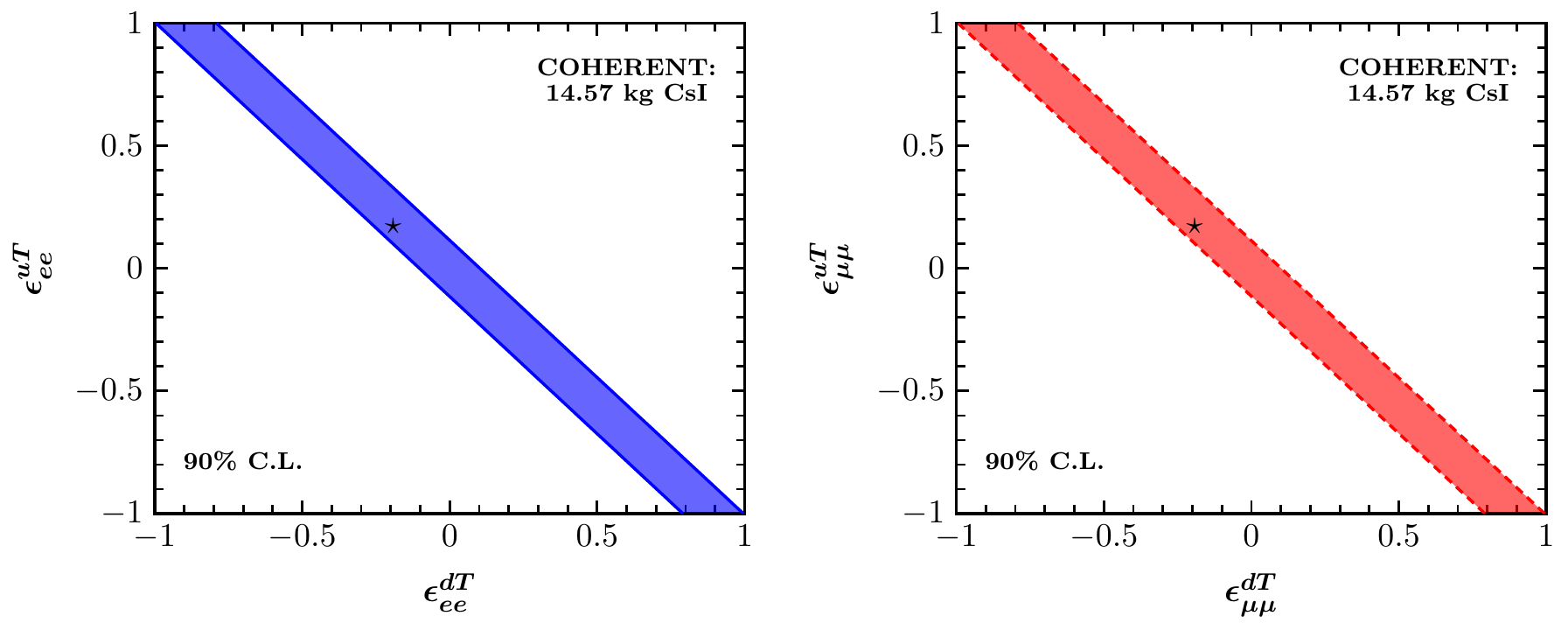}
\caption{Sensitivity contours in the vector (upper panel) and tensor (lower panel) NSI parameter space. The results are presented at 90\% C.L. assuming non-universal couplings only. The left (right) panel corresponds to the $\nu_e$ ($\nu_\mu + \bar{\nu}_\mu$) beam, while the best-fit points are indicated by an asterisk $\star$. Figure adapted from Ref.~\cite{Kosmas:2017tsq}  under the terms of the Creative Commons Attribution 4.0 International license. }
\label{fig:NSI}
\end{figure*}

There is a reach literature on NSI investigations with the recent COHERENT data. Assuming one nonvanishing coupling at a time, the authors of Ref.~\cite{Kosmas:2017tsq} focused on the non-universal terms and obtained the sensitivity profiles shown in the left panel of Fig.~\ref{fig:deltachi-NSI}, while the corresponding allowed regions resulting from a combined analysis of the NSI couplings are illustrated in the upper panel of Fig.~\ref{fig:NSI} at 90\% C.L. Regarding the future prospects of MINER, Ricochet, NUCLEUS and CONNIE, similar studies were conducted concentrating on the non-universal~\cite{Billard:2018jnl} and flavor-changing~\cite{Miranda:2019skf} terms. Indeed, a multitarget strategy can break degeneracies involved between up and down flavor-diagonal NSI terms that survives analysis of neutrino oscillation experiments~\cite{Dent:2017mpr}.  Constraints on the corresponding parameters arising from leptoquarks~\cite{Billard:2018jnl}, GNI~\cite{AristizabalSierra:2018eqm} and EFT~\cite{Altmannshofer:2018xyo, Bischer:2019ttk} have been also reported. NSI constraints from \cevns place meaningfull constraints excluding a large part of the existing CHARM constraints and overlap with results coming out of LHC monojet searches (see Ref.~\cite{Billard:2018jnl} for a usefull comparison).  Regarding the near-term future, a potential improvement on determination of the QF~\cite{Collar:2019ihs} may yield severe constraints. For example, updated bounds are possible by analyzing the number of events~\cite{Papoulias:2019txv}, the energy spectrum~\cite{Khan:2019cvi} as well as through a combined analysis of both energy and timing COHERENT data~\cite{Giunti:2019xpr}.

Novel tensor-type interactions are predicted in the general context of NSI~\cite{Barranco:2011wx} and GNI~\cite{AristizabalSierra:2018eqm} which induce terms of the form
\begin{equation}
\mathcal{O}_{\alpha \beta}^{q T} = \left(\bar{\nu}_\alpha \sigma^{\mu \nu } \nu_\beta \right) \left(\bar{q} \sigma_{\mu \nu} q\right) + \mathrm{H.c.} 
\end{equation}
Such interactions violate the chirality constraint allowing for a wide class of new interactions, e.g. relevant to neutrino EM properties (see Ref.~\cite{Healey:2013vka,Papoulias:2015iga}). Contrary to the vector NSI case, for tensorial interactions there is absence of interference with the SM interactions. In the approximation of a vector-type translation the corresponding tensor NSI charge has been expressed as~\cite{Barranco:2011wx}
\begin{equation}
\mathcal{Q}_{\mathrm{NSI}}^T =  (2 \epsilon_{\alpha \alpha}^{uT} + \epsilon_{\alpha \alpha}^{dT}) Z + (\epsilon_{\alpha \alpha}^{uT} + 2 \epsilon_{\alpha \alpha}^{dT} ) N \, ,
\end{equation}
while a more systematic interpetation has been carried out in Ref.~\cite{AristizabalSierra:2018eqm}.
To account for the new contributions in the presence of tensorial NSI, the \cevns cross is written~\cite{Kosmas:2017tsq}
\begin{equation}
\left(\frac{d \sigma}{dT_N} \right)_{\mathrm{SM+NSI_{tensor}}} = \mathcal{G}_{\mathrm{NSI}}^{T}(E_\nu,T_N)  \left(\frac{d \sigma}{d T_N} \right)_{\mathrm{SM}}\, ,
\end{equation}
with the tensor NSI factor defined as
\begin{equation}
\mathcal{G}_{\mathrm{NSI}}^{T} = 1 + 4 \left(\frac{\mathcal{Q}_{\mathrm{NSI}}^T}{\mathcal{Q}_W^V} \right)^2 \frac{1- \frac{M T_N}{4 E_\nu^2}}{1 - \frac{M T_N}{2 E_\nu^2}} \, .
\end{equation}
%
%
%
%

From the analysis of the COHERENT data, the  sensitivity profiles accounting to tensor NSIs, assuming one non-zero coupling at a time, are illustrated in the right panel of Fig.~\ref{fig:deltachi-NSI} (see also Ref.~\cite{Kosmas:2017tsq}). The corresponding allowed regions coming out from a two parameter analysis are presented in the lower panel of Fig.~\ref{fig:NSI} at 90\% C.L. The result is more stringent as compared to the analysis carried out in the framework of GNI for reasons discussed above. On the other hand, comparing with the vector NSI case the absence of SM-tensor NSI interference causes the allowed regions to appear with more narrow bands.
 
\subsection{The Novel NSI mediators $Z^{\prime}$ (vector) and $\phi$ (scalar)}\label{sec:NSI2}

Theories beyond the SM with an additional $\mathrm{U(1)}^\prime$ symmetry have been comprehensively investigated. Regarding \cevns related studies a novel massive mediator predicted in these concepts is expected to induce a detectable distortion to the nuclear recoil spectrum, provided that its mass is comparable to the momentum transfer. The study of models with new vector or scalar interactions that involve hidden sector particles may be also accessible at \cevns experiments~\cite{Datta:2018xty}.  Such frameworks are interesting since they may play a central role in  explaining anomalies with regards to $B$-meson decays at the LHCb experiment~\cite{Dalchenko:2017shg} and at DM searches~\cite{Bertuzzo:2017tuf}.

\begin{figure*}[t]
\centering
\begin{minipage}{0.45\textwidth}
\includegraphics[width= \linewidth]{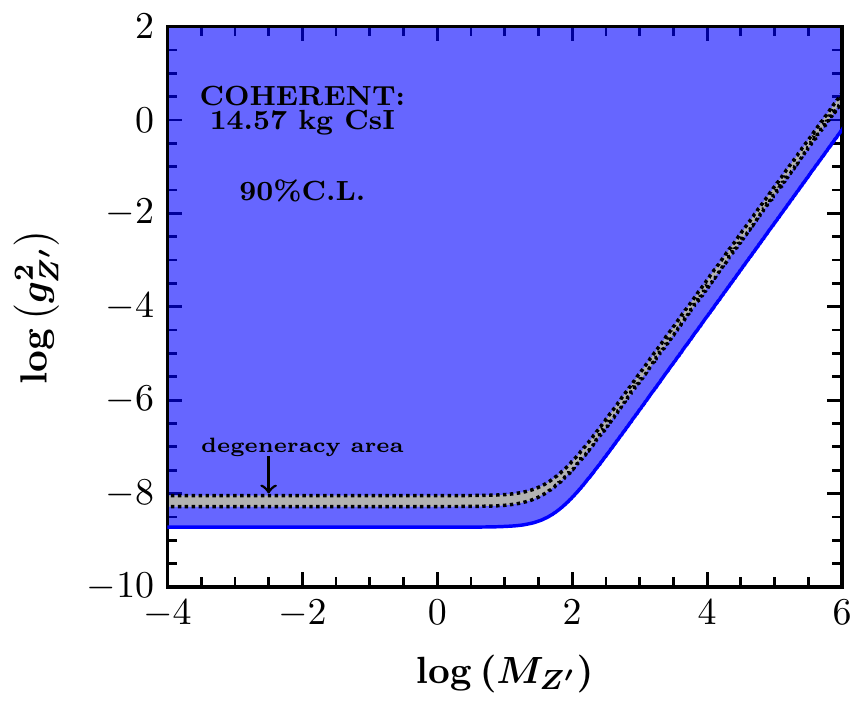}
\end{minipage}
\begin{minipage}{0.45\textwidth}
\includegraphics[width= \linewidth]{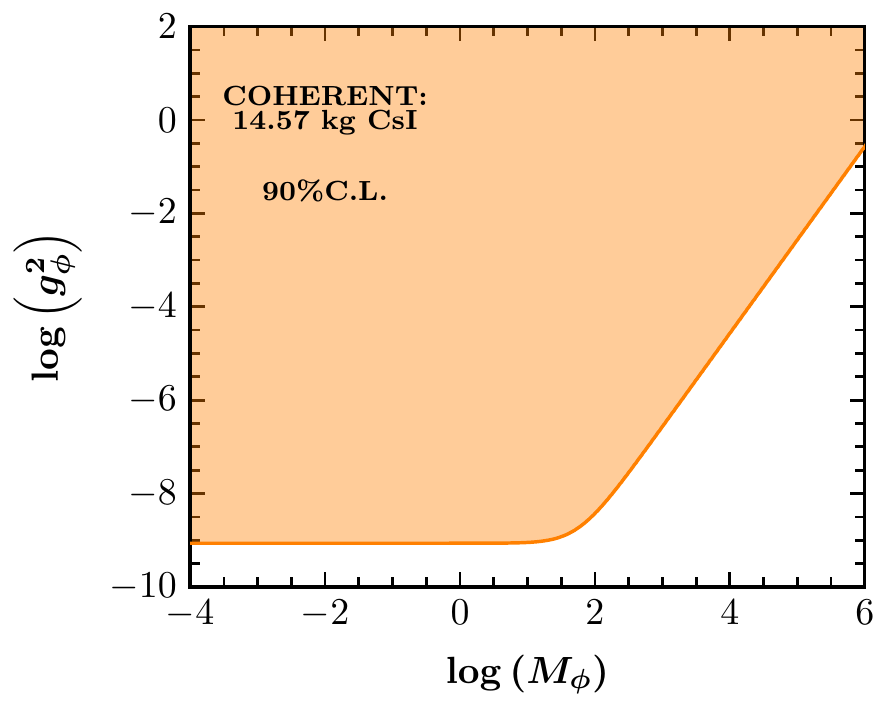}
\end{minipage}
\caption{Excluded regions at 90\% C.L. by the COHERENT experiment assuming simplified scenarios involving a $Z^\prime$ vector mediator (left panel) and a $\phi$ scalar mediator (right panel). The mediator masses are in units of MeV.  Figure adapted from Ref.~\cite{Kosmas:2017tsq}  under the terms of the Creative Commons Attribution 4.0 International license.}
\label{fig:zprime-scalar}
\end{figure*}

We first examine the case of a new massive vector boson $Z^\prime$. Restricting ourselves to the neutrino sector with only left-handed neutrinos the Lagrangian reads~\cite{Cerdeno:2016sfi}
\begin{equation}
\mathcal{L}_{\mathrm{vec}} =  Z^{\prime}_\mu \left(g_{Z^\prime}^{qV} \bar{q} \gamma^\mu q + g_{Z^\prime}^{\nu V} \bar{\nu}_L \gamma^\mu \nu_L\right) + \frac{1}{2} M_{Z^\prime}^2 Z^{\prime}_\mu Z^{\prime \mu} \, . 
\label{lagr:z-prime}
\end{equation}
The arising cross sections imply a re-scaling of the SM one according to the expression
\begin{equation}
\left( \frac{d \sigma}{dT_N}\right)_{\mathrm{SM} + Z^\prime} = \mathcal{G}_{Z^\prime}^2 (T_N) \left( \frac{d \sigma}{d T_N} \right)_{\mathrm{SM}} \, ,
\end{equation}
with the $Z^\prime$ factor taking the form 
\begin{equation}
\mathcal{G}_{Z^\prime} = 
1 - \frac{1}{2 \sqrt{2}G_F}\frac{\mathcal{Q}_{Z^\prime}}{\mathcal{Q}_W^V} \frac{g_{Z^\prime}^{\nu V}}{2 M T_N + M_{Z^\prime}^2} \, .
\label{eq:G_z-prime}
\end{equation}
Here, $g_{Z^\prime}^{\nu V}$ denotes the neutrino-vector coupling, while the respective $Z^\prime$ charge reads~\cite{Bertuzzo:2017tuf}
\begin{equation}
\mathcal{Q}_{Z^\prime} = \left(2 g^{uV}_{Z^\prime} + g^{dV}_{Z^\prime} \right)Z +  \left(g^{uV}_{Z^\prime} + 2g^{dV}_{Z^\prime} \right) N \, .
\end{equation}
However, in the general case the $\nu-Z^\prime$ coupling is flavor dependent $(g_{Z^\prime}^{\nu V})_{\alpha \beta}$. Ref.~\cite{Denton:2018xmq} has explored this  possibility and concluded that for a sufficiently small momentum transfer with respect to $M_{Z^\prime}$, the new physics contributions can be addressed in the form of NSIs 
\begin{equation}
\epsilon_{\alpha \beta}^{qV} = \frac{(g_{Z^\prime}^{\nu V})_{\alpha \beta} \, g^{qV}}{2 \sqrt{2} G_F M^2_{Z^\prime}} \, ,
\end{equation}
where the $Z^\prime$ has been integrated out. Unlike the NSI case that can only modify the energy spectrum by a global factor, the additional momentum dependence expected due to the new light mediators discussed here can be well encoded in the detected signature and subsequently lead to conclusive indications of the new physics nature.

We now turn our attention on new interactions induced by the presence of a   CP-even mediator. In particular, we consider a new real scalar boson $\phi$ with mass $M_\phi$, based on the Lagrangian~\cite{Cerdeno:2016sfi}
\begin{equation}
\mathcal{L}_{\mathrm{sc}} = \phi \left(g_{\phi}^{q S} \bar{q} q + g_{\phi}^{\nu S} \bar{\nu}_R \nu_L + \mathrm{H.c.}\right) -\frac{1}{2} M_\phi^2 \phi^2\, , 
\label{lagr:scalar}
\end{equation}
with $g_{\phi}^{q S}$ and $g_{\phi}^{\nu S}$ representing the scalar-quark and scalar-neutrino couplings, respectively. In this framework the SM \cevns cross section acquires an additive contribution due to the boson exchange that can be quantified in terms of the respective cross section 
\begin{equation}
\left(\frac{d \sigma}{d T_N} \right)_{\mathrm{scalar}} = \frac{G_F^2 M^2}{ 4 \pi} \frac{\mathcal{G}_\phi^2 M_\phi^4 T_N }{E_\nu^2 \left( 2 M T_N + M_\phi^2 \right)^2} F^2(T_N) \, ,
\end{equation}
with the scalar factor $\mathcal{G}_\phi$ being 
\begin{equation}
\mathcal{G}_\phi = \frac{g^{\nu S}_\phi \mathcal{Q}_\phi}{G_F M_\phi^2} \, .
\end{equation}
Analogously to the previous case, the corresponding scalar charge is defined as~\cite{Cerdeno:2016sfi}
\begin{equation}
\mathcal{Q}_\phi = \sum_{\mathcal{N},q} g^{q S}_\phi \frac{m_\mathcal{N}}{m_q} f_{T,q}^{(\mathcal{N})} \,  ,
\end{equation} 
where the form factors $f_{T,q}^{(\mathcal{N})}$ capture the effective low-energy coupling of $\phi$ to the nucleon $\mathcal{N}= \{p,n \}$ ($m_\mathcal{N}$ is the nucleon mass) for the quark $q$. 

As discussed previously, different new physics signatures are expected to leave different imprints on the event and recoil spectrum.
Contrary to the $Z^\prime$  scenario  discussed above, the Dirac structure of the  $\phi \bar{\nu \nu}$ vertex accounting for the scalar mediator is different (chirality-flipping) with respect to the SM one (chirality-conserving). Indeed, there is no interference between vector (or axial-vector) neutrino  interactions and (pseudo-)scalar, tensor neutrino interactions~\cite{Rodejohann:2017vup,Farzan:2018gtr}. Therefore, the absence of interference between SM-scalar interactions gives rise to an overall modification of the expected \cevns spectrum (see Ref.~\cite{Bertuzzo:2017tuf}). Moreover, comparing the vector and scalar cross sections it becomes evident that the corresponding signals are expected to be well distinguishable. The scalar effects are not pronounced at eV-thresholds, while on the contrary they are expected to be stronger at recoil energies of the order of keV. A thorough classification of the new physics signatures with respect to vector and scalar interactions is given in Ref.~\cite{AristizabalSierra:2019ykk} providing also key information on the possibility of breaking isospin-related degeneracies from combined measurements with different detector material.

Assuming universal couplings, one finds the equalities~\cite{Cerdeno:2016sfi}
\begin{equation}
g_{Z^\prime}^2 = \frac{g^{\nu V}_{Z^\prime} \mathcal{Q}_{Z^\prime}}{3 A}\, , \quad g_\phi^2 =  \frac{g^{\nu S}_\phi \mathcal{Q}_\phi}{ \left(14 A + 1.1 Z \right) } \, .
\end{equation}
Using the COHERENT data, bounds have been put on the parameter planes $(g_{Z^\prime}^2,M_{Z^\prime})$ and $(g_{\phi}^2,M_{\phi})$ for the vector and scalar mediators, respectively~\cite{Kosmas:2017tsq}. 
In the left panel of Fig.~\ref{fig:zprime-scalar}, the limits are shown at 90\% C.L., where a degenerate area  appears that  cannot be excluded by the data is found due to the cancellations involved in Eq.(\ref{eq:G_z-prime}). However as shown in Ref.~\cite{Liao:2017uzy} this degeneracy can be broken in the context of NSI, while for heavy mediator masses, $M_{Z^\prime} \gg \sqrt{2 M T_N} \sim 50~\mathrm{MeV}$, it remains unbroken and depends on the ratio
\begin{equation}
\frac{g^2_{Z^\prime}}{M^2_{Z^\prime}} \approx 2 \sqrt{2} G_F\frac{\mathcal{Q}_W^V}{3 A}  \, .
\end{equation}
For the case of light mediator masses $M_{Z^\prime} \ll \sqrt{2 M T_N}$, it holds
\begin{equation}
g^2_{Z^\prime} \approx 4 \sqrt{2}G_F\frac{\mathcal{Q}_W^V}{3 A}  M T_N\, ,
\end{equation}
which implies that the bound is only sensitive to the coupling. The latter could  be drastically improved by combining data from different detectors~\cite{Shoemaker:2017lzs}. The case of a scalar field mediating \cevns is explored in the right panel of Fig.~\ref{fig:zprime-scalar} where the extracted bounds in the $(M_\phi,g_\phi^2)$ plane are depicted at 90\% C.L. Significant improvements are possible through powerful analyses that are based on both energy and timing COHERENT data~\cite{Dutta:2019eml} as well as by taking into account  improved quenching factors~\cite{Papoulias:2019txv}. The future of \cevns experiments will offer a complementary probe to various existing limits  in the low- and high-energy regime. The currently best results for a low-energy light vector mediator of $M_{Z^\prime} < 10~\mathrm{MeV}$ have been recently reported by the CONNIE Collaboration~\cite{Aguilar-Arevalo:2019zme}. The attainable sensitivity is expected to be competitive with existing bounds from neutrino-electron scattering, dark photon searches at BaBar and LHCb results (see Refs.~\cite{Abdullah:2018ykz,Billard:2018jnl}). Before closing our discussion it is important to note that, very recently CP violating effects have been also analyzed with the current and future COHERENT data in the context of light  vector  mediator  scenarios~\cite{AristizabalSierra:2019ufd}. The latter have been also found to be applicable to reactor or solar/atmospheric neutrino searches with important implications on multi-ton dark matter detectors.


\subsection{ Studying electromagnetic neutrino interactions }

\begin{figure}[t]
\centering
\begin{minipage}{0.45\textwidth}\includegraphics[width= \linewidth]{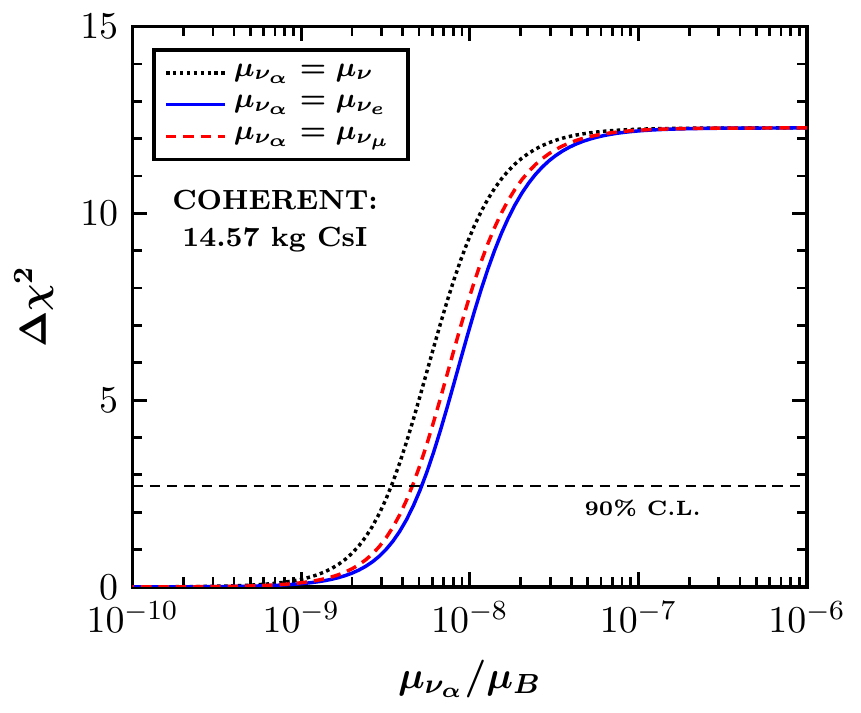}
\end{minipage}
\begin{minipage}{0.45\textwidth}
\includegraphics[width= \linewidth]{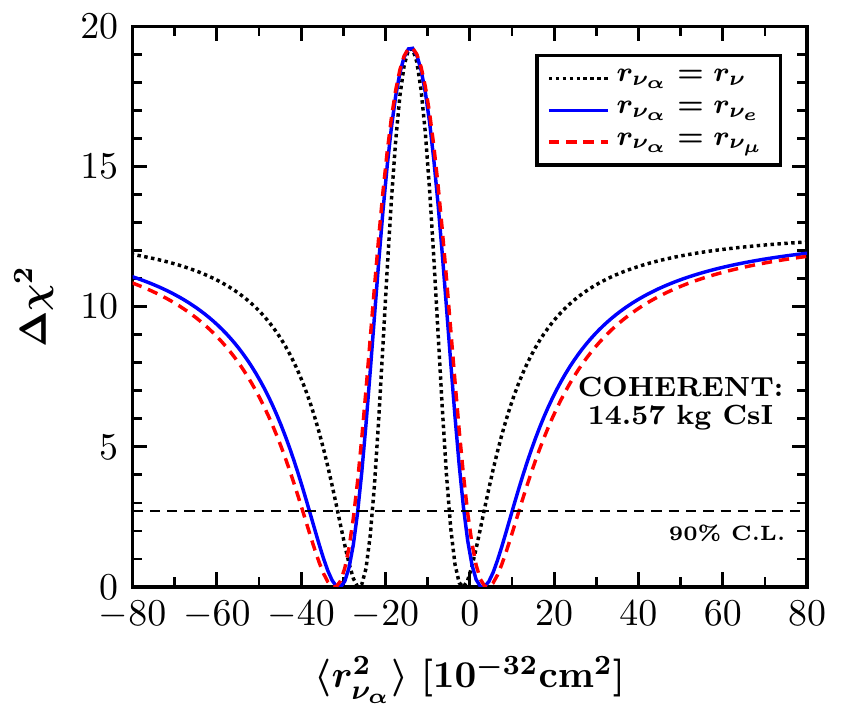}
\end{minipage}
\caption{Constraints on electromagnetic neutrino properties by the COHERENT experiment. Left: sensitivity to the effective neutrino magnetic moment $\mu_{\nu_\alpha}$. Right: sensitivity to the neutrino charge-radius $\langle r_{\nu_\alpha}^2\rangle$.  Figure adapted from Ref.~\cite{Kosmas:2017tsq}  under the terms of the Creative Commons Attribution 4.0 International license.}
\label{fig:charge-radius}
\end{figure}
%
\begin{figure*}[t]
\includegraphics[width=\textwidth]{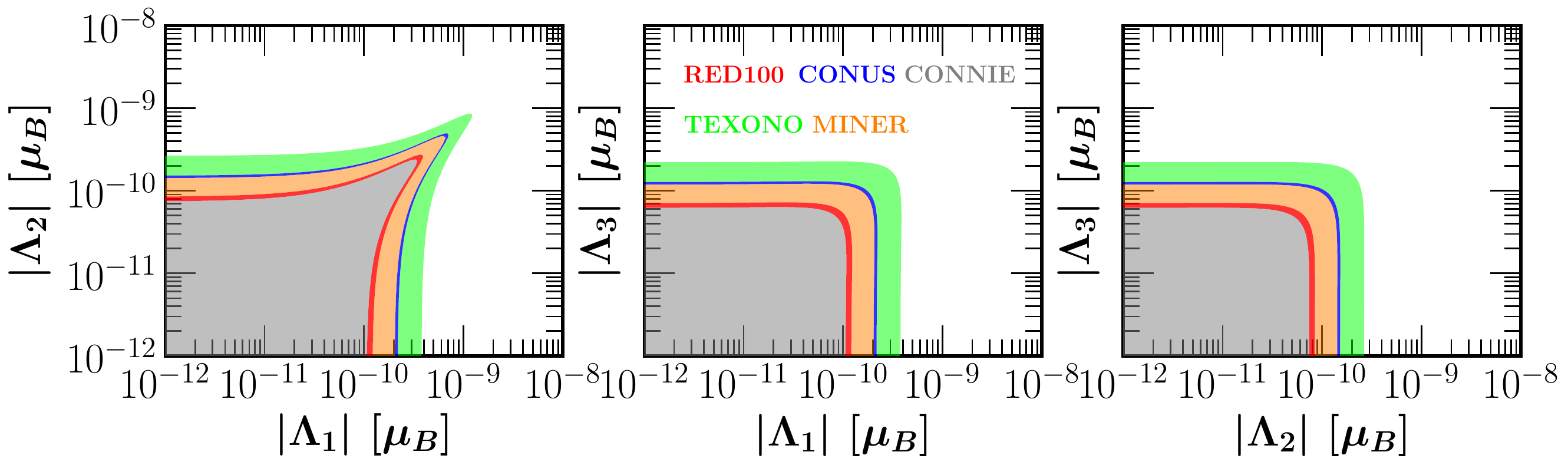}
\includegraphics[width=\textwidth]{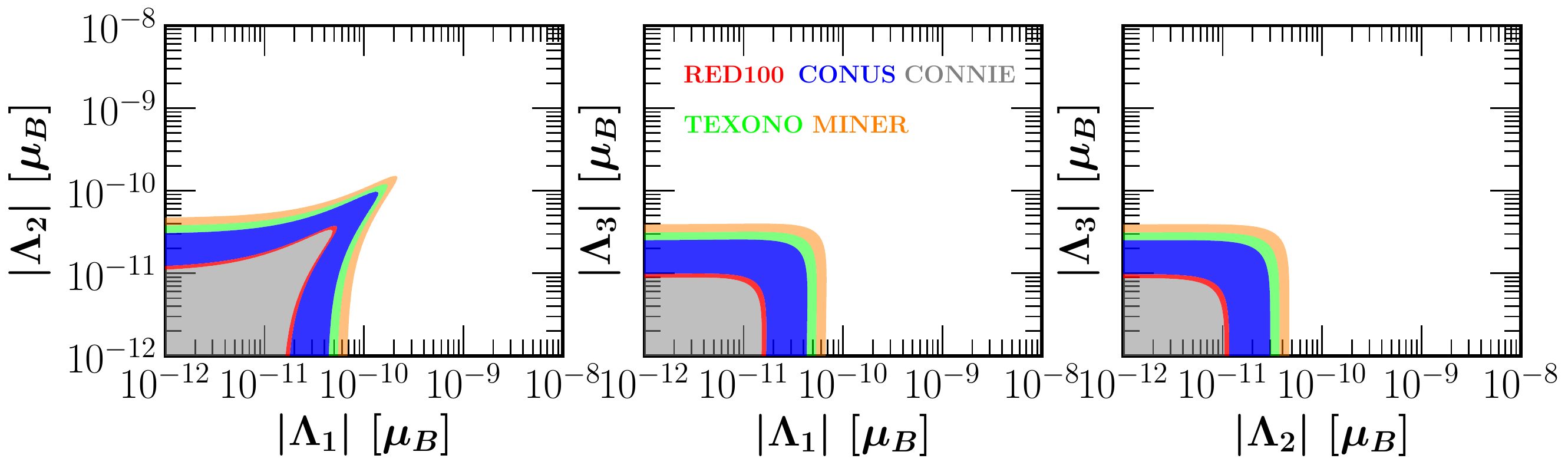}
\caption{Sensitivity contours at 90\% C.L.  in the $\left \vert \Lambda_i \right \vert - \left \vert \Lambda_j \right \vert$ parameter space that follow from the expected results of the current and future reactor based \cevns experiments (see the text). The calculation assumes vanishing $\left \vert \Lambda_k \right \vert$ and vanishing CP phases $\zeta_i$. Figure adapted from Ref.~\cite{Miranda:2019wdy}  under the terms of the Creative Commons Attribution 4.0 International license.}
\label{fig:Li_vs_Lj_reactors}
\end{figure*}
%
%
Non-trivial electromagnetic (EM) properties of massive neutrinos constitute an interesting probe to look for physics beyond the SM and at the same time they are crucial for distinguishing between the Dirac or Majorana nature of neutrinos~\cite{Giunti:2014ixa}. The two main phenomenological parameters observable at a neutrino experiment are the effective neutrino magnetic moment and the neutrino charge radius (the possibility of a neutrino millicharge is explored in Refs.~\cite{Parada:2019gvy,Cadeddu:2019eta}). Assuming Majorana neutrinos, the EM neutrino vertex is described by the electromagnetic field tensor $F_{\alpha \beta}$ of the effective Hamiltonian~\cite{Schechter:1981hw,Schechter:1981cv}
\begin{equation}
H_{\text{EM}}^{\mathrm{M}} = -\frac{1}{4} \nu_L^\mathsf{T} C^{-1} \lambda \sigma^{\alpha \beta} \nu_L F_{\alpha \beta} + \mathrm{H.c.} \, ,
\label{Hamiltonian:Majorana}
\end{equation}
while for the case of Dirac neutrinos one has
\begin{equation}
H_{\text{EM}}^{\mathrm{D}} = \frac{1}{2} \bar{\nu}_R \lambda \sigma^{\alpha \beta} \nu_L F_{\alpha \beta} + \mathrm{H.c.} 
\label{Hamiltonian:Dirac}
\end{equation}
It is important to note that for Majorana (Dirac) neutrinos  $\lambda = \mu - i \epsilon $ is an antisymmetric complex (general complex) 
matrix.  The two imaginary matrices $\mu$ (magnetic moment) and $\epsilon$ (electric dipole moment) obey the respective properties
$\mu^{\mathsf{T}} = -\mu$ ($\mu = \mu^\dagger$)   while $\epsilon^{\mathsf{T}} = -\epsilon$ ($\epsilon = \epsilon^\dagger$). It thus becomes evident that, unlike the Dirac case, for Majorana neutrinos the diagonal moments are vanishing  
$\mu^\mathrm{M}_{ii} = \epsilon^\mathrm{M}_{ii} = 0$.

For a low-energy neutrino scattering experiment the observable neutrino magnetic moment (flavor dependent) is in fact a combination of the neutrino transition magnetic moments (TMMs) discussed above. In the mass basis it reads~\cite{Grimus:2000tq}
\begin{equation}
\left(\mu_\nu^{M} \right)^2 = \tilde{\mathfrak{a}}_{-}^\dagger \tilde{\lambda}^\dagger \tilde{\lambda} \tilde{\mathfrak{a}}_{-} + \tilde{\mathfrak{a}}_{+}^\dagger \tilde{\lambda} \tilde{\lambda}^\dagger \tilde{\mathfrak{a}}_{+} \, .
\label{eq:TMM-mass}
\end{equation} 
In Eq.(\ref{eq:TMM-mass}) the following transformations have been introduced
\begin{equation}
\tilde{\mathfrak{a}}_{-} = U^\dagger \mathfrak{a}_{-}, \qquad  \tilde{\mathfrak{a}}_{+} = U^\mathsf{T} \mathfrak{a}_{+}, \qquad \tilde{\lambda} = U^\mathsf{T} \lambda U \, , 
\end{equation}
where the $3-$vectors $\mathfrak{a}_{+}$ and $\mathfrak{a}_{-}$ denote positive and negative helicity states respectively, while the magnetic moment matrix $\lambda$ ($\tilde{\lambda}$) in the flavor (mass) basis is written as~\cite{Tortola:2004vh}
\begin{equation}
\lambda = \left( \begin{array}{ccc}
0 & \Lambda_\tau & - \Lambda_\mu \\
- \Lambda_\tau &  0 & \Lambda_e \\
\Lambda_\mu & - \Lambda_e & 0
\end{array} \right), \qquad
\tilde{\lambda} = \left( \begin{array}{ccc}
0 & \Lambda_3 & - \Lambda_2 \\
- \Lambda_3 &  0 & \Lambda_1 \\
\Lambda_2 & - \Lambda_1 & 0
\end{array} \right) \, .
\label{NMM:matrix}
\end{equation}
with $\Lambda_\alpha = \left \vert \Lambda_\alpha \right \vert e^{i \zeta_\alpha}$ and $\Lambda_i = \left \vert \Lambda_i \right \vert e^{i \zeta_i}$ being the TMMs in the flavor and mass basis respectively, where $\zeta_\alpha$ and $\zeta_i$ denote the corresponding CP-phases.

The potential EM neutrino properties appear in the form of an effective neutrino magnetic moment that is conveniently expressed in the mass basis according to Eq.(\ref{eq:TMM-mass}) in terms of fundamental parameters (TMMs, CP-violating phases and neutrino mixing angles). The latter induce an additive contribution to the SM cross section~\cite{Vogel:1989iv}
\begin{equation}
\left(\frac{d \sigma}{dT_N} \right)_{\mathrm{SM+EM}} = \mathcal{G_{\mathrm{EM}}}(E_\nu, T_N) \left(\frac{d \sigma}{d T_N} \right)_{\mathrm{SM}} \, , 
\label{eq:EM-crossec}
\end{equation}
where the EM factor reads~\cite{Kosmas:2017tsq}
\begin{equation}
\mathcal{G}_\mathrm{EM} = 1 + \frac{1}{G_F^2 M}\left(\frac{\mathcal{Q}_{\mathrm{EM}}}{\mathcal{Q}_W^V} \right)^2 \frac{\frac{1- T_N/E_\nu}{T_N}}{1 - \frac{M T_N}{2 E_\nu^2}} \, .
\end{equation}
Here, the EM charge $\mathcal{Q}_{\mathrm{EM}}$  is written in terms of the fine structure constant $a_{\mathrm{EM}}$ and the effective neutrino magnetic moment, as~\cite{Scholberg:2005qs}
\begin{equation}
\mathcal{Q}_{\mathrm{EM}} = \frac{\pi a_{\mathrm{EM}} \mu_{\nu_\alpha}}{m_e} Z \, .
\label{eq:EM-charge}
\end{equation}
Moreover, the effect of a neutrino charge radius can be taken into consideration through a shift in the definition of the weak mixing angle~\cite{Hirsch:2002uv}
\begin{equation}
\sin^2 \theta_W \rightarrow \sin^2 \overline{\theta_W} + \frac{\sqrt{2} \pi a_{\mathrm{EM}}}{3 G_F} \langle r_{\nu_\alpha}^2 \rangle \, ,
\label{eq:charge-radius}
\end{equation}
where by $\sin^2 \overline{\theta_W}$ it is denoted the low energy value of the weak mixing angle, e.g. $\hat{s}_Z^2 = 0.2382$.

The presence of a neutrino magnetic moment is expected to yield a distortion in the recoil spectrum during the \cevns process, i.e. a detectable excess of events especially for low recoil energies. The left panel of Fig.~\ref{fig:charge-radius} shows the $\chi^2$ profile of the effective neutrino magnetic moment extracted by the first light of COHERENT data.  A similar analysis has been performed in order to quantify the sensitivity of COHERENT on the neutrino charge radius as shown in the right panel of the same figure. Note that, an essential improvement due to a more accurate treatment of the QF is possible (see Ref.~\cite{Papoulias:2019txv} for more details).

\begin{figure}[t]
\includegraphics[width= \linewidth]{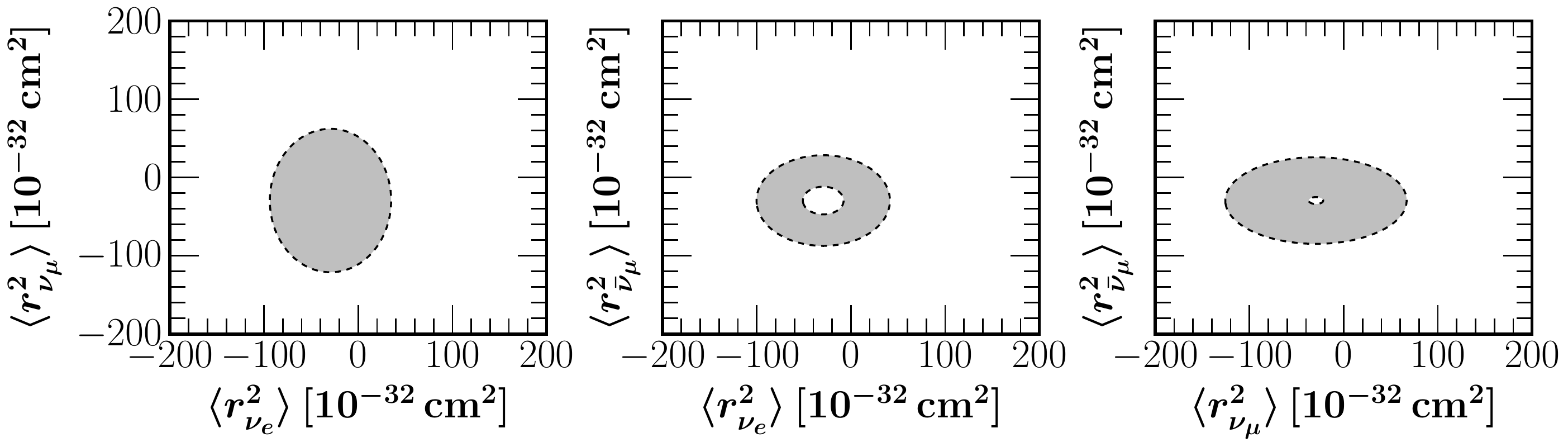}
\caption{Sensitivity in the $(\langle r_{\nu_\alpha}^2 \rangle, \langle r_{\nu_\beta}^2 \rangle)$ parameter space at 90\% C.L. from the analysis of the COHERENT data. Taken from Ref.~\cite{Papoulias:2019txv}. }
\label{fig:rv_combined}
\end{figure}

The authors of Ref.~\cite{Miranda:2019wdy} performed a comprehensive analysis on the sensitivity of various existing and future \cevns experiments and extracted constraints on the different components $\Lambda_i$ of the neutrino TMM matrix. In particular, their study focused on existing and next generation experimental setups of COHERENT as well as on the expected data from the future reactor-based experiments: CONUS, CONNIE, TEXONO, MINER and RED100. In a similar manner, Ref.~\cite{Billard:2018jnl} extracted constraints focusing on the NUCLEUS and Ricochet detectors at the Chooz NPP, however assuming the effective neutrino magnetic moment. 
Ref.~\cite{Miranda:2019wdy} performed a systematic  combined analysis with regards to the TMMs exploring also the effects of the CP violating phases of the complex matrix given in Eq.(\ref{NMM:matrix}).  As a concrete example, Fig.~\ref{fig:Li_vs_Lj_reactors} shows the contours in the ($\left \vert
\Lambda_i \right \vert$--$\left \vert
\Lambda_j \right \vert$) parameter plane for the case of current and next generation reactor-based \cevns experiments. It is worth mentioning that these bounds are comparable to existing ones from  low energy solar neutrino data at Borexino phase-II~\cite{Borexino:2017fbd}. Figure~\ref{fig:rv_combined} shows the sensitivity contours in the $(\langle r_{\nu_\alpha}^2 \rangle, \langle r_{\nu_\beta}^2 \rangle)$ plane that resulted from the COHERENT data.  A similar analysis is performed in Refs.~\cite{Papoulias:2019txv, Khan:2019cvi}, while for a comprehensive fit including energy and timing data the reader is referred to Ref.~\cite{Cadeddu:2019eta}.

%
%

\subsection{ The existence of the sterile neutrinos }

\begin{figure}
\includegraphics[width = 0.45 \textwidth]{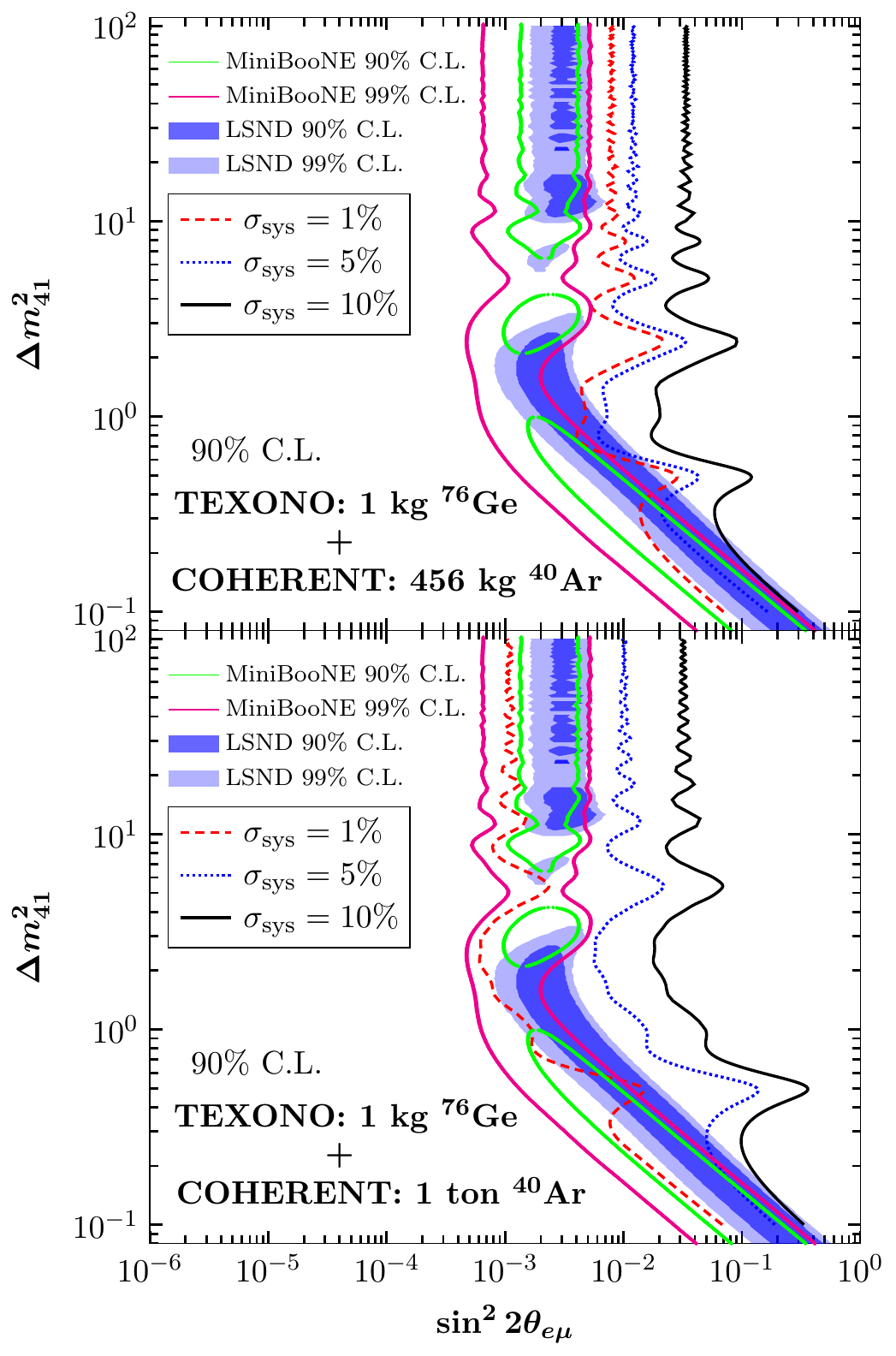}
\includegraphics[width = 0.45 \textwidth]{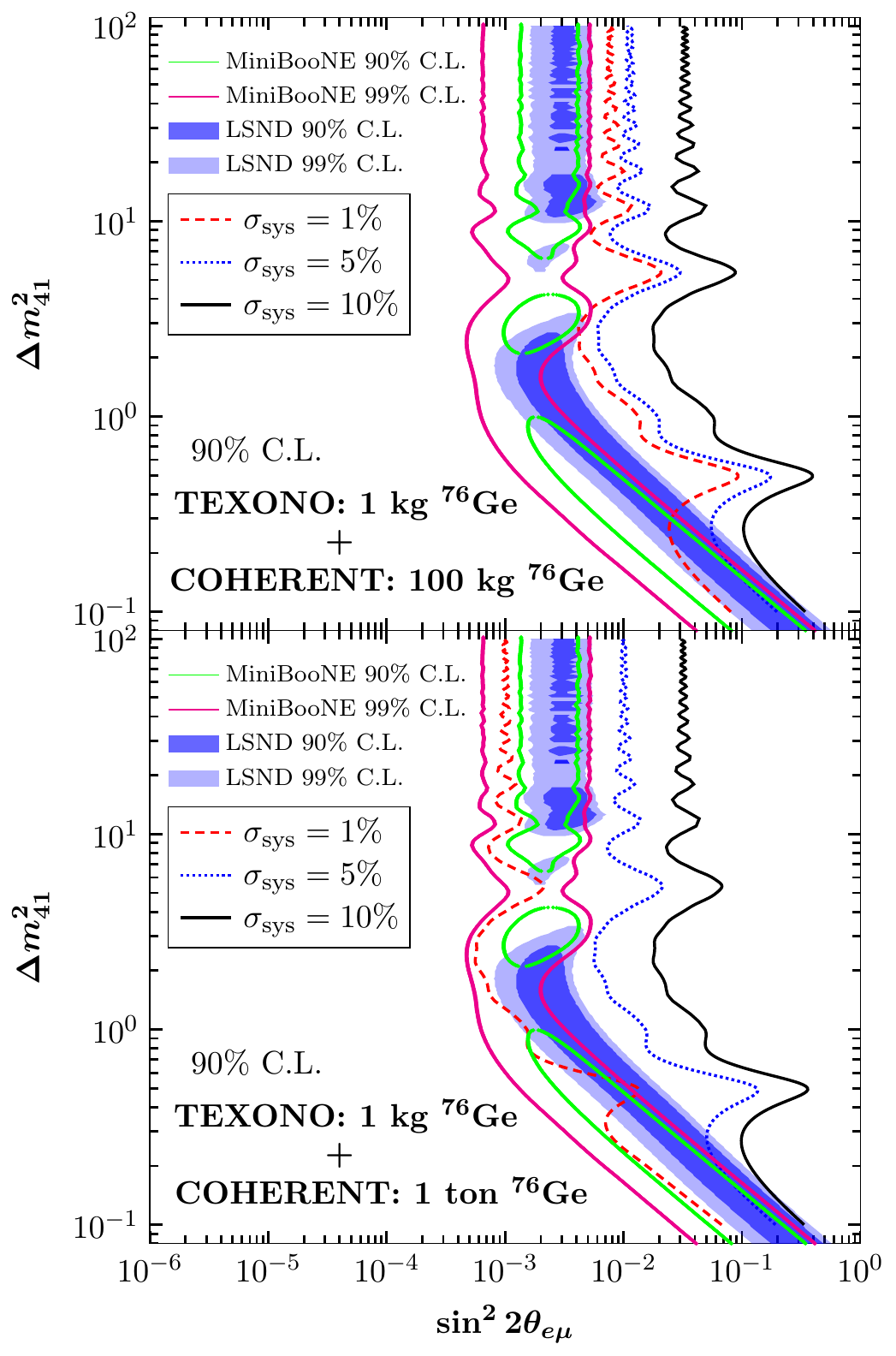}
\caption{Exclusion curves at 90\% C.L. in the  ($\Delta m^2_{41}$--$\sin^2 2 \theta_{e \mu}$)  parameter plane from a combined analysis of COHERENT and TEXONO experiments. The results are compared to existing constraints from MiniBooNE and LSND. Figure reproduced from Ref.~\cite{Kosmas:2017zbh}  with the permission of the American Physical Society and with updated results from MiniBooNE~\cite{Aguilar-Arevalo:2018gpe}.}
\label{fig:sterile}
\end{figure}

The three-neutrino paradigm has been put in rather solid grounds from the interpretation of solar and atmospheric oscillation data. On the other hand,  controversial anomalies such as those coming from recent reactor data as well as existing anomalies implied by the LSND and MiniBooNE experiments inspired a reach phenomenology beyond the three-neutrino oscillation picture, based on the existence of a fourth sterile neutrino state with  eV-scale mass ($m_{1,2,3} \ll m_4$) and tiny mixing angles.
To accommodate sterile neutrinos the lepton mixing matrix is minimally extended so that the flavor eigenstates $\nu_\alpha$, ($\alpha=e,\mu,\tau,s, \cdots $) are related to the mass eigenstates $\nu_i$  ($i= 1,2,3,4,\cdots $) by the  unitary transformation $\nu_\alpha =\sum_i U_{\alpha i} \nu_i$.
Then, for the short-baseline \cevns experiments the survival probability of an active neutrino at a distance $L$ is written~\cite{Blanco:2019vyp}
\begin{equation}
P_{\alpha \rightarrow e,\mu,\tau} = 1 - 4 \left \vert U_{\alpha 4} \right \vert^2 \left( 1-  \sum_{\beta = e, \mu, \tau} \left\vert U_{\beta 4} \right\vert^2 \right)  \sin^2 \left( \Delta \right) \, ,
\label{Eq.Pee}
\end{equation}
with the abbreviation $\Delta \equiv \Delta m^2 L /4 E_{\nu}$ and the mass splittings under the approximation $\Delta m^2_{41} \approx \Delta m^2_{42} \approx   \Delta m^2_{43} \equiv \Delta m^2$.  
At this point it should be stressed that neutrino-nucleus scattering experiments are favorable facilities to probe sterile neutrinos being complementary to dedicated experiments such as MINOS/MINOS+~\cite{Adamson:2017uda},  MiniBooNE~\cite{Aguilar-Arevalo:2018gpe}, Daya-Bay~\cite{Adey:2018zwh}, Juno~\cite{An:2015jdp} and NEOS~\cite{Ko:2016owz}. Indeed, due to the purely neutral-current character of the \cevns process it is not necessary to disentangle between active-sterile neutrino mixing~\cite{Formaggio:2011jt}.

The possibility of investigating sterile neutrinos in the simplest (3+1) scheme through the \cevns process was examined for the first time in Ref.~\cite{Anderson:2012pn}, relying on an SNS source. A combined sterile neutrino analysis was performed in Ref.~\cite{Kosmas:2017zbh} highlighting the complementarity between accelerator and reactor neutrino sources, by focusing on COHERENT and TEXONO experiments respectively (see Fig.~\ref{fig:sterile}). Moreover, a detailed study of various reactor-based \cevns proposals has been carried out in Ref.~\cite{Canas:2017umu}, showing how such future measurements can be exploited to solve the reactor antineutrino anomaly. After the first observation of \cevns by the COHERENT experiment, Ref.~\cite{Kosmas:2017tsq} reported the first constraints under the assumption of a universal new mixing angle, extracting the conclusion that the current sensitivity is rather poor. By exploiting timing data the potential of a future measurement at the next generation of COHERENT with a 100~kg CsI detector has been demonstrated in Ref.~\cite{Blanco:2019vyp}, concluding that the   prospects of  probing  the exclusion regions in the ($\Delta m^2_{41}$--$\sin^2 2 \theta_{e \mu}$) plane from the latest MiniBooNE~\cite{Aguilar-Arevalo:2018gpe} and LSND~\cite{Aguilar:2001ty} are promising. Finally, focusing at CONUS in Ref.~\cite{Berryman:2019nvr} it was shown that the complementarity between terrestrial-cosmological experiments may resolve the tension raised by astrophysical observations regarding the existence of sterile neutrinos.
%

\subsection{Summary of constraints}
Emphasis has been put on the physics beyond the SM by devoting a great part to the past and current research efforts  and by concentrating on the various channels contributing to \cevns processes and their interpretation. Through a $\chi^2$ sensitivity analysis based on the recoil or  timing spectra of the COHERENT data, the current limits are listed at 90\% C.L. in Table~\ref{tab:summary}. For a given parameter set $\mathcal{S}$, the best fit is found through the minimum value $\chi^2_\text{min}(\mathcal{S})$. The limits involve electroweak (weak-mixing angle), nuclear (nuclear radius), and physics beyond the SM (NSIs and EM neutrino properties). Significant improvements are expected through a more accurate determination of the QF and from a better control of the systematic uncertainties. The reported constraints on electroweak and NSIs have been extracted with various analysis methods, i.e. by combining existing APV measurements or global oscillation constraints with the recent COHERENT data, emphasizing the complementarity of \cevns data in the low energy regime.

\begin{table}[t!]
\begin{tabular}{l c c c}
\toprule
Parameter & dataset & Reference & Limit (90\% C.L.) \\
\hline
$\sin^2 \theta_W$~$^a$ & \multirow{2}{*}{COHERENT + APV} & \multirow{2}{*}{\cite{Cadeddu:2018izq}} & $0.239^{+0.006}_{-0.007}$ \\
$R_n$ &  & & $5.42^{+ 0.50}_{-0.50}$\\
\hline
$\epsilon_{ee}^{uV}$ & \multirow{4}{*}{COHERENT + oscillation} & \multirow{4}{*}{\cite{Coloma:2017egw}} & 0.028 -- 0.60 \\
$\epsilon_{ee}^{dV}$ &  & & 0.030 -- 0.55 \\
$\epsilon_{\mu \mu }^{uV}$ &   & & -0.088 -- 0.37 \\
$\epsilon_{\mu \mu}^{dV}$ &   & & -0.075 -- 0.33 \\
\hline
$\epsilon_{ee}^{uT}$ & \multirow{4}{*}{COHERENT (recoil)} & \multirow{4}{*}{\cite{Kosmas:2017tsq}} & -0.013 -- 0.013 \\
$\epsilon_{ee}^{dT}$ &   & & -0.011 -- 0.011 \\
$\epsilon_{\mu \mu }^{uT}$ &   & & -0.013 -- 0.013 \\
$\epsilon_{\mu \mu}^{dT}$ &   & &-0.011 -- 0.011 \\
\hline
$\mu_\nu$~ & \multirow{3}{*}{COHERENT (recoil)} & \multirow{3}{*}{\cite{Kosmas:2017tsq}} &  $< 43$   \\
$\mu_{\nu_e}$   &  & & $< 52$   \\
$\mu_{\nu_\mu}$ &  & & $< 46$   \\
\hline 
$\langle r^2_{\nu_e} \rangle$ & \multirow{5}{*}{COHERENT
 (timing and recoil)} & \multirow{5}{*}{\cite{Cadeddu:2018dux}} &  -63 -- 12  \\
$\langle r^2_{\nu_\mu} \rangle$         & & & -7 -- 9 \\
$\langle r^2_{\nu_{e\mu}} \rangle$      & & & $< 22$ \\
$\langle r^2_{\nu_{e\tau}} \rangle$     & & & $< 37$  \\
$\langle r^2_{\nu_{\mu \tau}} \rangle$~ & & & $< 26$  \\

\botrule
\footnotesize{$^a$ The limit is shown at 1$\sigma$.}
\end{tabular}

\caption{Constraints on electroweak, nuclear and new physics parameters at 90\% C.L. after the first \cevns measurement by the COHERENT experiment. The limits are presented in units of: fm for the nuclear rms radius, $10^{-10} \, \mu_B$ for the neutrino magnetic moment and  $10^{-32} \, \mathrm{cm^2}$ for the neutrino charge radius.}
\label{tab:summary}
\end{table}

\section{Connection of CE$\nu$NS with Dark Matter, cLFV processes and astrophysics}
\label{sect:connection}

Neutrino-nucleus scattering is one of the dominant processes taking place in SN environment and thus the emitted neutrinos can be an extremely useful tool for deep sky investigations.  Moreover, SN constitute an ideal source for flavor physics applications since all flavors are involved. Going beyond the SM, potential FCNC under stellar conditions can  modify the percentage of the neutrino flavors in the interior of massive stars~\cite{Amanik:2004vm}. The latter, may  drastically affect a plethora of other processes governing the explosive-stellar nucleosynthesis~\cite{Giannaka:2015sta}, causing significant alteration of the evolution phenomena~\cite{Amanik:2006ad,Stapleford:2016jgz}. If large enough, the modified neutrino energy-densities arriving at the terrestrial SN-neutrino detectors can be tested at \cevns experiments~\cite{Amaya:2011sn}. It should be stressed that a SN neutrino burst can be well detected by the current technology DM detectors.

\begin{figure}[t]
\includegraphics[width=\textwidth]{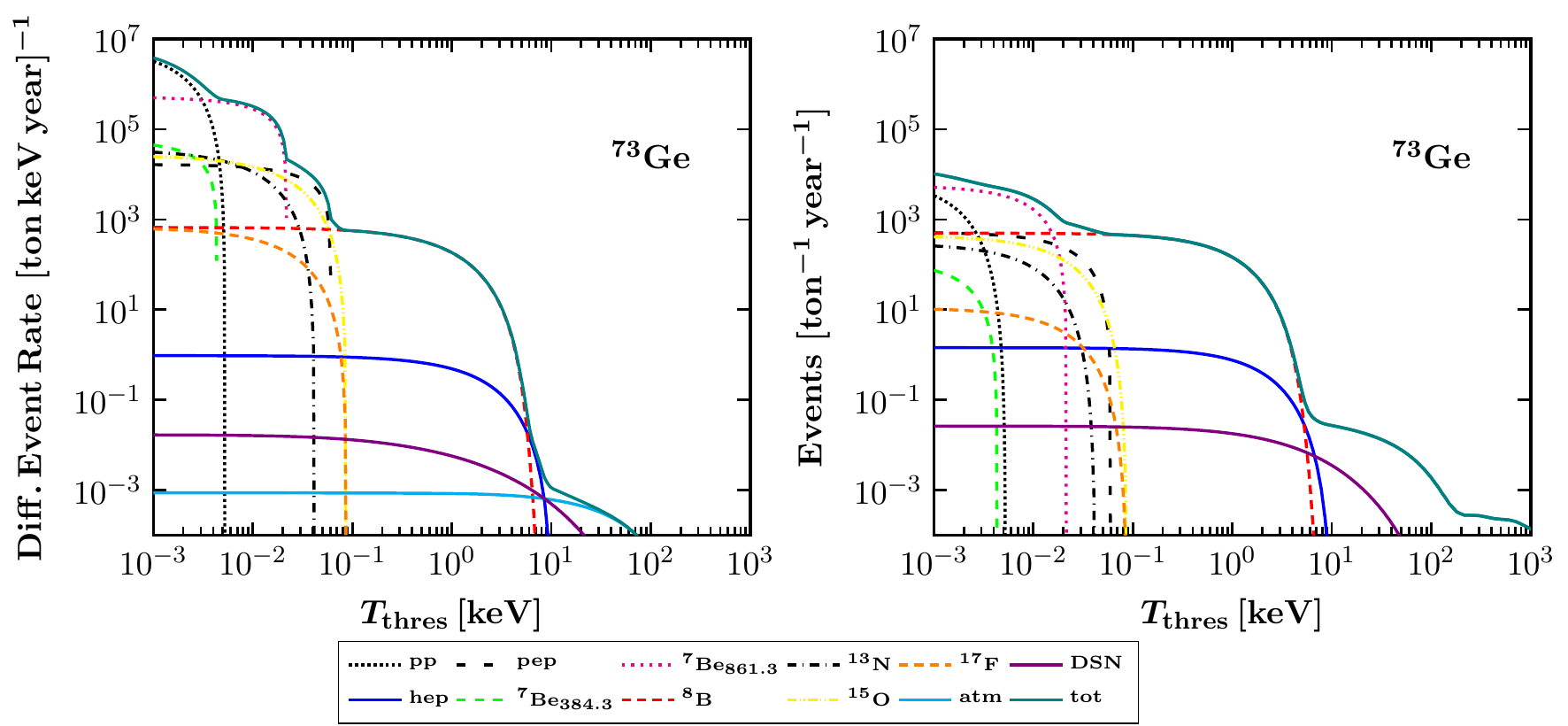}
\caption{Differential event rate (left) and total number of events above threshold (right) expected due to \cevns from solar, Atmospheric and DSNB neutrinos at a Germanium detector. Figure adapted from Ref.~\cite{Papoulias:2018uzy}  under the terms of the Creative Commons Attribution 4.0 International license.}
\label{Ge-DM-detector}
\end{figure}

Direct Dark Matter detection experiments are expected to be sensitive to astrophysical neutrinos from the Sun, the Atmosphere and from core-collapse SN (e.g. diffuse supernova background, DSNB)~\cite{Monroe:2007xp,OHare:2016pjy}. The neutrino-floor~\cite{Billard:2013qya}, being an irreducible background determines the criteria for using the appropriate detector material, threshold, mass, etc. Figure~\ref{Ge-DM-detector} illustrates the differential and integrated event rate of \cevns expected at a ton-scale DM detector, calculated in the framewrok of the DSM assuming only SM interactions. Future precision measurements at such rare-event facilities may become sensitive to nuclear structure effects which in principle can be explored by experiments looking for CE$\nu$NS. Therefore precise information on the nuclear form factors becomes very relevant for DM detectors especially for those involving multi-ton mass scale~\cite{AristizabalSierra:2019zmy}. For example, alterations are expected at high recoil energies of neutrino-induced interactions at direct DM detection searches~\cite{Papoulias:2018uzy} which on the other hand may be limited by the current uncertainties of the Atmospheric and DSNB neutrinos. Models involving light mediators are well testable at \cevns searches and may offer a key solution to LMA-Dark~\cite{Farzan:2015doa,Coloma:2017egw}  as well as implications to DM searches~\cite{Dent:2016wcr,Shoemaker:2017lzs} and to the neutrino floor~\cite{Boehm:2018sux}. In the same spirit, combined analyses of oscillation and \cevns data~\cite{Coloma:2017ncl,AristizabalSierra:2017joc} concluded that  the LMA-D solution is excluded at $3.1\sigma$ ($3.6\sigma$) for NSI with up (down) quarks. Finally, it has been recently pointed out that potential DM-induced signatures from dark photon decay could be also detectable at \cevns experiments,  explaining an excess in the timing distribution of the COHERENT signal~\cite{Dutta:2019nbn}.

In the case of cLFV processes of $\mu \to e$ transitions, especially coherent $\mu^-\to e^-$ conversion in the field of nuclei has attracted much interest in the context of new physics mechanisms discussed in this article~\cite{Kosmas:1994pt}. For example, $\mu \to e$ conversion has been studied in the context of the inverse seesaw~\cite{Deppisch:2005zm} and new $Z^\prime$ mediators~\cite{Farzan:2016wym}. It is given by
\begin{equation}
 \mu^{-} + (A,Z) \rightarrow e^{-} + (A,Z) \, ,
\label{mue}
\end{equation} 
which might have close relations to the process  given in Eq.(\ref{AHEP:neutrin-NSI}) in the neutral sector. When the final nuclear state coincides with the ground state, this process could be a coherent channel, which in fact dominates by its enhancement by a factor of the square of the number of nucleons  in nuclei. The cLFV processes are known to be highly suppressed in the SM even with lepton mixing due to the small neutrino masses, down to $\mathcal{O}(10^{-54})$ \cite{Petcov:1976ff}. However, many theoretical models involving NSI predict sizable rates which the future experiments could reach~\cite{Kuno:1999jp}. The future experiments aiming to search for $\mu^{-} \to e^{-}$ conversion are under preparation at J-PARC, Japan (COMET)~\cite{Lee:2018wcx} and Fermilab, in the USA (Mu2e)~\cite{Bernstein:2019fyh}. They expect to measure a  characteristic peak of outgoing electrons (at energy $E_e \approx m_\mu$) emitted from muonic atoms in a target. These experiments are aiming at sensitivities of the order of $\mathcal{O}(10^{-17})$  to $\mathcal{O}(10^{-18})$ , which is a factor of 10,000 or more improvement over the current experimental limits. Therefore they have excellent potential to establish or rule out the presence of new physics in the near future. 

It is important to notice that theoretically the $\mu^{-} \to e^{-}$ branching ratio depends on the nuclear form factor which can be probed from \cevns measurements as discussed in Sect~\ref{sect:constraints}. For the relevant nuclei such as $^{27}$Al and $^{48}$Ti the nuclear form factors at $q \approx m_\mu=0.53~\mathrm{fm^{-1}}$ have values 0.63 and 0.53 respectively i.e. well far from the approximation of point like nucleus (see Ref.~\cite{Kosmas:1988tz} for a detailed discussion). The incoherent channels of $\mu \to e$ conversion can be studied with the matrix elements described in Sect.~\ref{sect:theory} (see for example Refs.~\cite{Chiang:1993xz,Kosmas:1990tc}). Once $\mu^{-} \to e^{-}$ conversion is observed, it holds significant potential for constraining the parameters of the NSI Lagrangian of the lepton-nucleus interactions~\cite{Davidson:2003ha}. It may shed light on FCNC processes in the leptonic sector~\cite{Kitano:2002mt, Cirigliano:2009bz, Davidson:2018kud}, and particularly on the existence of the charged-lepton mixing which is analogous to neutrino oscillations at short baseline experiments.

\section{ Summary and conclusions}
\label{sect:conclusions}

In this review article, we made an attempt to summarize the main research efforts devoted to the conventional and  exotic neutrino-nucleus interactions, in the recent years. The standard process of neutral-current neutrino-nucleus scattering, mediated by the neutral $Z$-boson presents two channels: the elastic and inelastic scattering of neutrinos (and anti-neutrinos) off a nuclear isotope $(A,Z)$, with $A$ nucleons and $Z$ protons. In the elastic process, the initial and final states of the target nucleus are the same  and the detectable signal is an energy recoil, whereas in the case of the inelastic channel the final nucleus is an excited state with the signal being a de-excitation product (gammas). We have mainly concentrated on beyond the SM neutrino-nucleus interactions, and especially on the prospects of extracting new physics from the operating prominent rare-event detectors looking for the coherent elastic neutrino-nucleus scattering. Such channels may involve lepton LFV in neutral-currents. This is motivated by the recent measurements of CE$\nu$NS events at the COHERENT experiment, the analysis and interpretation of which  may imply  the necessity of including non-standard neutrino-nucleus interactions.  Towards this end, we discussed the impact of non-standard interactions and novel $Z^\prime$ or $\phi$ mediators to the CE$\nu$NS event rates providing an estimation of the attainable sensitivities at current and future experiments. With regards to neutrino oscillations constraints on NSIs from neutral current interactions at \cevns experiments are complementary since the former are only to sensitive to differences between the diagonal terms. It is  furthermore expected that the next generation of the currently operating experiments like the COHERENT, TEXONO, MINER, CONUS, RED100, vGEN, Ricochet, NUCLEUS etc., will be of benefit to unravel open issues of the leptonic sector. The studies covered in this review article have  evident connection with neutrino astronomy, SN physics, direct DM detection and  cLFV processes. To understand these new interactions the proton and neutron weak nuclear form factors play key roles. This opens up the necessity of measuring the neutron nuclear form factors by appropriately designed and appreciably sensitive experiments such as those looking for \cevns processes.

\acknowledgements 
\noindent
The authors are grateful to Valentina De Romeri and Jorge Terol Calvo for fruitful discussions as well as to M. Cadeddu and F. Dordei for useful correspondence. D.K.P. is supported by the Spanish grants SEV-2014-0398 and FPA2017-85216-P (AEI/FEDER, UE), PROMETEO/2018/165 (Generalitat
Valenciana) and the Spanish Red Consolider MultiDark FPA2017-90566-REDC.
Y.K. acknowledges support by the JSPS KAKENHI Grant No.\,18H04231.

\bibliographystyle{utphys}
\bibliography{review_CEvNS.bbl}

\providecommand{\href}[2]{#2}\begingroup\raggedright\begin{thebibliography}{100}

\bibitem{Ejiri:2019ezh}
H.~Ejiri, J.~Suhonen, and K.~Zuber, ``{Neutrino-nuclear responses for
  astro-neutrinos, single beta decays and double beta decays},''
  \href{http://dx.doi.org/10.1016/j.physrep.2018.12.001}{{\em Phys. Rept.}
  {\bfseries 797} (2019) 1--102}.

\bibitem{Schechter:1980gr}
J.~Schechter and J.~Valle, ``{Neutrino Masses in SU(2) x U(1) Theories},''
  \href{http://dx.doi.org/10.1103/PhysRevD.22.2227}{{\em Phys.Rev.} {\bfseries
  D22} (1980) 2227}.

\bibitem{Schechter:1981cv}
J.~Schechter and J.~Valle, ``{Neutrino Decay and Spontaneous Violation of
  Lepton Number},'' \href{http://dx.doi.org/10.1103/PhysRevD.25.774}{{\em
  Phys.Rev.} {\bfseries D25} (1982) 774}.

\bibitem{Kosmas:1996fh}
T.~Kosmas and E.~Oset, ``{Charged current neutrino nucleus reaction
  cross-sections at intermediate-energies},''
  \href{http://dx.doi.org/10.1103/PhysRevC.53.1409}{{\em Phys.Rev.} {\bfseries
  C53} (1996) 1409--1415}.

\bibitem{Ejiri:2000ps}
H.~Ejiri, ``{Nuclear spin isospin responses for low-energy neutrinos},''
  \href{http://dx.doi.org/10.1016/S0370-1573(00)00044-2}{{\em Phys.Rept.}
  {\bfseries 338} (2000) 265--351}.

\bibitem{Balasi:2015dba}
K.~Balasi, K.~Langanke, and G.~Mart{\'\i}nez-Pinedo,
  ``{Neutrino{\textendash}nucleus reactions and their role for supernova
  dynamics and nucleosynthesis},''
  \href{http://dx.doi.org/10.1016/j.ppnp.2015.08.001}{{\em
  Prog.Part.Nucl.Phys.} {\bfseries 85} (2015) 33--81},
  \href{http://arxiv.org/abs/1503.08095}{{\ttfamily arXiv:1503.08095
  [nucl-th]}}.

\bibitem{Freedman:1973yd}
D.~Z. Freedman, ``{Coherent Neutrino Nucleus Scattering as a Probe of the Weak
  Neutral Current},'' \href{http://dx.doi.org/10.1103/PhysRevD.9.1389}{{\em
  Phys.Rev.} {\bfseries D9} (1974) 1389--1392}.

\bibitem{Tubbs:1975jx}
D.~Tubbs and D.~Schramm, ``{Neutrino Opacities at High Temperatures and
  Densities},'' \href{http://dx.doi.org/10.1086/153909}{{\em Astrophys.J.}
  {\bfseries 201} (1975) 467--488}.

\bibitem{Drukier:1983gj}
A.~Drukier and L.~Stodolsky, ``{Principles and Applications of a Neutral
  Current Detector for Neutrino Physics and Astronomy},''
\href{http://dx.doi.org/10.1103/PhysRevD.30.2295}{{\em Phys. Rev.} {\bfseries
  D30} (1984) 2295}.

\bibitem{Akimov:2017ade}
{\bfseries COHERENT} Collaboration, D.~Akimov {\em et~al.}, ``{Observation of
  Coherent Elastic Neutrino-Nucleus Scattering},''
  \href{http://dx.doi.org/10.1126/science.aao0990}{{\em Science} {\bfseries
  357} (2017) 1123--1126}, \href{http://arxiv.org/abs/1708.01294}{{\ttfamily
  arXiv:1708.01294 [nucl-ex]}}.

\bibitem{Akimov:2018vzs}
{\bfseries COHERENT} Collaboration, D.~Akimov {\em et~al.}, ``{COHERENT
  Collaboration data release from the first observation of coherent elastic
  neutrino-nucleus scattering},''
  \href{http://arxiv.org/abs/1804.09459}{{\ttfamily arXiv:1804.09459
  [nucl-ex]}}.

\bibitem{AristizabalSierra:2019hcm}
D.~Aristizabal~Sierra {\em et~al.}, ``{Proceedings of The Magnificent CE$\nu$NS
  Workshop 2018},''
\href{http://arxiv.org/abs/1910.07450}{{\ttfamily arXiv:1910.07450 [hep-ex]}}.

\bibitem{Liao:2017uzy}
J.~Liao and D.~Marfatia, ``{COHERENT constraints on nonstandard neutrino
  interactions},'' \href{http://dx.doi.org/10.1016/j.physletb.2017.10.046}{{\em
  Phys.Lett.} {\bfseries B775} (2017) 54--57},
  \href{http://arxiv.org/abs/1708.04255}{{\ttfamily arXiv:1708.04255
  [hep-ph]}}.

\bibitem{Dent:2017mpr}
J.~B. Dent, B.~Dutta, S.~Liao, J.~L. Newstead, L.~E. Strigari, and J.~W.
  Walker, ``{Accelerator and reactor complementarity in coherent
  neutrino-nucleus scattering},''
  \href{http://dx.doi.org/10.1103/PhysRevD.97.035009}{{\em Phys.Rev.}
  {\bfseries D97} (2018) 035009},
  \href{http://arxiv.org/abs/1711.03521}{{\ttfamily arXiv:1711.03521
  [hep-ph]}}.

\bibitem{AristizabalSierra:2017joc}
D.~Aristizabal~Sierra, N.~Rojas, and M.~Tytgat, ``{Neutrino non-standard
  interactions and dark matter searches with multi-ton scale detectors},''
  \href{http://dx.doi.org/10.1007/JHEP03(2018)197}{{\em JHEP} {\bfseries 1803}
  (2018) 197}, \href{http://arxiv.org/abs/1712.09667}{{\ttfamily
  arXiv:1712.09667 [hep-ph]}}.

\bibitem{Denton:2018xmq}
P.~B. Denton, Y.~Farzan, and I.~M. Shoemaker, ``{Testing large non-standard
  neutrino interactions with arbitrary mediator mass after COHERENT data},''
  \href{http://dx.doi.org/10.1007/JHEP07(2018)037}{{\em JHEP} {\bfseries 1807}
  (2018) 037}, \href{http://arxiv.org/abs/1804.03660}{{\ttfamily
  arXiv:1804.03660 [hep-ph]}}.

\bibitem{Dutta:2019eml}
B.~Dutta, S.~Liao, S.~Sinha, and L.~E. Strigari, ``{Searching for Beyond the
  Standard Model Physics with COHERENT Energy and Timing Data},''
  \href{http://dx.doi.org/10.1103/PhysRevLett.123.061801}{{\em Phys. Rev.
  Lett.} {\bfseries 123} no.~6, (2019) 061801},
\href{http://arxiv.org/abs/1903.10666}{{\ttfamily arXiv:1903.10666 [hep-ph]}}.

\bibitem{Coloma:2017ncl}
P.~Coloma, M.~Gonzalez-Garcia, M.~Maltoni, and T.~Schwetz, ``{COHERENT
  Enlightenment of the Neutrino Dark Side},''
  \href{http://dx.doi.org/10.1103/PhysRevD.96.115007}{{\em Phys.Rev.}
  {\bfseries D96} (2017) 115007},
  \href{http://arxiv.org/abs/1708.02899}{{\ttfamily arXiv:1708.02899
  [hep-ph]}}.

\bibitem{Gonzalez-Garcia:2018dep}
M.~Gonzalez-Garcia, M.~Maltoni, Y.~F. Perez-Gonzalez, and
  R.~Zukanovich~Funchal, ``{Neutrino Discovery Limit of Dark Matter Direct
  Detection Experiments in the Presence of Non-Standard Interactions},''
  \href{http://dx.doi.org/10.1007/JHEP07(2018)019}{{\em JHEP} {\bfseries 1807}
  (2018) 019}, \href{http://arxiv.org/abs/1803.03650}{{\ttfamily
  arXiv:1803.03650 [hep-ph]}}.

\bibitem{Kosmas:2015sqa}
T.~Kosmas, O.~Miranda, D.~Papoulias, M.~Tortola, and J.~Valle, ``{Probing
  neutrino magnetic moments at the Spallation Neutron Source facility},''
  \href{http://dx.doi.org/10.1103/PhysRevD.92.013011}{{\em Phys.Rev.}
  {\bfseries D92} (2015) 013011},
  \href{http://arxiv.org/abs/1505.03202}{{\ttfamily arXiv:1505.03202
  [hep-ph]}}.

\bibitem{Kosmas:2017tsq}
D.~Papoulias and T.~Kosmas, ``{COHERENT constraints to conventional and exotic
  neutrino physics},'' \href{http://dx.doi.org/10.1103/PhysRevD.97.033003}{{\em
  Phys.Rev.} {\bfseries D97} (2018) 033003},
  \href{http://arxiv.org/abs/1711.09773}{{\ttfamily arXiv:1711.09773
  [hep-ph]}}.

\bibitem{Miranda:2019wdy}
O.~Miranda, D.~Papoulias, M.~T{\'o}rtola, and J.~Valle, ``{Probing neutrino
  transition magnetic moments with coherent elastic neutrino-nucleus
  scattering},'' \href{http://dx.doi.org/10.1007/JHEP07(2019)103}{{\em JHEP}
  {\bfseries 1907} (2019) 103},
  \href{http://arxiv.org/abs/1905.03750}{{\ttfamily arXiv:1905.03750
  [hep-ph]}}.

\bibitem{Parada:2019gvy}
A.~Parada, ``{New constraints on neutrino electric millicharge from elastic
  neutrino-electron scattering and coherent elastic neutrino-nucleus
  scattering},'' \href{http://arxiv.org/abs/1907.04942}{{\ttfamily
  arXiv:1907.04942 [hep-ph]}}.

\bibitem{Kosmas:2017zbh}
T.~Kosmas, D.~Papoulias, M.~Tortola, and J.~Valle, ``{Probing light sterile
  neutrino signatures at reactor and Spallation Neutron Source neutrino
  experiments},'' \href{http://dx.doi.org/10.1103/PhysRevD.96.063013}{{\em
  Phys.Rev.} {\bfseries D96} (2017) 063013},
  \href{http://arxiv.org/abs/1703.00054}{{\ttfamily arXiv:1703.00054
  [hep-ph]}}.

\bibitem{Canas:2017umu}
B.~Ca{\~n}as, E.~Garc{\'e}s, O.~Miranda, and A.~Parada, ``{The reactor
  antineutrino anomaly and low energy threshold neutrino experiments},''
  \href{http://dx.doi.org/10.1016/j.physletb.2017.11.074}{{\em Phys.Lett.}
  {\bfseries B776} (2018) 451--456},
  \href{http://arxiv.org/abs/1708.09518}{{\ttfamily arXiv:1708.09518
  [hep-ph]}}.

\bibitem{Blanco:2019vyp}
C.~Blanco, D.~Hooper, and P.~Machado, ``{Constraining Sterile Neutrino
  Interpretations of the LSND and MiniBooNE Anomalies with Coherent Neutrino
  Scattering Experiments},'' \href{http://arxiv.org/abs/1901.08094}{{\ttfamily
  arXiv:1901.08094 [hep-ph]}}.

\bibitem{AristizabalSierra:2019ufd}
D.~Aristizabal~Sierra, V.~De~Romeri, and N.~Rojas, ``{CP violating effects in
  coherent elastic neutrino-nucleus scattering processes},''
  \href{http://dx.doi.org/10.1007/JHEP09(2019)069}{{\em JHEP} {\bfseries 09}
  (2019) 069},
\href{http://arxiv.org/abs/1906.01156}{{\ttfamily arXiv:1906.01156 [hep-ph]}}.

\bibitem{Dent:2016wcr}
J.~B. Dent, B.~Dutta, S.~Liao, J.~L. Newstead, L.~E. Strigari, and J.~W.
  Walker, ``{Probing light mediators at ultralow threshold energies with
  coherent elastic neutrino-nucleus scattering},''
  \href{http://dx.doi.org/10.1103/PhysRevD.96.095007}{{\em Phys.Rev.}
  {\bfseries D96} (2017) 095007},
  \href{http://arxiv.org/abs/1612.06350}{{\ttfamily arXiv:1612.06350
  [hep-ph]}}.

\bibitem{Farzan:2018gtr}
Y.~Farzan, M.~Lindner, W.~Rodejohann, and X.-J. Xu, ``{Probing neutrino
  coupling to a light scalar with coherent neutrino scattering},''
  \href{http://dx.doi.org/10.1007/JHEP05(2018)066}{{\em JHEP} {\bfseries 1805}
  (2018) 066}, \href{http://arxiv.org/abs/1802.05171}{{\ttfamily
  arXiv:1802.05171 [hep-ph]}}.

\bibitem{Abdullah:2018ykz}
M.~Abdullah, J.~B. Dent, B.~Dutta, G.~L. Kane, S.~Liao, and L.~E. Strigari,
  ``{Coherent elastic neutrino nucleus scattering as a probe of a $Z^\prime$
  through kinetic and mass mixing effects},''
  \href{http://dx.doi.org/10.1103/PhysRevD.98.015005}{{\em Phys.Rev.}
  {\bfseries D98} (2018) 015005},
  \href{http://arxiv.org/abs/1803.01224}{{\ttfamily arXiv:1803.01224
  [hep-ph]}}.

\bibitem{Brdar:2018qqj}
V.~Brdar, W.~Rodejohann, and X.-J. Xu, ``{Producing a new Fermion in Coherent
  Elastic Neutrino-Nucleus Scattering: from Neutrino Mass to Dark Matter},''
  \href{http://dx.doi.org/10.1007/JHEP12(2018)024}{{\em JHEP} {\bfseries 1812}
  (2018) 024}, \href{http://arxiv.org/abs/1810.03626}{{\ttfamily
  arXiv:1810.03626 [hep-ph]}}.

\bibitem{Cadeddu:2017etk}
M.~Cadeddu, C.~Giunti, Y.~Li, and Y.~Zhang, ``{Average CsI neutron density
  distribution from COHERENT data},''
  \href{http://dx.doi.org/10.1103/PhysRevLett.120.072501}{{\em Phys.Rev.Lett.}
  {\bfseries 120} (2018) 072501},
  \href{http://arxiv.org/abs/1710.02730}{{\ttfamily arXiv:1710.02730
  [hep-ph]}}.

\bibitem{Ciuffoli:2018qem}
E.~Ciuffoli, J.~Evslin, Q.~Fu, and J.~Tang, ``{Extracting nuclear form factors
  with coherent neutrino scattering},''
  \href{http://dx.doi.org/10.1103/PhysRevD.97.113003}{{\em Phys.Rev.}
  {\bfseries D97} (2018) 113003},
  \href{http://arxiv.org/abs/1801.02166}{{\ttfamily arXiv:1801.02166
  [physics.ins-det]}}.

\bibitem{Huang:2019ene}
X.-R. Huang and L.-W. Chen, ``{Neutron Skin in CsI and Low-Energy Effective
  Weak Mixing Angle from COHERENT Data},''
  \href{http://arxiv.org/abs/1902.07625}{{\ttfamily arXiv:1902.07625
  [hep-ph]}}.

\bibitem{AristizabalSierra:2019zmy}
D.~Aristizabal~Sierra, J.~Liao, and D.~Marfatia, ``{Impact of form factor
  uncertainties on interpretations of coherent elastic neutrino-nucleus
  scattering data},'' \href{http://dx.doi.org/10.1007/JHEP06(2019)141}{{\em
  JHEP} {\bfseries 1906} (2019) 141},
  \href{http://arxiv.org/abs/1902.07398}{{\ttfamily arXiv:1902.07398
  [hep-ph]}}.

\bibitem{Papoulias:2019lfi}
D.~Papoulias, T.~Kosmas, R.~Sahu, V.~Kota, and M.~Hota, ``{Constraining nuclear
  physics parameters with current and future COHERENT data},''
  \href{http://arxiv.org/abs/1903.03722}{{\ttfamily arXiv:1903.03722
  [hep-ph]}}.

\bibitem{Arcadi:2019uif}
G.~Arcadi, M.~Lindner, J.~Martins, and F.~S. Queiroz, ``{New Physics Probes:
  Atomic Parity Violation, Polarized Electron Scattering and Neutrino-Nucleus
  Coherent Scattering},'' \href{http://arxiv.org/abs/1906.04755}{{\ttfamily
  arXiv:1906.04755 [hep-ph]}}.

\bibitem{Cadeddu:2019qmv}
M.~Cadeddu, F.~Dordei, C.~Giunti, K.~Kouzakov, E.~Picciau, and A.~Studenikin,
  ``{Potentialities of a low-energy detector based on $^4$He evaporation to
  observe atomic effects in coherent neutrino scattering and physics
  perspectives},'' \href{http://arxiv.org/abs/1907.03302}{{\ttfamily
  arXiv:1907.03302 [hep-ph]}}.

\bibitem{Papoulias:2018uzy}
D.~Papoulias, R.~Sahu, T.~Kosmas, V.~Kota, and B.~Nayak, ``{Novel
  neutrino-floor and dark matter searches with deformed shell model
  calculations},'' \href{http://dx.doi.org/10.1155/2018/6031362}{{\em Adv.High
  Energy Phys.} {\bfseries 2018} (2018) 6031362},
  \href{http://arxiv.org/abs/1804.11319}{{\ttfamily arXiv:1804.11319
  [hep-ph]}}.

\bibitem{Boehm:2018sux}
C.~B{\oe}hm, D.~Cerde{\~n}o, P.~N. Machado, A.~Olivares-Del~Campo, and E.~Reid,
  ``{How high is the neutrino floor?},''
  \href{http://dx.doi.org/10.1088/1475-7516/2019/01/043}{{\em JCAP} {\bfseries
  1901} (2019) 043}, \href{http://arxiv.org/abs/1809.06385}{{\ttfamily
  arXiv:1809.06385 [hep-ph]}}.

\bibitem{Link:2019pbm}
J.~M. Link and X.-J. Xu, ``{Searching for BSM neutrino interactions in dark
  matter detectors},'' \href{http://dx.doi.org/10.1007/JHEP08(2019)004}{{\em
  JHEP} {\bfseries 08} (2019) 004},
\href{http://arxiv.org/abs/1903.09891}{{\ttfamily arXiv:1903.09891 [hep-ph]}}.

\bibitem{Ge:2017mcq}
S.-F. Ge and I.~M. Shoemaker, ``{Constraining Photon Portal Dark Matter with
  Texono and Coherent Data},''
  \href{http://dx.doi.org/10.1007/JHEP11(2018)066}{{\em JHEP} {\bfseries 1811}
  (2018) 066}, \href{http://arxiv.org/abs/1710.10889}{{\ttfamily
  arXiv:1710.10889 [hep-ph]}}.

\bibitem{Ng:2017aur}
K.~C.~Y. Ng, J.~F. Beacom, A.~H.~G. Peter, and C.~Rott, ``{Solar Atmospheric
  Neutrinos: A New Neutrino Floor for Dark Matter Searches},''
  \href{http://dx.doi.org/10.1103/PhysRevD.96.103006}{{\em Phys.Rev.}
  {\bfseries D96} (2017) 103006},
  \href{http://arxiv.org/abs/1703.10280}{{\ttfamily arXiv:1703.10280
  [astro-ph.HE]}}.

\bibitem{Dutta:2019nbn}
B.~Dutta, D.~Kim, S.~Liao, J.-C. Park, S.~Shin, and L.~E. Strigari, ``{Dark
  matter signals from timing spectra at neutrino experiments},''
\href{http://arxiv.org/abs/1906.10745}{{\ttfamily arXiv:1906.10745 [hep-ph]}}.

\bibitem{Wong:2010zzc}
H.~T. Wong, ``{Neutrino-nucleus coherent scattering and dark matter searches
  with sub-keV germanium detector},''
  \href{http://dx.doi.org/10.1016/j.nuclphysa.2010.05.040}{{\em Nucl.Phys.}
  {\bfseries A844} (2010) 229C--233C}.

\bibitem{Aguilar-Arevalo:2016qen}
{\bfseries CONNIE} Collaboration, A.~Aguilar-Arevalo {\em et~al.}, ``{Results
  of the Engineering Run of the Coherent Neutrino Nucleus Interaction
  Experiment (CONNIE)},''
  \href{http://dx.doi.org/10.1088/1748-0221/11/07/P07024}{{\em JINST}
  {\bfseries 11} (2016) P07024},
  \href{http://arxiv.org/abs/1604.01343}{{\ttfamily arXiv:1604.01343
  [physics.ins-det]}}.

\bibitem{Agnolet:2016zir}
{\bfseries MINER} Collaboration, G.~Agnolet {\em et~al.}, ``{Background Studies
  for the MINER Coherent Neutrino Scattering Reactor Experiment},''
  \href{http://dx.doi.org/10.1016/j.nima.2017.02.024}{{\em Nucl.Instrum.Meth.}
  {\bfseries A853} (2017) 53--60},
  \href{http://arxiv.org/abs/1609.02066}{{\ttfamily arXiv:1609.02066
  [physics.ins-det]}}.

\bibitem{Belov:2015ufh}
V.~Belov {\em et~al.}, ``{The {\ensuremath{\nu}}GeN experiment at the Kalinin
  Nuclear Power Plant},''
  \href{http://dx.doi.org/10.1088/1748-0221/10/12/P12011}{{\em JINST}
  {\bfseries 10} (2015) P12011}.

\bibitem{conus}
Private communication with conus collaboration.

\bibitem{Billard:2016giu}
J.~Billard {\em et~al.}, ``{Coherent Neutrino Scattering with Low Temperature
  Bolometers at Chooz Reactor Complex},''
  \href{http://dx.doi.org/10.1088/1361-6471/aa83d0}{{\em J.Phys.} {\bfseries
  G44} (2017) 105101}, \href{http://arxiv.org/abs/1612.09035}{{\ttfamily
  arXiv:1612.09035 [physics.ins-det]}}.

\bibitem{Strauss:2017cuu}
R.~Strauss {\em et~al.}, ``{The $\nu$-cleus experiment: A gram-scale
  fiducial-volume cryogenic detector for the first detection of coherent
  neutrino-nucleus scattering},''
  \href{http://dx.doi.org/10.1140/epjc/s10052-017-5068-2}{{\em Eur.Phys.J.}
  {\bfseries C77} (2017) 506},
  \href{http://arxiv.org/abs/1704.04320}{{\ttfamily arXiv:1704.04320
  [physics.ins-det]}}.

\bibitem{Barranco:2005ps}
J.~Barranco, O.~Miranda, C.~Moura, and J.~Valle, ``{Constraining non-standard
  interactions in nu(e) e or anti-nu(e) e scattering},''
  \href{http://dx.doi.org/10.1103/PhysRevD.73.113001}{{\em Phys.Rev.}
  {\bfseries D73} (2006) 113001}.

\bibitem{Friedland:2004ah}
A.~Friedland, C.~Lunardini, and M.~Maltoni, ``{Atmospheric neutrinos as probes
  of neutrino-matter interactions},''
  \href{http://dx.doi.org/10.1103/PhysRevD.70.111301}{{\em Phys.Rev.}
  {\bfseries D70} (2004) 111301}.

\bibitem{Friedland:2004pp}
A.~Friedland, C.~Lunardini, and C.~Pena-Garay, ``{Solar neutrinos as probes of
  neutrino matter interactions},''
  \href{http://dx.doi.org/10.1016/j.physletb.2004.05.047}{{\em Phys.Lett.}
  {\bfseries B594} (2004) 347}.

\bibitem{Friedland:2005vy}
A.~Friedland and C.~Lunardini, ``{A Test of tau neutrino interactions with
  atmospheric neutrinos and K2K},''
  \href{http://dx.doi.org/10.1103/PhysRevD.72.053009}{{\em Phys.Rev.}
  {\bfseries D72} (2005) 053009}.

\bibitem{Miranda:2015dra}
O.~Miranda and H.~Nunokawa, ``{Non standard neutrino interactions: current
  status and future prospects},''
  \href{http://dx.doi.org/10.1088/1367-2630/17/9/095002}{{\em New J.Phys.}
  {\bfseries 17} (2015) 095002},
  \href{http://arxiv.org/abs/1505.06254}{{\ttfamily arXiv:1505.06254
  [hep-ph]}}.

\bibitem{Farzan:2017xzy}
Y.~Farzan and M.~Tortola, ``{Neutrino oscillations and Non-Standard
  Interactions},'' \href{http://dx.doi.org/10.3389/fphy.2018.00010}{{\em
  Front.in Phys.} {\bfseries 6} (2018) 10},
  \href{http://arxiv.org/abs/1710.09360}{{\ttfamily arXiv:1710.09360
  [hep-ph]}}.

\bibitem{Amanik:2004vm}
P.~S. Amanik, G.~M. Fuller, and B.~Grinstein, ``{Flavor changing supersymmetry
  interactions in a supernova},''
  \href{http://dx.doi.org/10.1016/j.astropartphys.2005.06.004}{{\em
  Astropart.Phys.} {\bfseries 24} (2005) 160--182}.

\bibitem{EstebanPretel:2009is}
A.~Esteban-Pretel, R.~Tomas, and J.~Valle, ``{Interplay between collective
  effects and non-standard neutrino interactions of supernova neutrinos},''
  \href{http://dx.doi.org/10.1103/PhysRevD.81.063003}{{\em Phys.Rev.}
  {\bfseries D81} (2010) 063003},
  \href{http://arxiv.org/abs/0909.2196}{{\ttfamily arXiv:0909.2196 [hep-ph]}}.

\bibitem{Akimov:2019wtg}
D.~{\relax Yu}. Akimov, V.~A. Belov, A.~Bolozdynya, {\relax Yu}.~V. Efremenko,
  A.~M. Konovalov, A.~V. Kumpan, D.~G. Rudik, V.~V. Sosnovtsev, A.~V. Khromov,
  and A.~V. Shakirov, ``{Coherent elastic neutrino scattering on atomic
  nucleus: recently discovered type of low-energy neutrino interaction},''
  \href{http://dx.doi.org/10.3367/UFNe.2018.05.038356,
  10.3367/UFNr.2018.05.038356}{{\em Phys. Usp.} {\bfseries 62} no.~2, (2019)
  166--178}.

\bibitem{Papoulias:2015vxa}
D.~Papoulias and T.~Kosmas, ``{Standard and Nonstandard Neutrino-Nucleus
  Reactions Cross Sections and Event Rates to Neutrino Detection
  Experiments},'' \href{http://dx.doi.org/10.1155/2015/763648}{{\em Adv.High
  Energy Phys.} {\bfseries 2015} (2015) 763648},
  \href{http://arxiv.org/abs/1502.02928}{{\ttfamily arXiv:1502.02928
  [nucl-th]}}.

\bibitem{Bednyakov:2018mjd}
V.~A. Bednyakov and D.~V. Naumov, ``{Coherency and incoherency in
  neutrino-nucleus elastic and inelastic scattering},''
  \href{http://dx.doi.org/10.1103/PhysRevD.98.053004}{{\em Phys.Rev.}
  {\bfseries D98} (2018) 053004},
  \href{http://arxiv.org/abs/1806.08768}{{\ttfamily arXiv:1806.08768
  [hep-ph]}}.

\bibitem{Bednyakov:2019dbl}
V.~A. Bednyakov and D.~V. Naumov, ``{On coherent neutrino and antineutrino
  scattering off nuclei},'' \href{http://arxiv.org/abs/1904.03119}{{\ttfamily
  arXiv:1904.03119 [hep-ph]}}.

\bibitem{Almosly:2019han}
W.~Almosly, B.~G. Carlsson, J.~Suhonen, and E.~Ydrefors, ``{Neutral-current
  supernova-neutrino cross sections for $^{204,206,208}$Pb calculated by Skyrme
  quasiparticle random-phase approximation},''
  \href{http://dx.doi.org/10.1103/PhysRevC.99.055801}{{\em Phys. Rev.}
  {\bfseries C99} no.~5, (2019) 055801}.

\bibitem{Chasioti:2009fby}
V.~Chasioti and T.~Kosmas, ``{A unified formalism for the basic nuclear matrix
  elements in semi-leptonic processes},''
  \href{http://dx.doi.org/10.1016/j.nuclphysa.2009.08.009}{{\em Nucl.Phys.}
  {\bfseries A829} (2009) 234--252}.

\bibitem{DeJager:1987qc}
H.~De~Vries, C.~W. De~Jager, and C.~De~Vries, ``{Nuclear charge and
  magnetization density distribution parameters from elastic electron
  scattering},'' \href{http://dx.doi.org/10.1016/0092-640X(87)90013-1}{{\em
  Atom. Data Nucl. Data Tabl.} {\bfseries 36} (1987) 495--536}.

\bibitem{Donnelly:1976fs}
T.~Donnelly and J.~Walecka, ``{Semileptonic Weak and Electromagnetic
  Interactions with Nuclei: Isoelastic Processes},''
  \href{http://dx.doi.org/10.1016/0375-9474(76)90209-8}{{\em Nucl.Phys.}
  {\bfseries A274} (1976) 368--412}.

\bibitem{Donnelly:1978tz}
T.~Donnelly and R.~Peccei, ``{Neutral Current Effects in Nuclei},''
  \href{http://dx.doi.org/10.1016/0370-1573(79)90010-3}{{\em Phys.Rept.}
  {\bfseries 50} (1979) 1}.

\bibitem{Lindner:2016wff}
M.~Lindner, W.~Rodejohann, and X.-J. Xu, ``{Coherent Neutrino-Nucleus
  Scattering and new Neutrino Interactions},''
  \href{http://dx.doi.org/10.1007/JHEP03(2017)097}{{\em JHEP} {\bfseries 1703}
  (2017) 097}, \href{http://arxiv.org/abs/1612.04150}{{\ttfamily
  arXiv:1612.04150 [hep-ph]}}.

\bibitem{Barranco:2005yy}
J.~Barranco, O.~Miranda, and T.~Rashba, ``{Probing new physics with coherent
  neutrino scattering off nuclei},''
  \href{http://dx.doi.org/10.1088/1126-6708/2005/12/021}{{\em JHEP} {\bfseries
  0512} (2005) 021}.

\bibitem{Tanabashi:2018oca}
{\bfseries Particle Data Group} Collaboration, M.~Tanabashi {\em et~al.},
  ``{Review of Particle Physics},''
  \href{http://dx.doi.org/10.1103/PhysRevD.98.030001}{{\em Phys.Rev.}
  {\bfseries D98} (2018) 030001}.

\bibitem{Papoulias:phd}
D.~K. Papoulias,
  \href{http://dx.doi.org/http://hdl.handle.net/10442/hedi/43131}{{\em {Exotic
  lepton flavour violating processes in the field of nucleus}}}.
\newblock PhD thesis, Ioannina U., 2016.

\bibitem{Angeli:2013epw}
I.~Angeli and K.~Marinova, ``{Table of experimental nuclear ground state charge
  radii: An update},'' \href{http://dx.doi.org/10.1016/j.adt.2011.12.006}{{\em
  Atom.Data Nucl.Data Tabl.} {\bfseries 99} (2013) 69--95}.

\bibitem{Kortelainen:2006rd}
M.~Kortelainen, J.~Suhonen, J.~Toivanen, and T.~Kosmas, ``{Event rates for CDM
  detectors from large-scale shell-model calculations},''
  \href{http://dx.doi.org/10.1016/j.physletb.2005.10.057}{{\em Phys.Lett.}
  {\bfseries B632} (2006) 226--232}.

\bibitem{Toivanen:2009zza}
P.~Toivanen, M.~Kortelainen, J.~Suhonen, and J.~Toivanen, ``{Large-scale
  shell-model calculations of elastic and inelastic scattering rates of
  lightest supersymmetric particles (LSP) on I-127, Xe-129, Xe-131, and Cs-133
  nuclei},'' \href{http://dx.doi.org/10.1103/PhysRevC.79.044302}{{\em
  Phys.Rev.} {\bfseries C79} (2009) 044302}.

\bibitem{Papoulias:2013gha}
D.~Papoulias and T.~Kosmas, ``{Nuclear aspects of neutral current non-standard
  $\nu$-nucleus reactions and the role of the exotic $\mu^-\to e^{-}$
  transitions experimental limits},''
  \href{http://dx.doi.org/10.1016/j.physletb.2013.12.028}{{\em Phys.Lett.}
  {\bfseries B728} (2014) 482--488},
  \href{http://arxiv.org/abs/1312.2460}{{\ttfamily arXiv:1312.2460 [nucl-th]}}.

\bibitem{Pirinen:2018gsd}
P.~Pirinen, J.~Suhonen, and E.~Ydrefors, ``{Neutral-current neutrino-nucleus
  scattering off Xe isotopes},''
  \href{http://dx.doi.org/10.1155/2018/9163586}{{\em Adv.High Energy Phys.}
  {\bfseries 2018} (2018) 9163586},
  \href{http://arxiv.org/abs/1804.08995}{{\ttfamily arXiv:1804.08995
  [nucl-th]}}.

\bibitem{Papoulias:2019txv}
D.~K. Papoulias, ``{COHERENT constraints after the Chicago-3 quenching factor
  measurement},'' \href{http://arxiv.org/abs/1907.11644}{{\ttfamily
  arXiv:1907.11644 [hep-ph]}}.

\bibitem{Kosmas:1992yxv}
T.~Kosmas and J.~Vergados, ``{Nuclear densities with fractional occupation
  probabilities of the states},''
  \href{http://dx.doi.org/10.1016/0375-9474(92)90246-G}{{\em Nucl.Phys.}
  {\bfseries A536} (1992) 72--86}.

\bibitem{Amanik:2009zz}
P.~Amanik and G.~McLaughlin, ``{Nuclear neutron form factor from neutrino
  nucleus coherent elastic scattering},''
  \href{http://dx.doi.org/10.1088/0954-3899/36/1/015105}{{\em J.Phys.}
  {\bfseries G36} (2009) 015105}.

\bibitem{Kosmas:1988tz}
T.~S. Kosmas and J.~D. Vergados, ``{Nuclear Matrix Elements for the Coherent
  $\mu - e$ Conversion Process},''
\href{http://dx.doi.org/10.1016/0370-2693(88)91341-X}{{\em Phys. Lett.}
  {\bfseries B215} (1988) 460--464}.

\bibitem{Kosmas:1990tc}
T.~Kosmas and J.~Vergados, ``{Study of the flavor violating (mu-, e-)
  conversion in nuclei},''
  \href{http://dx.doi.org/10.1016/0375-9474(90)90353-N}{{\em Nucl.Phys.}
  {\bfseries A510} (1990) 641--670}.

\bibitem{Helm:1956zz}
R.~H. Helm, ``{Inelastic and Elastic Scattering of 187-Mev Electrons from
  Selected Even-Even Nuclei},''
  \href{http://dx.doi.org/10.1103/PhysRev.104.1466}{{\em Phys.Rev.} {\bfseries
  104} (1956) 1466--1475}.

\bibitem{Piekarewicz:2016vbn}
J.~Piekarewicz, A.~Linero, P.~Giuliani, and E.~Chicken, ``{Power of two:
  Assessing the impact of a second measurement of the weak-charge form factor
  of $^{208}$Pb},'' \href{http://dx.doi.org/10.1103/PhysRevC.94.034316}{{\em
  Phys.Rev.} {\bfseries C94} (2016) 034316},
  \href{http://arxiv.org/abs/1604.07799}{{\ttfamily arXiv:1604.07799
  [nucl-th]}}.

\bibitem{Fricke:1995zz}
G.~Fricke {\em et~al.}, ``{Nuclear Ground State Charge Radii from
  Electromagnetic Interactions},''
  \href{http://dx.doi.org/10.1006/adnd.1995.1007}{{\em Atom.Data Nucl.Data
  Tabl.} {\bfseries 60} (1995) 177--285}.

\bibitem{Lewin:1995rx}
J.~Lewin and P.~Smith, ``{Review of mathematics, numerical factors, and
  corrections for dark matter experiments based on elastic nuclear recoil},''
  \href{http://dx.doi.org/10.1016/S0927-6505(96)00047-3}{{\em Astropart.Phys.}
  {\bfseries 6} (1996) 87--112}.

\bibitem{Sprung:1997}
D.~W.~L. Sprung and J.~Martorell, ``{The symmetrized Fermi function and its
  transforms},'' \href{http://dx.doi.org/10.1088/0305-4470/30/18/026}{{\em
  Journal of Physics A: Mathematical and General} {\bfseries 30} (1997)
  6525--6534}.

\bibitem{Klein:1999qj}
S.~Klein and J.~Nystrand, ``{Exclusive vector meson production in relativistic
  heavy ion collisions},''
  \href{http://dx.doi.org/10.1103/PhysRevC.60.014903}{{\em Phys.Rev.}
  {\bfseries C60} (1999) 014903}.

\bibitem{Akimov:2019rhz}
{\bfseries COHERENT} Collaboration, D.~Akimov {\em et~al.}, ``{First Constraint
  on Coherent Elastic Neutrino-Nucleus Scattering in Argon},''
\href{http://arxiv.org/abs/1909.05913}{{\ttfamily arXiv:1909.05913 [hep-ex]}}.

\bibitem{ALIANE2019162784}
A.~Aliane, , {\em et~al.}, ``First test of a li2wo4(mo) bolometric detector for
  the measurement of coherent neutrino-nucleus scattering,''
  \href{http://dx.doi.org/https://doi.org/10.1016/j.nima.2019.162784}{{\em
  Nuclear Instruments and Methods in Physics Research Section A: Accelerators,
  Spectrometers, Detectors and Associated Equipment} (2019) 162784}.
  \url{http://www.sciencedirect.com/science/article/pii/S0168900219312306}.

\bibitem{Bellenghi:2019vtc}
C.~Bellenghi, D.~Chiesa, L.~Di~Noto, M.~Pallavicini, E.~Previtali, and
  M.~Vignati, ``{Coherent elastic nuclear scattering of$^{51}$ Cr neutrinos},''
  \href{http://dx.doi.org/10.1140/epjc/s10052-019-7240-3}{{\em Eur. Phys. J.}
  {\bfseries C79} no.~9, (2019) 727},
\href{http://arxiv.org/abs/1905.10611}{{\ttfamily arXiv:1905.10611
  [physics.ins-det]}}.

\bibitem{Baxter:2019mcx}
D.~Baxter {\em et~al.}, ``{Coherent Elastic Neutrino-Nucleus Scattering at the
  European Spallation Source},''
\href{http://arxiv.org/abs/1911.00762}{{\ttfamily arXiv:1911.00762
  [physics.ins-det]}}.

\bibitem{Rich:2017lzd}
G.~C. Rich, {\em {Measurement of Low-Energy Nuclear-Recoil Quenching Factors in
  CsI[Na] and Statistical Analysis of the First Observation of Coherent,
  Elastic Neutrino-Nucleus Scattering}}.
\newblock PhD thesis, North Carolina U.,
2017.
\newblock

\bibitem{Akimov:2018ghi}
{\bfseries COHERENT} Collaboration, D.~Akimov {\em et~al.}, ``{COHERENT 2018 at
  the Spallation Neutron Source},''
  \href{http://arxiv.org/abs/1803.09183}{{\ttfamily arXiv:1803.09183
  [physics.ins-det]}}.

\bibitem{CCM}
R.~Thornton, ``{Searching for Muon Neutrino Disappearance at LSND Neutrino
  Energies with CCM}.''
  \url{https://indico.cern.ch/event/782953/contributions/3444547/}, 2019.

\bibitem{Hakenmuller:2019ecb}
J.~Hakenmuller {\em et~al.}, ``{Neutron-induced background in the CONUS
  experiment},'' \href{http://dx.doi.org/10.1140/epjc/s10052-019-7160-2}{{\em
  Eur. Phys. J.} {\bfseries C79} no.~8, (2019) 699},
\href{http://arxiv.org/abs/1903.09269}{{\ttfamily arXiv:1903.09269
  [physics.ins-det]}}.

\bibitem{Aguilar-Arevalo:2019jlr}
{\bfseries CONNIE} Collaboration, A.~Aguilar-Arevalo {\em et~al.}, ``{Exploring
  Low-Energy Neutrino Physics with the Coherent Neutrino Nucleus Interaction
  Experiment (CONNIE)},'' \href{http://arxiv.org/abs/1906.02200}{{\ttfamily
  arXiv:1906.02200 [physics.ins-det]}}.

\bibitem{Angloher:2019flc}
{\bfseries NUCLEUS} Collaboration, G.~Angloher {\em et~al.}, ``{Exploring
  CE$\nu$NS with NUCLEUS at the Chooz Nuclear Power Plant},''
  \href{http://arxiv.org/abs/1905.10258}{{\ttfamily arXiv:1905.10258
  [physics.ins-det]}}.

\bibitem{Akimov:2017hee}
D.~{\relax Yu}. Akimov {\em et~al.}, ``{Status of the RED-100 experiment},''
  \href{http://dx.doi.org/10.1088/1748-0221/12/06/C06018}{{\em JINST}
  {\bfseries 12} no.~06, (2017) C06018}.

\bibitem{Wong:2005vg}
H.~T. Wong, H.-B. Li, J.~Li, Q.~Yue, and Z.-Y. Zhou,
  \href{http://dx.doi.org/10.1088/1742-6596/39/1/064}{``{Research program
  towards observation of neutrino-nucleus coherent scattering},''} vol.~39,
  pp.~266--268.
\newblock 2006.

\bibitem{Cadeddu:2018izq}
M.~Cadeddu and F.~Dordei, ``{Reinterpreting the weak mixing angle from atomic
  parity violation in view of the Cs neutron rms radius measurement from
  COHERENT},'' \href{http://dx.doi.org/10.1103/PhysRevD.99.033010}{{\em
  Phys.Rev.} {\bfseries D99} (2019) 033010},
  \href{http://arxiv.org/abs/1808.10202}{{\ttfamily arXiv:1808.10202
  [hep-ph]}}.

\bibitem{Canas:2018rng}
B.~Ca{\~n}as, E.~Garc{\'e}s, O.~Miranda, and A.~Parada, ``{Future perspectives
  for a weak mixing angle measurement in coherent elastic neutrino nucleus
  scattering experiments},''
  \href{http://dx.doi.org/10.1016/j.physletb.2018.07.049}{{\em Phys.Lett.}
  {\bfseries B784} (2018) 159--162},
  \href{http://arxiv.org/abs/1806.01310}{{\ttfamily arXiv:1806.01310
  [hep-ph]}}.

\bibitem{Collar:2019ihs}
J.~Collar, A.~Kavner, and C.~Lewis, ``{Response of CsI[Na] to Nuclear Recoils:
  Impact on Coherent Elastic Neutrino-Nucleus Scattering (CE$\nu$NS)},''
  \href{http://arxiv.org/abs/1907.04828}{{\ttfamily arXiv:1907.04828
  [nucl-ex]}}.

\bibitem{Patton:2012jr}
K.~Patton, J.~Engel, G.~C. McLaughlin, and N.~Schunck, ``{Neutrino-nucleus
  coherent scattering as a probe of neutron density distributions},''
  \href{http://dx.doi.org/10.1103/PhysRevC.86.024612}{{\em Phys.Rev.}
  {\bfseries C86} (2012) 024612},
  \href{http://arxiv.org/abs/1207.0693}{{\ttfamily arXiv:1207.0693 [nucl-th]}}.

\bibitem{Horowitz:2012tj}
C.~Horowitz {\em et~al.}, ``{Weak charge form factor and radius of 208Pb
  through parity violation in electron scattering},''
  \href{http://dx.doi.org/10.1103/PhysRevC.85.032501}{{\em Phys.Rev.}
  {\bfseries C85} (2012) 032501},
  \href{http://arxiv.org/abs/1202.1468}{{\ttfamily arXiv:1202.1468 [nucl-ex]}}.

\bibitem{Cadeddu:2019eta}
M.~Cadeddu, F.~Dordei, C.~Giunti, Y.~F. Li, and Y.~Y. Zhang, ``{Neutrino,
  Electroweak and Nuclear Physics from COHERENT Elastic Neutrino-Nucleus
  Scattering with Refined Quenching Factor},''
  \href{http://arxiv.org/abs/1908.06045}{{\ttfamily arXiv:1908.06045
  [hep-ph]}}.

\bibitem{Dev:2019anc}
P.~S. Bhupal~Dev {\em et~al.}, ``{Neutrino Non-Standard Interactions: A Status
  Report},'' \href{http://arxiv.org/abs/1907.00991}{{\ttfamily arXiv:1907.00991
  [hep-ph]}}.
\url{http://lss.fnal.gov/archive/2019/conf/fermilab-conf-19-299-t.pdf}.

\bibitem{Babu:2019mfe}
K.~S. Babu, P.~S.~B. Dev, S.~Jana, and A.~Thapa, ``{Non-Standard Interactions
  in Radiative Neutrino Mass Models},''
\href{http://arxiv.org/abs/1907.09498}{{\ttfamily arXiv:1907.09498 [hep-ph]}}.

\bibitem{Deppisch:2005zm}
F.~Deppisch, T.~Kosmas, and J.~Valle, ``{Enhanced mu- - e- conversion in nuclei
  in the inverse seesaw model},''
  \href{http://dx.doi.org/10.1016/j.nuclphysb.2006.06.032}{{\em Nucl.Phys.}
  {\bfseries B752} (2006) 80--92}.

\bibitem{Malinsky:2008qn}
M.~Malinsky, T.~Ohlsson, and H.~Zhang, ``{Non-Standard Neutrino Interactions
  from a Triplet Seesaw Model},''
  \href{http://dx.doi.org/10.1103/PhysRevD.79.011301}{{\em Phys. Rev.}
  {\bfseries D79} (2009) 011301},
\href{http://arxiv.org/abs/0811.3346}{{\ttfamily arXiv:0811.3346 [hep-ph]}}.

\bibitem{Forero:2011pc}
D.~Forero, S.~Morisi, M.~Tortola, and J.~Valle, ``{Lepton flavor violation and
  non-unitary lepton mixing in low-scale type-I seesaw},''
  \href{http://dx.doi.org/10.1007/JHEP09(2011)142}{{\em JHEP} {\bfseries 1109}
  (2011) 142}, \href{http://arxiv.org/abs/1107.6009}{{\ttfamily arXiv:1107.6009
  [hep-ph]}}.

\bibitem{Das:2012ii}
S.~Das, F.~Deppisch, O.~Kittel, and J.~Valle, ``{Heavy Neutrinos and Lepton
  Flavour Violation in Left-Right Symmetric Models at the LHC},''
  \href{http://dx.doi.org/10.1103/PhysRevD.86.055006}{{\em Phys.Rev.}
  {\bfseries D86} (2012) 055006},
  \href{http://arxiv.org/abs/1206.0256}{{\ttfamily arXiv:1206.0256 [hep-ph]}}.

\bibitem{Petrov:2013vka}
A.~A. Petrov and D.~V. Zhuridov, ``{Lepton flavor-violating transitions in
  effective field theory and gluonic operators},''
  \href{http://dx.doi.org/10.1103/PhysRevD.89.033005}{{\em Phys.Rev.}
  {\bfseries D89} (2014) 033005},
  \href{http://arxiv.org/abs/1308.6561}{{\ttfamily arXiv:1308.6561 [hep-ph]}}.

\bibitem{Amanik:2006ad}
P.~S. Amanik and G.~M. Fuller, ``{Stellar Collapse Dynamics With Neutrino
  Flavor Changing Neutral Currents},''
  \href{http://dx.doi.org/10.1103/PhysRevD.75.083008}{{\em Phys.Rev.}
  {\bfseries D75} (2007) 083008}.

\bibitem{Scholberg:2005qs}
K.~Scholberg, ``{Prospects for measuring coherent neutrino-nucleus elastic
  scattering at a stopped-pion neutrino source},''
  \href{http://dx.doi.org/10.1103/PhysRevD.73.033005}{{\em Phys.Rev.}
  {\bfseries D73} (2006) 033005}.

\bibitem{AristizabalSierra:2018eqm}
D.~Aristizabal~Sierra, V.~De~Romeri, and N.~Rojas, ``{COHERENT analysis of
  neutrino generalized interactions},''
  \href{http://dx.doi.org/10.1103/PhysRevD.98.075018}{{\em Phys.Rev.}
  {\bfseries D98} (2018) 075018},
  \href{http://arxiv.org/abs/1806.07424}{{\ttfamily arXiv:1806.07424
  [hep-ph]}}.

\bibitem{Altmannshofer:2018xyo}
W.~Altmannshofer, M.~Tammaro, and J.~Zupan, ``{Non-standard neutrino
  interactions and low energy experiments},''
  \href{http://arxiv.org/abs/1812.02778}{{\ttfamily arXiv:1812.02778
  [hep-ph]}}.

\bibitem{Bischer:2019ttk}
I.~Bischer and W.~Rodejohann, ``{General Neutrino Interactions from an
  Effective Field Theory Perspective},''
  \href{http://dx.doi.org/10.1016/j.nuclphysb.2019.114746}{{\em Nucl. Phys.}
  {\bfseries B947} (2019) 114746},
  \href{http://arxiv.org/abs/1905.08699}{{\ttfamily arXiv:1905.08699
  [hep-ph]}}.

\bibitem{Davidson:2019iqh}
S.~Davidson and M.~Gorbahn, ``{Charged lepton flavour change and Non-Standard
  neutrino Interactions},''
\href{http://arxiv.org/abs/1909.07406}{{\ttfamily arXiv:1909.07406 [hep-ph]}}.

\bibitem{Billard:2018jnl}
J.~Billard, J.~Johnston, and B.~J. Kavanagh, ``{Prospects for exploring New
  Physics in Coherent Elastic Neutrino-Nucleus Scattering},''
  \href{http://dx.doi.org/10.1088/1475-7516/2018/11/016}{{\em JCAP} {\bfseries
  1811} (2018) 016}, \href{http://arxiv.org/abs/1805.01798}{{\ttfamily
  arXiv:1805.01798 [hep-ph]}}.

\bibitem{Miranda:2019skf}
O.~G. Miranda, G.~Sanchez~Garcia, and O.~Sanders, ``{Coherent elastic
  neutrino-nucleus scattering as a precision test for the Standard Model and
  beyond: the COHERENT proposal case},''
  \href{http://dx.doi.org/10.1155/2019/3902819}{{\em Adv. High Energy Phys.}
  {\bfseries 2019} (2019) 3902819},
\href{http://arxiv.org/abs/1902.09036}{{\ttfamily arXiv:1902.09036 [hep-ph]}}.

\bibitem{Khan:2019cvi}
A.~N. Khan and W.~Rodejohann, ``{New physics from COHERENT data with improved
  Quenching Factors},'' \href{http://arxiv.org/abs/1907.12444}{{\ttfamily
  arXiv:1907.12444 [hep-ph]}}.

\bibitem{Giunti:2019xpr}
C.~Giunti, ``{General COHERENT Constraints on Neutrino Non-Standard
  Interactions},'' \href{http://arxiv.org/abs/1909.00466}{{\ttfamily
  arXiv:1909.00466 [hep-ph]}}.

\bibitem{Barranco:2011wx}
J.~Barranco, A.~Bolanos, E.~Garces, O.~Miranda, and T.~Rashba, ``{Tensorial NSI
  and Unparticle physics in neutrino scattering},''
  \href{http://dx.doi.org/10.1142/S0217751X12501473}{{\em Int.J.Mod.Phys.}
  {\bfseries A27} (2012) 1250147},
  \href{http://arxiv.org/abs/1108.1220}{{\ttfamily arXiv:1108.1220 [hep-ph]}}.

\bibitem{Healey:2013vka}
K.~J. Healey, A.~A. Petrov, and D.~Zhuridov, ``{Nonstandard neutrino
  interactions and transition magnetic moments},''
  \href{http://dx.doi.org/10.1103/PhysRevD.89.059904}{{\em Phys.Rev.}
  {\bfseries D87} (2013) 117301},
  \href{http://arxiv.org/abs/1305.0584}{{\ttfamily arXiv:1305.0584 [hep-ph]}}.

\bibitem{Papoulias:2015iga}
D.~Papoulias and T.~Kosmas, ``{Neutrino transition magnetic moments within the
  non-standard neutrino{\textendash}nucleus interactions},''
  \href{http://dx.doi.org/10.1016/j.physletb.2015.06.039}{{\em Phys.Lett.}
  {\bfseries B747} (2015) 454--459},
  \href{http://arxiv.org/abs/1506.05406}{{\ttfamily arXiv:1506.05406
  [hep-ph]}}.

\bibitem{Datta:2018xty}
A.~Datta, B.~Dutta, S.~Liao, D.~Marfatia, and L.~E. Strigari, ``{Neutrino
  scattering and B anomalies from hidden sector portals},''
  \href{http://dx.doi.org/10.1007/JHEP01(2019)091}{{\em JHEP} {\bfseries 01}
  (2019) 091},
\href{http://arxiv.org/abs/1808.02611}{{\ttfamily arXiv:1808.02611 [hep-ph]}}.

\bibitem{Dalchenko:2017shg}
M.~Abdullah {\em et~al.}, ``{Bottom-quark fusion processes at the LHC for
  probing $Z^\prime$ models and $B$ -meson decay anomalies},''
  \href{http://dx.doi.org/10.1103/PhysRevD.97.075035}{{\em Phys.Rev.}
  {\bfseries D97} (2018) 075035},
  \href{http://arxiv.org/abs/1707.07016}{{\ttfamily arXiv:1707.07016
  [hep-ph]}}.

\bibitem{Bertuzzo:2017tuf}
E.~Bertuzzo, F.~F. Deppisch, S.~Kulkarni, Y.~F. Perez~Gonzalez, and
  R.~Zukanovich~Funchal, ``{Dark Matter and Exotic Neutrino Interactions in
  Direct Detection Searches},''
  \href{http://dx.doi.org/10.1007/JHEP04(2017)073}{{\em JHEP} {\bfseries 1704}
  (2017) 073}, \href{http://arxiv.org/abs/1701.07443}{{\ttfamily
  arXiv:1701.07443 [hep-ph]}}.

\bibitem{Cerdeno:2016sfi}
D.~G. Cerde{\~n}o, M.~Fairbairn, T.~Jubb, P.~A.~N. Machado, A.~C. Vincent, and
  C.~B{\oe}hm, ``{Physics from solar neutrinos in dark matter direct detection
  experiments},'' \href{http://dx.doi.org/10.1007/JHEP05(2016)118}{{\em JHEP}
  {\bfseries 1605} (2016) 118},
  \href{http://arxiv.org/abs/1604.01025}{{\ttfamily arXiv:1604.01025
  [hep-ph]}}.

\bibitem{Rodejohann:2017vup}
W.~Rodejohann, X.-J. Xu, and C.~E. Yaguna, ``{Distinguishing between Dirac and
  Majorana neutrinos in the presence of general interactions},''
  \href{http://dx.doi.org/10.1007/JHEP05(2017)024}{{\em JHEP} {\bfseries 05}
  (2017) 024},
\href{http://arxiv.org/abs/1702.05721}{{\ttfamily arXiv:1702.05721 [hep-ph]}}.

\bibitem{AristizabalSierra:2019ykk}
D.~Aristizabal~Sierra, B.~Dutta, S.~Liao, and L.~E. Strigari, ``{Coherent
  elastic neutrino-nucleus scattering in multi-ton scale dark matter
  experiments: Classification of vector and scalar interactions new physics
  signals},''
\href{http://arxiv.org/abs/1910.12437}{{\ttfamily arXiv:1910.12437 [hep-ph]}}.

\bibitem{Shoemaker:2017lzs}
I.~M. Shoemaker, ``{COHERENT search strategy for beyond standard model neutrino
  interactions},'' \href{http://dx.doi.org/10.1103/PhysRevD.95.115028}{{\em
  Phys.Rev.} {\bfseries D95} (2017) 115028},
  \href{http://arxiv.org/abs/1703.05774}{{\ttfamily arXiv:1703.05774
  [hep-ph]}}.

\bibitem{Aguilar-Arevalo:2019zme}
{\bfseries CONNIE} Collaboration, A.~Aguilar-Arevalo {\em et~al.}, ``{Light
  vector mediator search in the low-energy data of the CONNIE reactor neutrino
  experiment},''
\href{http://arxiv.org/abs/1910.04951}{{\ttfamily arXiv:1910.04951 [hep-ex]}}.

\bibitem{Giunti:2014ixa}
C.~Giunti and A.~Studenikin, ``{Neutrino electromagnetic interactions: a window
  to new physics},'' \href{http://dx.doi.org/10.1103/RevModPhys.87.531}{{\em
  Rev.Mod.Phys.} {\bfseries 87} (2015) 531},
  \href{http://arxiv.org/abs/1403.6344}{{\ttfamily arXiv:1403.6344 [hep-ph]}}.

\bibitem{Schechter:1981hw}
J.~Schechter and J.~Valle, ``{Majorana Neutrinos and Magnetic Fields},''
  \href{http://dx.doi.org/10.1103/PhysRevD.24.1883}{{\em Phys.Rev.} {\bfseries
  D24} (1981) 1883--1889}.

\bibitem{Grimus:2000tq}
W.~Grimus and T.~Schwetz, ``{Elastic neutrino electron scattering of solar
  neutrinos and potential effects of magnetic and electric dipole moments},''
  \href{http://dx.doi.org/10.1016/S0550-3213(00)00451-X}{{\em Nucl.Phys.}
  {\bfseries B587} (2000) 45--66}.

\bibitem{Tortola:2004vh}
M.~Tortola, ``{Constraining neutrino magnetic moment with solar and reactor
  neutrino data},'' \href{http://arxiv.org/abs/hep-ph/0401135}{{\ttfamily
  hep-ph/0401135}}.

\bibitem{Vogel:1989iv}
P.~Vogel and J.~Engel, ``{Neutrino Electromagnetic Form-Factors},''
  \href{http://dx.doi.org/10.1103/PhysRevD.39.3378}{{\em Phys.Rev.} {\bfseries
  D39} (1989) 3378}.

\bibitem{Hirsch:2002uv}
M.~Hirsch, E.~Nardi, and D.~Restrepo, ``{Bounds on the tau and muon neutrino
  vector and axial vector charge radius},''
  \href{http://dx.doi.org/10.1103/PhysRevD.67.033005}{{\em Phys.Rev.}
  {\bfseries D67} (2003) 033005}.

\bibitem{Borexino:2017fbd}
{\bfseries Borexino} Collaboration, M.~Agostini {\em et~al.}, ``{Limiting
  neutrino magnetic moments with Borexino Phase-II solar neutrino data},''
  \href{http://dx.doi.org/10.1103/PhysRevD.96.091103}{{\em Phys. Rev.}
  {\bfseries D96} no.~9, (2017) 091103},
  \href{http://arxiv.org/abs/1707.09355}{{\ttfamily arXiv:1707.09355
  [hep-ex]}}.

\bibitem{Aguilar-Arevalo:2018gpe}
{\bfseries MiniBooNE} Collaboration, A.~A. Aguilar-Arevalo {\em et~al.},
  ``{Significant Excess of ElectronLike Events in the MiniBooNE Short-Baseline
  Neutrino Experiment},''
  \href{http://dx.doi.org/10.1103/PhysRevLett.121.221801}{{\em Phys. Rev.
  Lett.} {\bfseries 121} no.~22, (2018) 221801},
  \href{http://arxiv.org/abs/1805.12028}{{\ttfamily arXiv:1805.12028
  [hep-ex]}}.

\bibitem{Adamson:2017uda}
{\bfseries MINOS+} Collaboration, P.~Adamson {\em et~al.}, ``{Search for
  sterile neutrinos in MINOS and MINOS+ using a two-detector fit},''
  \href{http://dx.doi.org/10.1103/PhysRevLett.122.091803}{{\em Phys. Rev.
  Lett.} {\bfseries 122} no.~9, (2019) 091803},
  \href{http://arxiv.org/abs/1710.06488}{{\ttfamily arXiv:1710.06488
  [hep-ex]}}.

\bibitem{Adey:2018zwh}
{\bfseries Daya Bay} Collaboration, D.~Adey {\em et~al.}, ``{Measurement of the
  Electron Antineutrino Oscillation with 1958 Days of Operation at Daya Bay},''
  \href{http://dx.doi.org/10.1103/PhysRevLett.121.241805}{{\em Phys. Rev.
  Lett.} {\bfseries 121} no.~24, (2018) 241805},
  \href{http://arxiv.org/abs/1809.02261}{{\ttfamily arXiv:1809.02261
  [hep-ex]}}.

\bibitem{An:2015jdp}
{\bfseries JUNO} Collaboration, F.~An {\em et~al.}, ``{Neutrino Physics with
  JUNO},'' \href{http://dx.doi.org/10.1088/0954-3899/43/3/030401}{{\em J.
  Phys.} {\bfseries G43} no.~3, (2016) 030401},
  \href{http://arxiv.org/abs/1507.05613}{{\ttfamily arXiv:1507.05613
  [physics.ins-det]}}.

\bibitem{Ko:2016owz}
{\bfseries NEOS} Collaboration, Y.~Ko {\em et~al.}, ``{Sterile Neutrino Search
  at the NEOS Experiment},''
  \href{http://dx.doi.org/10.1103/PhysRevLett.118.121802}{{\em Phys.Rev.Lett.}
  {\bfseries 118} (2017) 121802},
  \href{http://arxiv.org/abs/1610.05134}{{\ttfamily arXiv:1610.05134
  [hep-ex]}}.

\bibitem{Formaggio:2011jt}
J.~A. Formaggio, E.~Figueroa-Feliciano, and A.~Anderson, ``{Sterile Neutrinos,
  Coherent Scattering and Oscillometry Measurements with Low-temperature
  Bolometers},'' \href{http://dx.doi.org/10.1103/PhysRevD.85.013009}{{\em
  Phys.Rev.} {\bfseries D85} (2012) 013009},
  \href{http://arxiv.org/abs/1107.3512}{{\ttfamily arXiv:1107.3512 [hep-ph]}}.

\bibitem{Anderson:2012pn}
A.~Anderson {\em et~al.}, ``{Measuring Active-to-Sterile Neutrino Oscillations
  with Neutral Current Coherent Neutrino-Nucleus Scattering},''
  \href{http://dx.doi.org/10.1103/PhysRevD.86.013004}{{\em Phys.Rev.}
  {\bfseries D86} (2012) 013004},
  \href{http://arxiv.org/abs/1201.3805}{{\ttfamily arXiv:1201.3805 [hep-ph]}}.

\bibitem{Aguilar:2001ty}
{\bfseries LSND} Collaboration, A.~Aguilar-Arevalo {\em et~al.}, ``{Evidence
  for neutrino oscillations from the observation of anti-neutrino(electron)
  appearance in a anti-neutrino(muon) beam},''
  \href{http://dx.doi.org/10.1103/PhysRevD.64.112007}{{\em Phys.Rev.}
  {\bfseries D64} (2001) 112007}.

\bibitem{Berryman:2019nvr}
J.~M. Berryman, ``{Constraining Sterile Neutrino Cosmology with Terrestrial
  Oscillation Experiments},''
  \href{http://dx.doi.org/10.1103/PhysRevD.100.023540}{{\em Phys. Rev.}
  {\bfseries D100} no.~2, (2019) 023540},
  \href{http://arxiv.org/abs/1905.03254}{{\ttfamily arXiv:1905.03254
  [hep-ph]}}.

\bibitem{Coloma:2017egw}
P.~Coloma, P.~B. Denton, M.~Gonzalez-Garcia, M.~Maltoni, and T.~Schwetz,
  ``{Curtailing the Dark Side in Non-Standard Neutrino Interactions},''
  \href{http://dx.doi.org/10.1007/JHEP04(2017)116}{{\em JHEP} {\bfseries 1704}
  (2017) 116}, \href{http://arxiv.org/abs/1701.04828}{{\ttfamily
  arXiv:1701.04828 [hep-ph]}}.

\bibitem{Cadeddu:2018dux}
M.~Cadeddu, C.~Giunti, K.~Kouzakov, Y.~Li, A.~Studenikin, and Y.~Zhang,
  ``{Neutrino Charge Radii from COHERENT Elastic Neutrino-Nucleus
  Scattering},'' \href{http://dx.doi.org/10.1103/PhysRevD.98.113010}{{\em
  Phys.Rev.} {\bfseries D98} (2018) 113010},
  \href{http://arxiv.org/abs/1810.05606}{{\ttfamily arXiv:1810.05606
  [hep-ph]}}.

\bibitem{Giannaka:2015sta}
P.~Giannaka and T.~Kosmas, ``{Electron capture cross sections for stellar
  nucleosynthesis},'' \href{http://dx.doi.org/10.1155/2015/398796}{{\em
  Adv.High Energy Phys.} {\bfseries 2015} (2015) 398796},
  \href{http://arxiv.org/abs/1502.07225}{{\ttfamily arXiv:1502.07225
  [nucl-th]}}.

\bibitem{Stapleford:2016jgz}
C.~J. Stapleford, D.~J. Väänänen, J.~P. Kneller, G.~C. McLaughlin, and B.~T.
  Shapiro, ``{Nonstandard Neutrino Interactions in Supernovae},''
  \href{http://dx.doi.org/10.1103/PhysRevD.94.093007}{{\em Phys. Rev.}
  {\bfseries D94} no.~9, (2016) 093007},
\href{http://arxiv.org/abs/1605.04903}{{\ttfamily arXiv:1605.04903 [hep-ph]}}.

\bibitem{Amaya:2011sn}
M.~Biassoni and C.~Martinez, ``{Study of supernova {\ensuremath{\nu}} -nucleus
  coherent scattering interactions},''
  \href{http://dx.doi.org/10.1016/j.astropartphys.2012.05.009}{{\em
  Astropart.Phys.} {\bfseries 36} (2012) 151--155},
  \href{http://arxiv.org/abs/1110.3536}{{\ttfamily arXiv:1110.3536
  [astro-ph.HE]}}.

\bibitem{Monroe:2007xp}
J.~Monroe and P.~Fisher, ``{Neutrino Backgrounds to Dark Matter Searches},''
  \href{http://dx.doi.org/10.1103/PhysRevD.76.033007}{{\em Phys.Rev.}
  {\bfseries D76} (2007) 033007},
  \href{http://arxiv.org/abs/0706.3019}{{\ttfamily arXiv:0706.3019
  [astro-ph]}}.

\bibitem{OHare:2016pjy}
C.~A. O'Hare, ``{Dark matter astrophysical uncertainties and the neutrino
  floor},'' \href{http://dx.doi.org/10.1103/PhysRevD.94.063527}{{\em Phys.
  Rev.} {\bfseries D94} no.~6, (2016) 063527},
  \href{http://arxiv.org/abs/1604.03858}{{\ttfamily arXiv:1604.03858
  [astro-ph.CO]}}.

\bibitem{Billard:2013qya}
J.~Billard, L.~Strigari, and E.~Figueroa-Feliciano, ``{Implication of neutrino
  backgrounds on the reach of next generation dark matter direct detection
  experiments},'' \href{http://dx.doi.org/10.1103/PhysRevD.89.023524}{{\em
  Phys. Rev.} {\bfseries D89} no.~2, (2014) 023524},
\href{http://arxiv.org/abs/1307.5458}{{\ttfamily arXiv:1307.5458 [hep-ph]}}.

\bibitem{Farzan:2015doa}
Y.~Farzan, ``{A model for large non-standard interactions of neutrinos leading
  to the LMA-Dark solution},''
  \href{http://dx.doi.org/10.1016/j.physletb.2015.07.015}{{\em Phys.Lett.}
  {\bfseries B748} (2015) 311--315},
  \href{http://arxiv.org/abs/1505.06906}{{\ttfamily arXiv:1505.06906
  [hep-ph]}}.

\bibitem{Kosmas:1994pt}
T.~Kosmas and J.~Vergados, ``{(mu-, e-) conversion: A Symbiosis of particle and
  nuclear physics},''
  \href{http://dx.doi.org/10.1016/0370-1573(95)00041-0}{{\em Phys.Rept.}
  {\bfseries 264} (1996) 251--266}.

\bibitem{Farzan:2016wym}
Y.~Farzan and J.~Heeck, ``{Neutrinophilic nonstandard interactions},''
  \href{http://dx.doi.org/10.1103/PhysRevD.94.053010}{{\em Phys. Rev.}
  {\bfseries D94} no.~5, (2016) 053010},
\href{http://arxiv.org/abs/1607.07616}{{\ttfamily arXiv:1607.07616 [hep-ph]}}.

\bibitem{Petcov:1976ff}
S.~T. Petcov, ``{The Processes mu --> e Gamma, mu --> e e anti-e, Neutrino' -->
  Neutrino gamma in the Weinberg-Salam Model with Neutrino Mixing},'' {\em Sov.
  J. Nucl. Phys.} {\bfseries 25} (1977) 340.
[Erratum: Yad. Fiz.25,1336(1977)].

\bibitem{Kuno:1999jp}
Y.~Kuno and Y.~Okada, ``{Muon decay and physics beyond the standard model},''
  \href{http://dx.doi.org/10.1103/RevModPhys.73.151}{{\em Rev.Mod.Phys.}
  {\bfseries 73} (2001) 151--202}.

\bibitem{Lee:2018wcx}
M.~Lee, ``{COMET Muon Conversion Experiment in J-PARC},''
\href{http://dx.doi.org/10.3389/fphy.2018.00133}{{\em Front.in Phys.}
  {\bfseries 6} (2018) }.

\bibitem{Bernstein:2019fyh}
{\bfseries Mu2e} Collaboration, R.~H. Bernstein, ``{The Mu2e Experiment},''
  \href{http://dx.doi.org/10.3389/fphy.2019.00001}{{\em Front.in Phys.}
  {\bfseries 7} (2019) 1},
\href{http://arxiv.org/abs/1901.11099}{{\ttfamily arXiv:1901.11099
  [physics.ins-det]}}.

\bibitem{Chiang:1993xz}
H.~Chiang, E.~Oset, T.~Kosmas, A.~Faessler, and J.~Vergados, ``{Coherent and
  incoherent (mu-, e-) conversion in nuclei},''
  \href{http://dx.doi.org/10.1016/0375-9474(93)90259-Z}{{\em Nucl.Phys.}
  {\bfseries A559} (1993) 526--542}.

\bibitem{Davidson:2003ha}
S.~Davidson, C.~Pena-Garay, N.~Rius, and A.~Santamaria, ``{Present and future
  bounds on nonstandard neutrino interactions},''
  \href{http://dx.doi.org/10.1088/1126-6708/2003/03/011}{{\em JHEP} {\bfseries
  0303} (2003) 011}.

\bibitem{Kitano:2002mt}
R.~Kitano, M.~Koike, and Y.~Okada, ``{Detailed calculation of lepton flavor
  violating muon electron conversion rate for various nuclei},''
  \href{http://dx.doi.org/10.1103/PhysRevD.66.096002}{{\em Phys.Rev.}
  {\bfseries D66} (2002) 096002}.

\bibitem{Cirigliano:2009bz}
V.~Cirigliano, R.~Kitano, Y.~Okada, and P.~Tuzon, ``{On the model
  discriminating power of mu ---> e conversion in nuclei},''
  \href{http://dx.doi.org/10.1103/PhysRevD.80.013002}{{\em Phys. Rev.}
  {\bfseries D80} (2009) 013002},
\href{http://arxiv.org/abs/0904.0957}{{\ttfamily arXiv:0904.0957 [hep-ph]}}.

\bibitem{Davidson:2018kud}
S.~Davidson, Y.~Kuno, and M.~Yamanaka, ``{Selecting $\mu \to e$ conversion
  targets to distinguish lepton flavour-changing operators},''
  \href{http://dx.doi.org/10.1016/j.physletb.2019.01.042}{{\em Phys. Lett.}
  {\bfseries B790} (2019) 380--388},
\href{http://arxiv.org/abs/1810.01884}{{\ttfamily arXiv:1810.01884 [hep-ph]}}.

\bibitem{Edmonds}
A.~R. Edmonds, {\em {Angular Momentum in Quantum Mechanics}}.
\newblock Princeton University Press, reissue~ed., 1, 1996.
\newblock {ISBN: 9780691025896}.

\end{thebibliography}\endgroup
 \appendix
\section{}

\subsection{Multipole operators}
\label{app:multiple-operators}
The Donnelly-Walecka multipole decomposition method yields a set of eight linearly independent irreducible tensor operators which are typically expressed in terms of the Spherical Bessel functions, $j_l$, and combined with the Spherical Harmonics, $Y^L_M$, or the vector Spherical Harmonics, ${\bf Y}^{(L,1)J}_M$~\cite{Edmonds}

\begin{equation}
\label{proj1M-M}
M^J_M(\kappa{\bf r}) \,=\,\delta_{LJ} j_L(\kappa r)Y^L_M(\hat r),
\end{equation}
\begin{equation}
 {\bf M}_M^{(L1)J}(\kappa {\bf r}) \, =\,
j_L(\kappa r){{\bf Y}}_M^{(L1)J}(\hat r) 
\label{proj2M-M} \, ,
\end{equation}
with
\begin{equation}
{\bf Y}^{(L,1)J}_M (\hat r) \, = \, \sum_{M_L, \lambda} \langle L M_L 1 \lambda
\vert J M\rangle Y^L_{M_L}(\hat r) \mathbf{e}_\lambda \, .
\end{equation}
As a consequence of the V-A structure of electroweak interactions
\begin{equation}
\hat{\mathcal{J}}_\mu  =  {\hat J}_\mu - {\hat J}^5_\mu  =
({\hat \rho}, \hat {\bf J})  - ({\hat \rho}^5, \hat {\bf J}^5) \, ,
\label{V-A-curr}
\end{equation}
four operators are associated to the vector component $\hat{J}_{\lambda} = (\hat{\rho},\bf{\hat{J}})$ and four to the axial-vector component $\hat{J}_{\lambda}^{5} = (\hat{\rho}^{5},\bf{\hat{J}^{5}})$ of the hadronic current
\begin{eqnarray}
\label{M-ten-mult-op}
\hat{\mathcal{ M}}_{JM}(\kappa) &=& \hat M_{JM}^{coul} - \hat M_{JM}^{coul5}
= \int d{\bf r} M^J_M(\kappa {\bf r}) \hat{\mathcal{J}}_0 ({\bf r}),\\
\label{L-ten-mult-op}
\hat{\mathcal{ L}}_{JM}(\kappa) &=& \hat L_{JM} - \hat
L_{JM}^5 = i\int d{\bf r}\left(\frac{1}{\kappa} \boldsymbol{\nabla} M^J_M(\kappa {\bf
r})\right) \cdot \hat{\mathcal{J}} ({\bf r}),\\
\label{Tel-ten-mult-op}
\hat{\mathcal{T}}_{JM}^{el}(\kappa) &=& \hat T_{JM}^{el} -
\hat T_{JM}^{el5} = \int d{\bf r}\left(\frac{1}{q} \boldsymbol{\nabla}
\times {\bf M}^{JJ}_M(\kappa {\bf r})\right) \cdot
\hat{\mathcal{J}}({\bf r}),\\
\label{Tmag-ten-mult-op}
\hat{\mathcal{T}}_{JM}^{mag}(\kappa) &=&  \hat T_{JM}^{mag} -
\hat T_{JM}^{mag5}  =  \int d{\bf r} {\bf M}^{JJ}_M(\kappa {\bf r})
\cdot \hat{\mathcal{J}}({\bf r})\, ,
\end{eqnarray}
where $\kappa = \vert \mathbf{q} \vert$ denotes the 3-momentum transfer.  Note that, the vector component yields the Coulomb $M_{JM}^{coul}$, longitudinal $L_{JM}$, transverse electric $T^{el}_{JM}$ (normal parity $\pi = (-)^{J}$) and transverse magnetic $T^{mag}_{JM}$ (abnormal parity $\pi = (-)^{J+1}$), while regarding the axial-vector component $M_{JM}^{coul5}$, $L_{JM}^{5}$, $T^{el5}_{JM}$ have abnormal parity and $T^{mag5}_{JM}$ has normal parity. The matrix elements of the above operators involve momentum dependence of the nucleon form factors $F_{X}(Q^2)$, $X=1,A,P$ and $\mu^{V}(Q^2)$
\begin{eqnarray}
\hspace{-1.0cm} \hat {M}_{JM}^{coul}(\kappa {\bf r})  &=& F_1^V(Q^2)M^J_M(\kappa {\bf r})\, , 
\label{Coul-op1} \\
\hspace{-1.0cm} \hat{L}_{JM}(\kappa {\bf r})  &=& \frac{q_0}{\kappa}\hat{M}_{JM}^{coul}(\kappa {\bf r})\, , 
\label{Long-op1} \\
\hspace{-1.0cm} \hat{T}_{JM}^{el}(\kappa {\bf r})   &= & 
\frac{\kappa}{m_N}\left[F_1^V(Q^2){\Delta ^{\prime}}^J_{M}(\kappa {\bf r}) +\frac{1}{2}\mu^V(Q^2){\Sigma}^J_{M}(\kappa {\bf r})\right]\, ,
\label{Tr-el-op1} \\
\hspace{-1.0cm} i\hat T_{JM}^{mag}(\kappa {\bf r})  &= & \frac{\kappa}{m_N}\left[F_1^V(Q^2)\Delta ^J_{M}(\kappa {\bf r})
-\frac{1}{2}\mu^V(Q^2){\Sigma^{\prime}}^J_{M}(\kappa {\bf r})\right]\, ,
\label{Tr-ma-op1} \\
\hspace{-1.0cm} i\hat{M}_{JM}^5(\kappa {\bf r})   &= & \frac{\kappa}{m_N}\Biggr[F_A(Q^2)\Omega^J_{M}(\kappa {\bf r}) + \frac{1}{2}\left( F_A(Q^2) +q_0F_P(Q^2) \right){\Sigma^{\prime\prime}}^J_{M}(\kappa {\bf r})\Biggl] \, ,
\label{Coul5-op1} \\
\hspace{-1.0cm} -i\hat{L}_{JM}^5(\kappa {\bf r})  & = & \left[F_A(Q^2)-\frac{\kappa^2}{2 m_N}F_P(Q^2)\right]
{\Sigma^{\prime\prime}}^J_{M}(\kappa {\bf r})\, , 
\label{Long5-op1} \\
\hspace{-1.0cm} -i\hat{T}_{JM}^{el5}(\kappa {\bf r}) & = & F_A(Q^2){\Sigma^{\prime}}^J_{M}(\kappa {\bf r})\, , 
\label{Tr-el5-op1}\\
\hspace{-1.0cm} \hat{T}_{JM}^{mag5}(\kappa {\bf r})  &= & F_A(Q^2){\Sigma}^J_{M}(\kappa {\bf r})\, . 
\label{Tr-ma5-op1} 
\end{eqnarray}
It becomes evident that only seven are linearly independent
\begin{eqnarray}
\label{opM} 
T_1^{JM} \, &\equiv & \, M^J_M(\kappa {\bf r}) \, = \, \delta_{LJ}\,
j_L(\kappa {\bf r})Y^L_M(\hat r),\\
 T_2^{JM} \, &\equiv & \, {\Sigma}^J_{M}(\kappa {\bf r}) \,=\,{{\bf
M}}^{JJ}_{M}\cdot \mbox{\boldmath $ \sigma $}, \\
\label{opSig} 
 T_3^{JM} \, &\equiv & \, {\Sigma^{\prime}}^J_{M}(\kappa {\bf r}) \,= \,
-i\left[\frac{1}{\kappa} \boldsymbol{\nabla} \times {{\bf M}}^{JJ}_{M}(\kappa {\bf
r})\right] \cdot \mbox{\boldmath $ \sigma $},\\
\label{opSigp} 
T_4^{JM} \, &\equiv & \,{\Sigma ''}^J_{M}(\kappa {\bf r}) \, = \,
\Big[\frac{1}{\kappa} \boldsymbol{\nabla} M^J_{M}(\kappa {\bf r})\Big]\cdot\mbox{\boldmath
$\sigma$},\\
\label{opSigpp} 
T_5^{JM} \, &\equiv & \,{\Delta}^J_{M}(\kappa {\bf r}) \,= \, {{\bf
M}}^{JJ}_{M}(\kappa {\bf r}) \cdot\frac{1}{\kappa} \boldsymbol{\nabla},\\
\label{opDel}
 T_6^{JM} \, &\equiv & \,{\Delta ^{ \prime}}^J_{M}(\kappa {\bf r}) \,=
\, -i\Big[\frac{1}{\kappa} \boldsymbol{\nabla} \times {{\bf M}}^{JJ}_{M}(\kappa {\bf r})\Big]
\cdot \frac{1}{\kappa} \boldsymbol{\nabla}, \\
\label{opDelp}
T_7^{JM} \, &\equiv & \, \Omega^J_{M}(\kappa {\bf r}) \, = \,
M^J_{M}(\kappa {\bf r}) \mbox{\boldmath $ \sigma $}
\cdot\frac{1}{\kappa} \boldsymbol{\nabla} \, .
\label{opOm} 
\end{eqnarray}

In the proton-neutron representation,  $T_i^{JM}(\kappa {\bf r})$, $i=1, 2,\cdots, 7$ can be  written in closed form~\cite{Chasioti:2009fby}
\begin{equation}
\label{compact}
\langle j_{1} \vert \vert T_{i}^{J} \vert \vert j_{2} \rangle=e^{-y} y^{\beta/2} \sum_{\mu=0}^{n_{max}} \mathcal{P}_{\mu}^{i,\,J} y^{\mu}, \qquad i=1,\cdots,7.
\end{equation}
%


\subsection{Coefficients for calculating the charge density distribution and form factors in the context of FOP}
\label{app:coeff-FOP}

The coefficients $\theta_\lambda$ of the polynomial $\Phi\left(\vert \mathbf{q} \vert \,b,Z \right)$ are evaluated, as
\begin{equation}
\theta_\lambda = \frac{\sqrt{\pi}}{4^\lambda} \sum_{\begin{subarray}{c}  (n,l)j \\  (2n+l >\lambda) \end{subarray}}^{N_{\mathrm{max}}} \sum_{m=s}^{2n} \frac{(2j+1)n! C_{nl}^m \Lambda_\lambda(m+l,0)(l+m)!}{2 \Gamma(n+l+ \frac{3}{2})} \, .
\end{equation}
In the latter expression, $\Gamma(x)$ denotes the Gamma function while the definition of the index $s$ is 
\begin{equation}
s = \begin{cases} 
   0, & \text{if } \lambda-l \leq 0 \\
   \lambda-l       & \text{if } \lambda-l > 0
  \end{cases}\, ,
\end{equation}
and
\begin{equation}
 \Lambda_k (n,l)= \frac{(-)^k}{k!} \left( \begin{array}{c}
    {n+l+1/2} \\ {n-k}  \end{array} \right), \quad C_{nl}^m = \sum_{k = 0}^m \Lambda_{m-k}(n,l)\Lambda_k(n,l)\, .
\label{Eq:Lambda}    
\end{equation}
The corresponding coefficients $f_{\lambda}$ are written as
\begin{equation}
f_{\lambda}=\sum_{(n,l)j} \frac{\pi^{1/2}(2j+1)n! \,C_{n l}^{\lambda-l} }{2 \Gamma\left( n+l+\frac{3}{2} \right)} \, .
\end{equation}
As a concrete example the coefficients  $\theta_{\lambda}$ and $f_{\lambda}$ for even-even nuclei up to $^{50}$Sn are listed in Table~\ref{table:theta-lambda}.  
%
\begin{table}[t!]
\centering
\begin{tabular}{cccccccccccc}
\toprule
$Z$ ($N$) & $(nl)j$ & \multicolumn{2}{c}{$\lambda = 0$} & \multicolumn{2}{c}{$\lambda = 1$} & \multicolumn{2}{c}{$\lambda = 2$} & \multicolumn{2}{c}{$\lambda = 3$} &  \multicolumn{2}{c}{$\lambda = 4$} \\
\hline
~2 & $0s_{1/2}$ &  $2$ & $(~2)$ &		           &&	&	  &         & & & \\
~6 & $0p_{3/2}$ &  $2$ & $(~6)$ & $\frac{8}{3}$ & $(~-\frac{2}{3})$ &	&	&	   & & & \\
~8 & $0p_{1/2}$ &  $2$ & $(~8)$ & $4$ & $(~-1)$            &           &    & & & & \\
14 & $0d_{5/2}$ &  $2$ & $(14)$ & $4$ & $(~-3)$           & $\frac{8}{5}$ & $(\frac{1}{10})$ & & & & \\
18 & $0d_{3/2}$ &  $2$ & $(18)$ & $4$ & $(-\frac{13}{3})$   & $\frac{8}{3}$ & $(~\frac{1}{6})$ & & & & \\
20 & $1s_{1/2}$ &  $5$ & $(20)$ & $0$ & $(~-5)$          & $4$ & $(~\frac{1}{4})$           & & & &\\
22 & $1p_{1/2}$ &  $5$ & $(22)$ & $\frac{10}{3}$ & $(~-6)$ & $\frac{4}{3}$ & $(\frac{13}{3})$ & $\frac{8}{15}$ & $(-\frac{1}{120})$   & & \\
30 & $0f_{7/2}$ &  $5$ & $(30)$ & $\frac{10}{3}$ & $(-10)$ & $\frac{4}{3}$ & $(~\frac{5}{6})$ & $\frac{8}{7}$ & $(~-\frac{1}{56})$     & & \\
34 & $1p_{3/2}$ &  $5$ & $(34)$ & $10$ & $(-12)$          & $-4$ & $(~\frac{6}{5})$          & $\frac{232}{105}$ & $(-\frac{29}{840})$ & & \\
40 & $0f_{5/2}$ &  $5$ & $(40)$ & $10$ & $(-15)$           & $-4$ & $(~\frac{3}{2})$          & $\frac{8}{3}$ & $(~-\frac{1}{24})$    & & \\
50 & $0h_{9/2}$ &  $5$ & $(50)$ & $10$ & $(-\frac{65}{3})$          & $-4$ & $(~\frac{5}{2})$         & $\frac{8}{3}$ & $(~-\frac{5}{56})$   & $\frac{32}{189}$ & $(\frac{1}{1512})$\\
\botrule
\end{tabular}
\caption{Calculated coefficients $f_\lambda$ $(\theta_\lambda)$ for the determination of the proton/neutron density distributions (nuclear form factors). Table adapted from Ref.~\cite{Papoulias:2015vxa} under the terms of the Creative Commons Attribution 4.0 International license.}
\label{table:theta-lambda}
\end{table}

\end{document}